\documentclass[12pt]{article}

\makeatletter
\def\@maketitle{%
  \newpage
  \null
  \vskip 2em%
  \begin{center}%
  \let\footnote\thanks
    {\LARGE \@title \par}%
    \vskip 0.1em
    {\large
      \lineskip .5em%
      \begin{tabular}[t]{c}%
        \@author
      \end{tabular}\par}%
    \vskip 1em%
    {\large \@date}%
  \end{center}%
  \par
  \vskip 1.5em}
\makeatother

\title{Disclosure under Noisy Information Processing\thanks{We thank Pierpaolo Battigalli, Eti Einhorn, Pingyang Gao, Navin Kartik, Xiumin Martin, Ivan Marinovic, Mario Milone, and Richard Saouma for valuable comments. We also thank our discussants Hui Chen, Ilan Guttman, and Elyashiv Wiedman, and seminar participants at Washington University, the Accounting and Economics Society, the Zurich Accounting Research Conference, and the Carnegie Mellon Accounting Mini Conference.}}

\date{October 19, 2026}

\author{Jeremy Bertomeu\\[0.2em]
\small Washington University\\[-0.2em]\small in St. Louis
\and Edwige Cheynel\\[0.2em]
\small Washington University\\[-0.2em]\small in St. Louis
\and Peicong (Keri) Hu\\[0.2em]
\small The University of Hong Kong}

\usepackage{tikz} 
\usetikzlibrary{trees, calc} 
\usepackage{mathtools}
\usepackage{amsmath,amssymb}
\usepackage{amsfonts}

\usepackage{amsthm}
\usepackage{appendix}
\usepackage{bm}
\usepackage[round]{natbib}
\usepackage{setspace}
\usepackage[hidelinks,pdfauthor={Jeremy Bertomeu, Edwige Cheynel, Peicong (Keri) Hu},pdfcreator={},pdftitle={Disclosure under Noisy Information Processing},pdfsubject={},pdfkeywords={},pdfproducer={}]{hyperref}

\usepackage{graphicx}
\graphicspath{{figures/}}
\usepackage{mathrsfs}
\usepackage{mathabx}
\usepackage[utf8]{inputenc}
\usepackage{csquotes}
\usepackage{thmtools}
\usepackage{thm-restate}
\usepackage{accents}
\usepackage{subcaption}
\usepackage{caption}
\usepackage[margin=1in]{geometry}
\usepackage{pst-node,pst-plot}
\usetikzlibrary{trees,arrows.meta,positioning}
\usepackage{pstricks,pst-node,pst-plot}
\usepackage{float}
\usepackage{needspace}

\makeatletter
\renewcommand\@seccntformat[1]{\csname the#1\endcsname\csname #1@cntsep\endcsname}
\newcommand\section@cntsep{.\hspace{0.5em}}
\newcommand\subsection@cntsep{\hspace{0.5em}}
\newcommand\subsubsection@cntsep{\hspace{0.5em}}
\renewcommand\section{\@startsection{section}{1}{\z@}%
  {-1.1\baselineskip \@plus -1ex \@minus -.2ex}%
  {0.55\baselineskip}%
  {\centering\normalfont\normalsize\bfseries\MakeUppercase}}
\renewcommand\subsection{\@startsection{subsection}{2}{\z@}%
  {-0.9\baselineskip \@plus -1ex \@minus -.2ex}%
  {0.45\baselineskip}%
  {\normalfont\normalsize\bfseries}}
\renewcommand\subsubsection{\@startsection{subsubsection}{3}{\z@}%
  {-0.8\baselineskip \@plus -1ex \@minus -.2ex}%
  {0.4\baselineskip}%
  {\normalfont\normalsize\bfseries\itshape}}
\makeatother

\providecommand{\U}[1]{\protect\rule{.1in}{.1in}}

\newtheorem{corollary}{Corollary}

\newtheorem{definition}{Definition}

\newtheorem{lemma}{Lemma}

\newtheorem{proposition}{Proposition}

\DeclareMathOperator{\E}{\mathbb{E}}
\makeatletter

\newcommand{\Rmnum}[1]{\expandafter\@slowromancap\romannumeral #1@}

\makeatother

\DeclareSymbolFont{arrows}{OMS}{cmsy}{m}{n}
\DeclareMathSymbol{\leftrightarrow}{\mathrel}{arrows}{"24}
\DeclareMathSymbol{\leftarrow}{\mathrel}{arrows}{"20}

\DeclareMathSymbol{\rightarrow}{\mathrel}{arrows}{"21}
\let\to=\rightarrow
\DeclareMathSymbol{\mapstochar}{\mathrel}{arrows}{"37}
\DeclareMathSymbol{\relbardash}{\mathbin}{arrows}{"00}

\usepackage[no-math]{fontspec}
\begin{document}

\maketitle  

  \begin{abstract}
\normalsize

Capital markets process disclosures with noise. Adding this single friction overturns the standard characterization of costly voluntary disclosure. Firms may unravel to full disclosure despite strictly positive disclosure costs and, when unravelling fails, the model generically admits two equilibria rather than one. In the lower-threshold equilibrium, disclosure becomes more frequent as the cost rises and as reports become more accurate, falls with fundamental volatility when it is below the processing noise, but approaches full disclosure again when volatility grows large. These results help explain the mixed empirical evidence on the proprietary cost hypothesis and generate conditional predictions that distinguish high and low disclosure regimes. The model also explains why ESG ratings can be noisy, costly, and widely purchased, while still influencing market prices.
\bigskip
\end{abstract}

\vspace{0.8em}

\vspace{0.4em}

\newpage

\bigskip

           \section{Introduction}
           \vspace{-0.3cm}
           \paragraph{} 
In practice, the information flowing between managers, auditors, and investors is imperfect, incomplete, and open to some degree of interpretation. Forecasts, management commentary, environmental ratings, and even disclosures during earnings forecasts are imprecise or selectively framed, and verification, through audits, due diligence, or external assurance, is costly. Users of accounting information must therefore decide how much to trust management claims, while preparers anticipate the market responses that their messages will attract. These frictions suggest that the processing of information by external parties is a noisy signal of the information voluntarily reported by the firm. We develop a model that captures the tension between strategic communication and processing frictions.

The model features noisy communication, in which the firm can issue a verified message at a cost but, at the time of disclosure, does not fully know how users will process that message. We thereby relax the standard assumption that disclosures reach users intact. We ask whether noise dampens market responses to disclosed information, weakens the firm's incentive to disclose, and reshapes the equilibrium pricing of certification mechanisms such as environmental, social and governance (ESG) ratings, financial audits, credit ratings, and equity research reports.

Contrary to conventional models (\citealt{beycohlys10,sto13}), real-world disclosure practices suggest that information is rarely processed in a perfectly ``noiseless" manner. Many firms report substantial uncertainty about how the market will interpret or react to their disclosures. While earnings announcements often convey quantitative metrics, such as earnings or revenue, voluntary disclosures about business models, strategies, and risks introduce a higher degree of complexity. These qualitative narratives are often as important as, if not more important than, the quantitative information itself. The result is a state of information overload, where, in the words of the former SEC Chair Mary Jo White (\href{https://www.sec.gov/newsroom/speeches-statements/spch101513mjw}{``The Path Forward on Disclosure,'' October~15, 2013}), \textit{``ever-increasing amounts of disclosure make it difficult for an investor to wade through the volume of information she receives to ferret out the information that is most relevant."} Given the existence of partial processing, it is unlikely that firms can anticipate precisely which elements of their communications investors will successfully process or emphasize.
Corporate filings also illustrate uncertainty about market reactions to disclosure. For instance, in its July 28, 2009 Amendment No. 4 to Form S-1 regarding the acquisition of eRx, Emdeon states: \textit{``We cannot predict how investors will react when we file these financial statements or what effect it will have on the market price of our Class A common stock.''}\footnote{Emdeon Inc., \href{https://www.sec.gov/Archives/edgar/data/1444598/000095012309025946/y01809a4sv1za.htm}{Amendment No. 4 to Form S-1}, filed July 28, 2009, p.~33, under ``Risks Related to This Offering and our Class A Common Stock''; SEC accession no.~0000950123-09-025946. Other filings acknowledge uncertainty about investors'
interpretation of financial reporting changes and the accuracy of
information reaching them. Discussing the requirement to expense
stock-based compensation, Business Objects S.A. stated that
``we cannot predict how investors will view this additional expense''
(Form 10-Q for the quarter ended September 30, 2005, p.~40).
Symyx Technologies, Inc., discussing stock-option expensing and
possible changes to its option program, similarly stated:
``We cannot determine how investors will view this change and how
this change may impact our stock price'' (Form 10-Q for the quarter
ended September 30, 2005, p.~36). Snap Inc. also warned that media
coverage may misreport statements by its officers or employees or
become misleading through omitted information (Form S-1,
February 2, 2017, p.~40).}

Beyond financial reporting, we use two applications as leading examples. Each illustrates an institutional feature that is puzzling within standard models. The first is the proprietary cost hypothesis, under which firms incur real costs when publicly revealing information that competitors, regulators, or parties whose interests are not aligned with shareholders can exploit. These costs can be substantial relative to the benefits of moving short-term prices, especially when the same information will eventually be disclosed through mandatory channels. Yet, in the selected 2004--2015 U.S. firm-year sample of \citet[Table~2]{bermamar20}, the mean annual management-forecast frequency is 54 percent, and many firms supplement their mandatory filings with substantial additional discussion and context. Compounding the puzzle, empirical measures of the intensity of proprietary costs do not consistently correlate with the probability of disclosure. Our model delivers two testable implications. First, the sign of the correlation between disclosure and cost depends on the disclosure environment, and proprietary costs \emph{increase} disclosure in the lower-threshold equilibrium. A higher cost requires greater separation between disclosers and nondisclosers, which is most efficiently achieved in the tail of the distribution, where only firms with extremely negative news withhold. Second, when the disclosure environment is not directly observable, our model implies a joint test under which, within a regime, the disclosure-cost correlation is opposite in sign to the disclosure-noise correlation and, when volatility is below the noise, to the disclosure-volatility correlation.

Our second application concerns the rise of ESG certification. Such certifications have been widely criticized as expensive and unreliable, given that performance is intrinsically difficult to measure and interpret. Despite these criticisms, many large firms obtain some form of costly ESG certification, in part because nonparticipation sends an adverse signal. We derive the unique least-cost equilibrium and show that the demand for ratings may lead to high-noise, high-adoption, and high-fee regimes. From a regulatory perspective, controls on rating quality can discipline the industry and may, counter-intuitively, lower voluntary adoption.

Interestingly, we show that noisy information processing alters the standard insights on voluntary disclosure. In the canonical model with no processing noise, a unique equilibrium exists and features partial disclosure. That result need not survive once processing noise is introduced. When verification costs are small \textit{but positive}, the model unravels to full disclosure, because processing noise induces marginal firms to disclose more, which lets lower types mimic higher disclosing types more effectively. This incentive outweighs small verification costs. Perhaps even more strikingly, a tiny
amount of processing  noise implies the existence of a near full-disclosure equilibrium regardless of the cost.

Any partial-disclosure equilibrium is generically not unique. As verification costs rise beyond the threshold at which unravelling occurs, the model admits two distinct equilibria with different levels of disclosure, and the equilibrium with more communication may involve more disclosure as costs increase. In the limit, full disclosure emerges even for arbitrarily high verification costs. The presumption that frictions necessarily reduce disclosure may therefore fail within an otherwise standard model of voluntary disclosure.

The intuition behind these somewhat surprising results is that the common heuristic that higher costs reduce incentives to disclose is inconsistent with the game-theoretic logic underlying disclosure equilibria. A disclosure game is not a single-agent decision problem in which the firm simply reduces information supply when disclosure becomes more costly, as if the firm cut its consumption of an expensive verification ``good'' until marginal utility equated cost. Rather, as in the classic market for lemons, market prices endogenously adjust to the implied adverse selection, so that the demand for information, and thus the equilibrium disclosure incentive, does not remain constant when costs change. 

In an equilibrium with higher verification costs, the benefits of disclosure must also be higher, by construction, to justify incurring the cost. With noisy information processing, this benefit becomes large over low disclosure thresholds, where disclosures are frequent. The marginal discloser is then pooled with more favorable information, whereas withholding would reveal the firm to belong to the lower tail of the distribution. An increase in the verification cost must therefore be offset by a corresponding increase in the benefit of disclosure relative to silence. That adjustment requires the disclosure threshold to fall, which produces even more disclosure until the benefit again equates the cost. Processing noise thus reverses the conventional view that higher costs necessarily reduce disclosure.

Ultimately, whether these effects arise depends on whether the equilibrium features frequent or infrequent disclosures. In settings characterized by low overall disclosure, or where firms can coordinate on less informative prices or equilibria with low verification costs, the model delivers the standard cost comparative static, with disclosure decreasing in costs, increasing with noisier reports, and increasing with fundamental volatility under the conditions in Corollary~\ref{dog}. In environments with high baseline levels of disclosure, or where disclosure decisions are primarily shaped by investor demand for information, as is likely in modern financial markets, the opposite pattern emerges. In these settings, higher costs may induce more disclosure, while noisier or less informative signals may be disclosed less frequently. The model thus suggests that the direction of comparative statics in disclosure behavior depends on the prevailing level of communication in the market.

The mechanism can survive bounded firm values, deviations from normality, and access to alternate information sources. With alternate sources, preliminary results suggest that competition between multiple intermediaries may raise disclosure costs and reduce disclosure. A firm that privately controls processing noise prefers a noisy system ex ante, which may, counter-intuitively, leave that firm facing third parties charging a high price for uninformative certifications. Under pure uncertainty about information endowment, finally, the equilibrium is unique and features a positive relationship between disclosure and noise, offering further scope to separate theories.

Our approach connects to several strands of the literature that share a focus on verified communication.\footnote{This body of work has examined, for instance, the dynamic revelation of evidence \citep{gutkreskr14, aghan21, kreschskr22}, the presence of alternative information channels \citep{ein18, fregutkre20, licwek23}, the bounds on misreporting penalties \citep{cas26}, and the interaction between disclosure and investment \citep{bendeklip18, gutmen21}. We differ from these studies by focusing specifically on processing noise and its implications for strategic disclosure and market pricing.} Within this literature, \citet{sui07} is closely related to our work, analyzing a setting in which the sender cannot anticipate the receiver's reaction. His framework and its underlying mechanism nonetheless differ substantially from ours. There, even absent frictions that prevent full unraveling, the sender may optimally withhold information because investors' allocation decisions are a non-linear function of disclosure, so that withholding hedges against adverse investor responses. Our model contains no such mechanism, since the firm evaluates the average posterior belief and the friction arises because investors observe the disclosure with noise. This is why, while both frameworks incorporate uncertainty about market responses, they generate distinct equilibrium outcomes.

Several earlier studies identify a channel through which noise in disclosed
information can encourage voluntary disclosure. The basic intuition, common
to this literature, is that noise pools the marginal discloser with higher
types and thereby raises the expected market reaction to disclosure relative
to silence. \citet{liz99} establishes this mechanism in a model of
certification intermediaries, showing that a profit-maximizing certifier may
rationally garble the signal to increase demand for certification.\footnote{
This literature bridges into the broader theory of audit and assurance, by
showing how voluntary noisy audits can create adverse incentives, further
contributing to models focusing on the strategic pricing incentives of
auditors (\citealt{launew10,frimah21}).
}
Surprisingly, greater noise reduces disclosure at a given cost in our
lower-threshold regime, reversing the effect emphasized in the concurrent
study by \citet{ein25}, which also obtains costly full disclosure.\footnote{
Several other studies show that imprecision can increase voluntary
disclosure. \citet{berchecia21} obtain a related result in a different
information structure, where imprecision in asset measurement expands the
set of types that disclose. \citet{berhuliu23} document a similar pattern
theoretically and empirically when investors only partially process
disclosed information.
}
We further show that endogenous certification pricing can select this regime
even under a least-cost criterion and that firms and investors can prefer
different equilibria. \citet{marsri15} also study certification pricing in a
costly disclosure environment, where an intermediary produces a signal with
additive noise; their results imply that greater noise expands disclosure. Our analysis, however, avoids their pointwise Taylor expansion for approximating the firm’s expectation over noise (Assumption 2, p. 33), which they acknowledge to be valid only for arbitrarily large noise.\footnote{The approximation is constructed pointwise at a single signal realization but then applied as a functional approximation for all signal values on the real line, so its quality deteriorates in the tails. We show in our setting that several of their conclusions do not survive in the exact problem, since unravelling equilibria can exist, the partial-disclosure equilibrium is generically not unique, and the comparative statics of the equilibrium set differ from those obtained under the approximation.}

A related conclusion that lower disclosure costs need not increase disclosure appears in \citet{karleesue26}, in a multi-sender environment with disclosure costs. Under their conditions, senders' disclosures are strategic substitutes, so adding a sender or lowering one sender's cost can reduce other senders' disclosure. Their mechanism rests on strategic interaction across senders while ours rests on the noisy processing of a single sender's report, yet the comparative-static conclusion parallels the one we obtain at the lower threshold.

Several other studies consider the value of certification, from the perspective of interactions between voluntary and mandatory channels. \cite{jiayan17} and \cite{bervayxue21} show that a mandatory disclosure of bad news, as in conservative accounting systems, can reduce the equilibrium signalling costs incurred voluntarily to credibly communicate good news. Other studies such as \cite{ver21} and \cite{jiaxinxio23} describe environments in which voluntary choices to certify can provide information above and beyond a disclosure mandate. While we also examine certification, the intuitions from these previously-documented mechanisms would carry over to our setting, so we restrict attention to voluntary channels.

A growing literature studies ESG disclosure and certification, one of the two main applications of our model. \citet{frihei16} and \citet{friheilun24} examine how investors with non-pecuniary preferences price socially responsible firms and the resulting incentives for disclosure, while \citet{sal25} considers the equilibrium implications of investor heterogeneity for the credibility of ESG reporting. Closer to our focus on the structure of voluntary ESG choices, \citet{smi23} studies climate-risk disclosure as a risk-sharing device, \citet{xue25} examines how market forces shape ESG disclosure and real investment efficiency, and \citet{aghan24} compare mandatory and voluntary ESG regimes. These papers primarily emphasize the demand side, real effects, or regulatory comparison. Our contribution complements them on the supply side, by characterizing how a certifier's fee, together with the multiplicity of disclosure equilibria induced by processing noise, can rationalize the coexistence of widely-purchased ratings that are simultaneously perceived as noisy and expensive.\footnote{While this literature is too broad to fully discuss, our model shares with several earlier studies an environment where the sender may not fully know the information observed by the receiver at the point of disclosure, such as models featuring arrival of information (\citealt{ein18,fregutkre20,libmicwie23,fregutkre24}). These studies focus on the effects of external (third-party) signals conditional on non-disclosure, whereas in our framework, miscommunication arises from the garbling of the sender's own message.}

           \vspace{-0.3cm}
        
        \section{The Model}
                   \vspace{-0.3cm}
        \paragraph{}
We describe below a model in which a privately informed firm decides whether to make a costly public disclosure that, once made, is processed by investors with noise. Throughout, we refer to this action as ``disclosure'' to describe a costly action to provide public verifiable information. To make the intuitions as clear-cut as possible, we assume that the firm has no other means to credibly disclose its information. The quality of the public disclosure is exogenously given but, throughout, is only a noisy signal of what the firm actually observes. The cost of disclosure is also exogenous in this section. Our baseline interpretation is that the firm incurs a proprietary cost when disclosing, for example because competitors or other parties whose interests are not aligned with shareholders can exploit the disclosed information, so that no third party collects the cost as revenue. Alternative interpretations include preparation, audit, or certification fees paid to an external party. 

Formally, the firm privately observes its realized value, denoted $\tilde{v} \sim N(0, \sigma^2)$ with mean normalized to zero. The firm has an information set strictly more precise than what investors will observe. Specifically, the firm can incur a cost $c\geq 0$, in which case a signal $\tilde{x}=\tilde{v} + \tilde{\varepsilon}$ is publicly released, where $\tilde{\varepsilon}\sim N(0,s^2)$ with $s>0$ is normally-distributed white noise independent of $\tilde{v}$, the noiseless case $s=0$ serving only as a limiting benchmark. The noise captures imperfect transmission or processing of the firm's information. Even when the firm communicates truthfully, investors typically receive only an imprecise representation of what the firm knows. Otherwise, no disclosure is made, denoted by the symbol $\emptyset$. We use here the normal distribution because of its regularity, with $\Phi(.)$ and $\phi(.)$ indicating the cdf and pdf of the standard normal, showing later that the primary insights do not depend on normality.

The firm's objective is to maximize the expected price net of disclosure cost: for any realization of the firm value $v$,\vspace{-0.5cm}
        \begin{equation*}
            d(v)\in   \arg\max_{d\in \{0,1\}} d \mathbb{E}  P(v+\tilde{\varepsilon})+(1-d)P(\emptyset)-dc, \vspace{-0.5cm}
\end{equation*}
where $P(\tilde{x})$ is the disclosure price and $P(\emptyset)$ is the no-disclosure price.

When possible (i.e., over non-zero probability events), the price $P(.)$ must satisfy Bayes' rule as the expectation of $\tilde{v}$ conditional on the public information available to investors, that is, $P(x)=\mathbb{E}(\tilde{v}|d(\tilde{v})=1,x)$ and $P(\emptyset)=\mathbb{E}(\tilde{v}|d(\tilde{v})=0)$. We consider threshold ``sanitization'' equilibria, such that the firm discloses $d(v)=1$ if and only if $v\geq \tau$ (\citealt{shi94,gutkreskr14}).\footnote{That the equilibrium is a threshold is a consequence of the monotone likelihood ratio property (MLRP), since MLRP is preserved by truncation. That is, $P(x)=\mathbb{E}(v|v\in A,x)$ is increasing in $x$ for a non-zero probability disclosure set $A$, so that the expected disclosure price $\mathbb{E}(P(v+\tilde{\varepsilon})|v,A)$ must also be increasing in $v$. A formal proof is given in the Online Appendix.} The firm observes the realization $v$, chooses $d(v)$, and then prices are formed. The optimal disclosure strategy then reduces to an indifference condition for the marginal discloser at the threshold, \vspace{-0.5cm}
\begin{equation}
\label{dv2}
\mathbb{E}  P(\tau+\tilde{\varepsilon})-c=P(\emptyset),\vspace{-0.5cm}
\end{equation}
 with associated market prices \vspace{-0.5cm}
\begin{equation}\label{Bayes2}
P(x)=\mathbb{E}(\tilde{v}|\tilde{v}\geq \tau,x),\hspace{1cm}P(\emptyset)=\mathbb{E}(\tilde{v}|\tilde{v}< \tau).\vspace{-0.5cm}
\end{equation}


\begin{definition}
\label{d1}
For a given cost $c\geq 0$, an equilibrium is a price function $P(.)$ and a threshold $\tau$, with values in $\overline{\mathbb{R}}$, such that:\vspace{-0.3cm}
\begin{itemize}
\item[(i)] the equilibrium conditions  (\ref{dv2})-(\ref{Bayes2}) are satisfied;\vspace{-0.3cm}
\item[(ii)] if $\tau=-\infty$ (resp., $\tau=\infty$) involves full disclosure (resp., no disclosure) almost surely, then the threshold firm would also prefer this choice at any finite threshold.\footnote{Formally:  for any $\hat{\tau}\in \mathbb{R}$ and price $\hat{P}(x)=\mathbb{E}(\tilde{v}|\tilde{v}\geq \hat{\tau},x)$,
$\Delta(\hat{\tau})=\mathbb{E}\hat{P}(\hat{\tau}+\tilde{\varepsilon})-c-\mathbb{E}(\tilde{v}|\tilde{v}< \hat{\tau})$ is positive if $\tau=-\infty$ and negative if $\tau=\infty$. Symmetrically, the definition rules out no disclosure caused by punishing expectations $P(x)=-\infty$ after any disclosure.}\end{itemize}
\end{definition}
In the benchmark where the noise $s$ is exactly zero (that is, setting $\mathbb{E}(P(\tilde{x})|v)=v$), the firm's problem reduces to standard costly disclosure. Under log-concavity, the marginal-discloser benefit increases in $\tau$ faster than $P(\emptyset)$ \citep{bagber05}, yielding a single interior crossing.
With noise, there are situations in which there is no partial-disclosure equilibrium despite a positive cost. One remaining issue is that, by imposing only (i), full disclosure with $\tau=P(\emptyset)=-\infty$ is always trivially an equilibrium. To address this, condition (ii) requires full disclosure to occur only if it can be rationalized as a strict incentive to disclose for \textit{any} conjectured threshold, ensuring that unravelling is a natural implication of the model rather than an artifact of unbounded skeptical off-equilibrium beliefs.
\vspace{-0.3cm}

\vspace{-0.5cm}

        \section{Optimal Disclosure} \label{sec:analysisF}
   \vspace{-0.3cm}
\subsection{Prices under Noisy Disclosure}   
   \vspace{-0.3cm}
\paragraph{}
Unlike traditional disclosure frameworks \`a la \cite{ver83} and \cite{dye85}, investors in the model never directly observe the firm's private information, even when a disclosure occurs. Instead, they form a Bayesian inference $P(x)$ based on their conjectured disclosure strategy, using the observed message $x$ to extract information about the true firm value $v$. 

First, the marginal discloser $v = \tau(c)$ is, by construction, the firm with the lowest value among those choosing to disclose. Observing a disclosure therefore reveals that $\tilde{v} \geq \tau(c)$, prompting a positive revision in investors' posterior beliefs. However, because the signal is noisy, investors cannot fully disentangle the marginal type from higher types and thus tend to update beliefs too optimistically. Second, higher-value firms that disclose are no longer precisely identified. That is, if the disclosure threshold $\tau(c)$ is low, investors may respond with excessive skepticism, attributing high signals partly to noise rather than strong fundamentals.  The noise therefore attenuates the price response to disclosure, flattening the mapping between disclosures and market valuations.


The first step of our analysis establishes a technical lemma that, using condition (ii) in Definition~\ref{d1}, rules out equilibria in which disclosure occurs only off the equilibrium path. Such equilibria are implausible because, in principle, a firm with sufficiently favorable private information should be willing to pay a finite disclosure cost, even when the signal is noisy.

\vspace{-0.1cm}
\begin{lemma}\label{L1}
There exists an equilibrium, and any such equilibrium must satisfy $\tau<\infty$, i.e., cannot feature no disclosure with probability one.
\end{lemma}
\vspace{-0.1cm}   

Having shown that disclosure is on the equilibrium path, the market price conditional on observing a noisy signal $x$ can be written explicitly as   \vspace{-0.3cm}      
        \begin{equation} \label{eq:price}
                P(x) = \E[\tilde{v}|x, \tilde{v} \ge \tau] = \frac{\sigma^2}{\sigma^2 + s^2} x + \frac{\sigma s}{\sqrt{\sigma^2 + s^2}}  \frac{\phi\left(\frac{\tau - \frac{\sigma^2}{\sigma^2 + s^2} x}{\frac{\sigma s}{\sqrt{\sigma^2 + s^2}}}\right)}{1 - \Phi\left(\frac{\tau - \frac{\sigma^2}{\sigma^2 + s^2} x}{\frac{\sigma s}{\sqrt{\sigma^2 + s^2}}}\right)}.
        \end{equation}
        
        \paragraph{}
The right-hand side follows from standard properties of truncated normal distributions applied to the bivariate normal vector $(\tilde{v}, \tilde{x})$; see, for example, \cite{hor05}. 
In the absence of disclosure, investors receive no additional information, so the price as a function of the disclosure threshold  is   \vspace{-0.3cm}               \begin{equation*}    
                        P (\emptyset) =  \E[\tilde{v}|\tilde{v} < \tau] = - \sigma \frac{\phi(\frac{\tau}{\sigma})}{\Phi(\frac{\tau}{\sigma})}.\vspace{-0.2cm}   
                \end{equation*}

The firm does not know the realization of the noise when choosing whether to disclose, so the firm’s expected price following disclosure is obtained by taking expectations in  (\ref{eq:price}) over all realizations of $\tilde{x}|v \sim N(v, s^2)$ given the privately-observed value $v$:  \vspace{-0.3cm}   
        \begin{equation*}  
                \mathbb{E}[P(\tilde{x})|v]= \frac{1}{s} \int P(x)\phi(\frac{x-v}{s}) d x.
  \vspace{-0.3cm}      \end{equation*}   
Given that the expression for $P(x)$ is cumbersome, and the integral does not admit a closed-form expression, it is convenient to write this equation in terms of the expectations of a transformed standard normal.

\vspace{-0.3cm}
\Needspace{8\baselineskip}
\begin{lemma}
\label{zebra}
The expected price evaluated for the marginal discloser $v=\tau$ is
\vspace{-0.3cm}\begin{equation}\label{EQ}
\mathbb{E}(P(\tau+\varepsilon)|\tau)=a(s)\,\tau + b(s)\,\mathbb{E}\bigl[\lambda(\theta_s(\tilde{Y}))\bigr],\vspace{-0.3cm}
\end{equation}
where $\tilde{Y}\sim N(0,1)$ is a standard normal random variable and
\vspace{-0.3cm}\[
a(s)\equiv\frac{\sigma^{2}}{\sigma^{2}+s^{2}}, \quad  
b(s)\equiv \frac{s\sigma}{\sqrt{\sigma^{2}+s^{2}}}, \quad  \lambda(z)\equiv \frac{\phi(z)}{1-\Phi(z)}, \quad
\theta_s(\tilde{Y})\equiv \frac{\tau s/\sigma-\sigma \tilde{Y}}{\sqrt{\sigma^{2}+s^{2}}}.
\vspace{-0.3cm}\]
Further, $\partial \mathbb{E}(P(\tau+\varepsilon)|\tau)/\partial \tau\in (0,1)$.
\end{lemma}
Lemma~\ref{zebra} will be useful in proving our main results but also offers critical intuition about how noise affects the payoff of the marginal discloser. The expected price in (\ref{EQ}) consists of two components. The first component, weighted by $a(s)\in (0,1)$, captures how a disclosure reflects the firm's private information. This term is linear in $\tau$, implying that the expected payoff deteriorates uniformly as $\tau$ decreases, regardless of the threshold. 

 Without noise, the weight $a(s)$ equals one, so that the marginal effect of a decrease in the threshold on the firm's payoff remains exactly one. In comparison, from the log-concavity of the normal distribution (and many other ``regular" distributions), $P(\emptyset)$ is an average of values below the threshold, and thus decreases more slowly than one as $\tau$ falls. The difference 
$\mathbb{E}[P(\tau+\varepsilon)\mid\tau] - P(\emptyset)$
shrinks as $\tau$ decreases, leading to the well-known unique equilibrium.  

With noise, however, equation~(\ref{EQ}) includes an additional term weighted by $b(s)$. This second component reflects the market's residual uncertainty. After a disclosure, investors know that $v \geq \tau$ but cannot determine with certainty the firm's private information. In the limit of infinite noise, for example, the equation simplifies to 
$\mathbb{E}[ P(\tau+\varepsilon)] = \mathbb{E}(v \mid v \geq \tau),$ since the market can only infer that, on the equilibrium path, some type above $\tau$ made the disclosure. At finite noise, the expected price of the marginal discloser is likewise a weighted average of the potential disclosers with $v \geq \tau$. This averaging effect dampens the sensitivity of expected prices to changes in the threshold, so that the slope is now strictly less than one.

The net benefit of disclosure, $\mathbb{E}[P(\tau+\varepsilon)]-c - P(\emptyset)$, is no longer monotonic because the disclosure payoff is flattened by the weight $b(s)$. Intuitively, the slope of each component with respect to $\tau$ depends on the extent of averaging. When $\tau$ is large, $\mathbb{E}[P(\tau+\varepsilon)]$ is averaged over a smaller set of values $v \geq \tau$, whereas $P(\emptyset)$ is averaged over a larger set of values $v < \tau$. The former term therefore increases more steeply in $\tau$. Conversely, when $\tau$ is small, the situation is reversed, and the net benefit of disclosure \textit{increases} as $\tau$ shrinks.
\vspace{-0.2cm}

\begin{lemma}
\label{hyena}
$\lim_{|\tau|\rightarrow \infty} \mathbb{E}[P(\tau+\varepsilon)]-c-P(\emptyset)=\infty$.
\end{lemma}
Lemma~\ref{hyena} shows that the net benefit of disclosure remains strictly positive relative to non-disclosure for values of $\tau$ sufficiently close to full disclosure ($\tau \rightarrow -\infty$).\footnote{This property holds for any $c$ (even large) due to the unbounded support of the normal distribution, though an analogous result can be established for other distributions with unbounded support. If the support is bounded, the same conclusion applies over a range of positive $c$ (whereas it would be zero in the absence of noise). Intuitively, when the distribution of firm values is bounded from below, a cost above $\mathbb{E}(\tilde{v})-\underline{v}$ rules out full disclosure, since that difference is the most the worst type can gain by separating from the non-disclosure pool. Partial disclosure can nonetheless survive at such costs, because high-value firms may still find it worthwhile to separate. }  As a consequence, there is greater scope for equilibria in which unravelling coexists with positive disclosure costs, since disclosure introduces upside uncertainty, with the market inferring that the firm may be better than the marginal type. In fact, if the difference $\mathbb{E}\!\left[P(\tau+\varepsilon)\right] - P(\emptyset)$ does not shrink sufficiently as $\tau$ increases, there are positive costs at which the only equilibrium in the subgame unravels to full disclosure, with even arbitrarily low-value firms choosing to pay a positive cost to disclose.

\vspace{-0.3cm}

\subsection{Equilibrium Disclosure}
   \vspace{-0.3cm}

\begin{figure}[H]
    \centering
    \includegraphics[width=.76\linewidth]{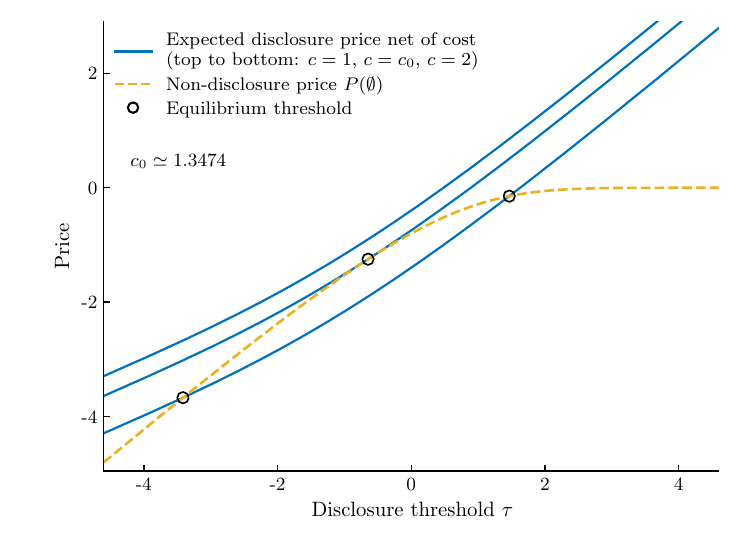}
    \caption{Expected disclosure net of costs $\mathbb{E}[P(x)] - c$ in blue and $P(\emptyset)$ for different thresholds, with equilibrium indicated as a circle ($\sigma=s=1$).}
    \label{LM1}
\end{figure}

\paragraph{}
To gain intuition, Figure \ref{LM1} plots an illustrative example in which the conjectured threshold $\tau$ is varied to characterize equilibria satisfying the indifference condition. The price $P(\emptyset)$ does not depend on the cost $c$ or the noise level $s$, and is a concave function that increases at a rate less than one from $-\infty$ to the unconditional mean $\mathbb{E}(v)=0$. The marginal discloser payoff $\mathbb{E}[P(\tau+\varepsilon)]-c$ is an average of a mirror-image upper truncation $v\geq \tau$, and is convex
with a slope converging to one as $\tau \rightarrow \infty$. Raising the cost $c$ then shifts the net benefit from disclosure uniformly downward.
      
Starting from a sufficiently low cost, the two curves do not intersect, so the net benefit of disclosure remains strictly above zero. Intuitively, the marginal discloser strictly benefits from revealing information by pooling over $v \geq \tau$, and this benefit exceeds the disclosure cost. Interestingly, this unravelling outcome does not require quantitatively extreme assumptions about the economic primitives, since it is the unique equilibrium when costs and noise are both set equal to one standard deviation of the fundamental uncertainty.

 As the cost increases, the disclosure-benefit curve shifts downward until it becomes tangent to $P(\emptyset)$, at which point the benefit from disclosure exactly offsets the cost. Beyond this critical value, the two curves intersect twice, giving rise to two equilibria. The equilibrium with the lower threshold features more frequent disclosure but more skeptical beliefs conditional on non-disclosure. The equilibrium with the higher threshold instead features less frequent disclosure. This threshold is typically preferred by firms, since they obtain a higher price when not disclosing and are pooled with better types when they disclose.

\begin{proposition}\label{mosttwo}
There is a cost level $c_0>0$ such that:
\begin{itemize}\vspace{-0.25cm}

\item[(i)] if $c<c_0$, the unique subgame equilibrium is full disclosure, with threshold $\tau(c)=-\infty$;\vspace{-0.2cm}

\item[(ii)] if $c=c_0$, the unique subgame equilibrium features partial disclosure threshold $\tau(c_0)<0$;\vspace{-0.2cm}

\item[(iii)] if $c>c_0$, the game admits exactly two subgame equilibria. These are characterized by finite thresholds $\underline{\tau}(c)$ and $\overline{\tau}(c)$, continuous in $c$. As $c$ increases from $c_0$ to $\infty$, the lower threshold $\underline{\tau}(c)$ falls from $\tau(c_0)$ to $-\infty$, while the upper threshold $\overline{\tau}(c)$ rises from $\tau(c_0)$ to $\infty$.
\end{itemize}
\vspace{-0.3cm}
\end{proposition}

Proposition \ref{mosttwo} contrasts sharply with the classical noiseless benchmark of \citet{ver83}, in which the equilibrium exhibits partial disclosure for every positive disclosure cost. Once information processing is noisy, however, full disclosure can arise for a range of strictly positive costs, and for higher costs the game generically admits two partial-disclosure equilibria, one of which features more disclosure as the cost increases.\footnote{The exact count of two equilibria in case (iii) may appear to violate the oddness-of-equilibria result. The boundary equilibrium absent from the count is full disclosure ($\tau=-\infty$) sustained by skeptical off-path beliefs, ruled out by Definition~\ref{d1}(ii); without that restriction, the equilibrium set would also include this boundary point, for a total of three.}

At first glance, one might expect a higher disclosure cost to \textit{necessarily} reduce the probability of disclosure, since a higher cost lowers the net benefit of revealing information. As shown in Figure \ref{plot:eqXY}, that intuition need not hold, because the effect depends on the degree of market skepticism. 
Analogous to the classic market for lemons (\citealt{ake70}), the firm faces investors adjusting their beliefs to correct for adverse selection. The demand for a verified disclosure need not slope downward in the cost if investors anticipate more severe adverse selection. Absent any disclosure noise ($s = 0$), the net benefit of disclosure $\tau- c - P(\emptyset)$ is strictly increasing in the threshold. Hence, adverse selection and the probability of disclosure cannot simultaneously increase.  

With noisy disclosure, on the other hand, the net benefit of disclosure is non-monotonic. At the upper threshold $\overline{\tau}(c)$, where the probability of disclosure is low, disclosures tend to be more precise because investors already know that $v \geq \overline{\tau}(c)$. This corresponds to the conventional comparative static in which higher costs lead to less disclosure. At the lower threshold $\underline{\tau}(c)$, however, when investors interpret non-disclosure skeptically, greater disclosure raises the net benefit of disclosing, because it becomes harder to distinguish marginal disclosers with $v \geq \underline{\tau}(c)$ while non-disclosure becomes more revealing. Higher costs therefore call for a higher benefit of disclosure over the non- disclosure price, driving the equilibrium threshold further into the lower tail. The probability of disclosure then \textit{increases} with the cost of disclosure. Indeed, in the limit as $c \rightarrow \infty$, firms unravel to full disclosure with an arbitrarily costly disclosure.

The duality of subgame equilibria arises from a fundamental trade-off, i.e., whether the gap between disclosure and non-disclosure stems from more precise disclosure at $\overline{\tau}(c)$ or from more precise non-disclosure at $\underline{\tau}(c)$. Given a higher cost, this gap must increase to balance the cost. When the threshold is expected to be high, i.e., in environments with low levels of disclosure, disclosure events ``$v \geq \tau$'' tend to more precisely indicate good news, so that a higher threshold tends to increase the post-disclosure expectation more sharply. The higher cost must be balanced with an increase in the post-disclosure expectation, leading to a higher disclosure threshold (and a lower probability of disclosure).  When investors instead expect the threshold to be low, i.e., disclosure levels are high, a disclosure conveys less information than a non-disclosure $``\tilde{v} < \tau."$ For a higher cost, the gap must then be increased by making the non-disclosure more precise, which is achieved by further decreasing $\tau$, increasing the probability of disclosure.

Joining both intuitions, the firm is not reducing its consumption of a disclosure good akin to a standard demand model, because its benefit to disclose is not exogenously decreasing in the conjectured level of disclosure. Instead, this benefit is determined by the informational value of the disclosed signal. It is a non-monotonic function of the disclosure threshold, highest in the lower and upper tails when either a disclosure or non-disclosure best separates extreme signals.\footnote{The force behind these counterintuitive comparative statics does not rely on unbounded support of firm value. Unboundedness governs only tail behavior, allowing the lower threshold to keep shifting down as the cost increases, because non-disclosure can be made arbitrarily punitive. With bounded support, the same mechanism operates over an intermediate cost range, because non-disclosure cannot be made more punitive once the threshold nears the lower bound, and, with support bounded at both ends, sufficiently high fees leave no disclosure. Section \ref{sec:bound} treats the bounded case.}

In our model, the two subgame equilibria exhibit distinct properties that can be more or less descriptive depending on features of the empirical setting. On the one hand, the lower threshold $\underline{\tau}(c)$ features more frequent communication. With noisy disclosure, however, disclosure frequency alone does not rank the information available to investors or determine which equilibrium investors prefer; Section~\ref{sec:hetinv} provides this comparison for mean-variance investors. Truth-leaning equilibria (\citealt{harkreper17,rap25}) tend to select more communication. Similarly, the common emphasis on fully revealing equilibria in signalling games, or on the most informative equilibria in cheap-talk models (\citealt{fissto01,chen2008}), implicitly assumes that communications gravitate toward more communication, even when equilibria less costly to the sender, such as pooling or babbling, exist.
  
  On the other hand, the upper threshold $\overline{\tau}(c)$ is preferred by firms, as lower investor skepticism enables them to disclose less and achieve higher payoffs net of disclosure costs. Firms would attain this equilibrium if they could credibly communicate neologisms to investors (\citealt{bercia18}), which ultimately depends on their ability to be heard and to influence market expectations. This trade-off between the informativeness of communication and its strategic cost arises in other settings as well. Signalling models, for instance, can feature less informative equilibria alongside lower signalling costs (see, e.g., \citealt{stover04,gutkadkan06}), as do models in the real-effects literature (\citealt{kan07}).\footnote{For now, we remain agnostic about which equilibrium prevails given that, in a rational-expectations setting, firms do not choose an equilibrium directly but instead respond to expectations about price formation. Coordination on a disclosure-efficient equilibrium is unlikely to be straightforward in a large, decentralized capital market, where many firms and investors lack obvious mechanisms for communication or pre-commitment. Equilibrium multiplicity is also not unique to our setting. It is common in communication games, including cheap talk (\citealt{crasob82,fissto01,morsto03}), signaling (\citealt{gutkadkan06}), standard-setting (\citealt{cheyan23}), and even verifiable-disclosure models (\citealt{aghsmi23}). Global games are a notable exception because they are designed to model coordination failures in large decentralized settings; see, e.g., \cite{stozha23} for an application to financial reporting.}

\vspace{-0.3cm}
\subsection{Comparative Statics}   
\begin{figure}[H]
\begin{minipage}[t]{0.48\linewidth}
    \centering
    \includegraphics[width=\linewidth]{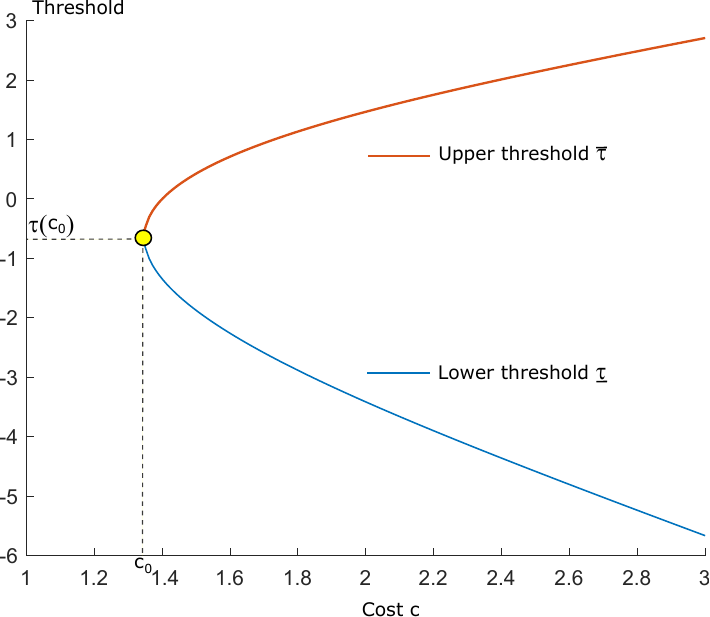}
    \caption{The two equilibrium thresholds $\underline{\tau}(c)$ and $\overline{\tau}(c)$ vary with $c$ ($\sigma=s=1$).}
    \label{plot:eqXY}
\end{minipage}
\hfill
\begin{minipage}[t]{0.48\linewidth}
    \centering
    \includegraphics[width=\linewidth]{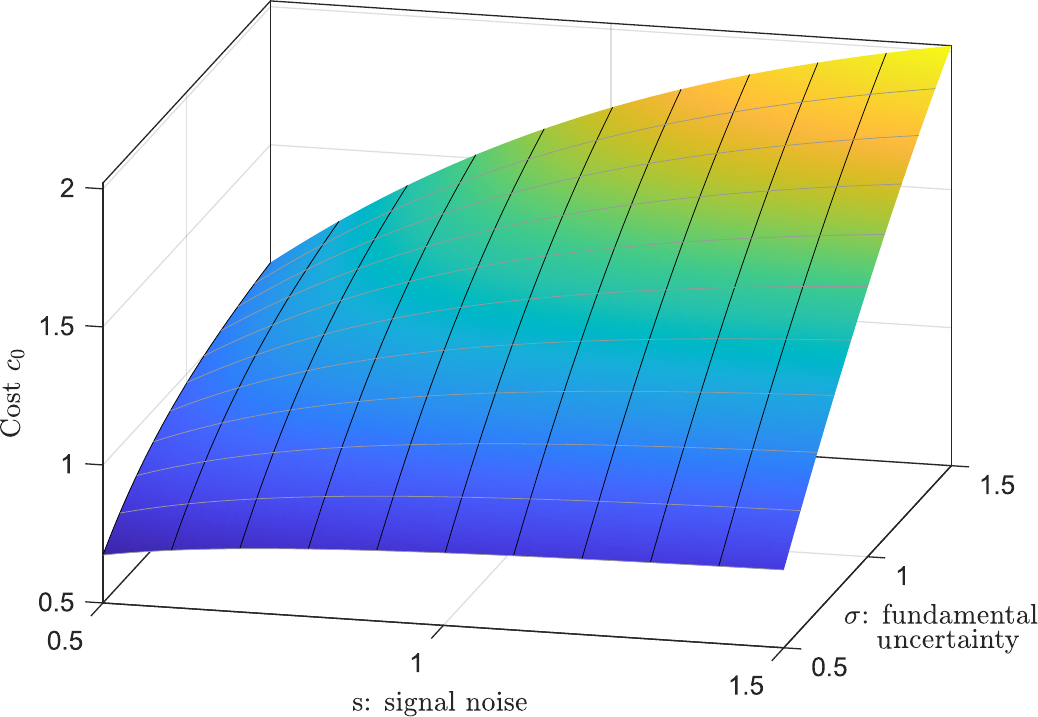}
    \caption{Maximum cost $c_0$ consistent with full disclosure.}
    \label{plot:c0}
\end{minipage}
\end{figure}

\vspace{-0.3cm}   
\paragraph{}
Our next step is to examine the comparative statics of the maximal cost $c_0$, below which firms' disclosures unravel to full disclosure.\vspace{-0.3cm}
\begin{corollary}\label{C1}
As the noise $s$ or the fundamental uncertainty $\sigma$ increases from $0$ to $\infty$, the unravelling bound $c_0$ increases, with $\lim_{\sigma\rightarrow 0} c_0=\lim_{s\rightarrow 0} c_0=0$.\footnote{In the Online Appendix, we also show that $c_0$ converges to a finite value when either $\sigma$ or $s$ becomes large.} Further, $\tau(c_0)<0$.
\end{corollary}\vspace{-0.3cm}
The effect of the information environment in implementing a full-disclosure outcome is consistent with several earlier intuitions. We plot the bound $c_0$ as a function of $(s, \sigma)$ in Figure \ref{plot:c0}.
The additional noise $s$ flattens the marginal discloser's price, preventing investors from precisely identifying the marginal discloser and increasing the set of costs for which the net benefit of disclosure exceeds zero. How a higher $\sigma$ affects this trade-off further matches the intuition in \cite{ver90}, since more information is transmitted per unit of cost, which raises the firm's disclosure propensity and nudges the model toward full disclosure. The tangency point $\tau(c_0)<0$ lies below the unconditional mean, suggesting that the discontinuous jump  to partial disclosure when $c_0$ is attained cannot be too large and must initially affect only unconditionally unfavorable information. 

The comparative statics in the volatility of fundamentals $\sigma$ require more caution than those in the noise $s$, because the threshold is only a description of the strategy while the probability of disclosure is jointly affected by a shift in the threshold and a change in the distribution of fundamentals. To address this, we examine the probability of disclosure $\Pr(v\geq \tau(c))=1-\Phi(\tau(c)/\sigma)$, which is monotone in the threshold for a change in noise $s$ but otherwise requires a total derivative of the numerator and denominator of $\tau(c)/\sigma$.\footnote{It is problematic to focus on comparative statics on the threshold $\tau(c)$, unless the threshold has a monotone mapping with the probability of disclosure. Consider the following example in \cite{dye85} and \cite{junkwo88}. Suppose that the measurement is changed from US dollars to Chinese yuans, at a rate of 1 for 7. It is clear that this unit of measurement should have no effects on any meaningful economics of the problem (and will not change the probability of disclosure), despite the fact that the actual threshold in yuans will be higher, at seven times the threshold in dollars. From an economic standpoint, the object of analysis is the final allocations, i.e., the probability of disclosure and distribution of prices.}  

\Needspace{6\baselineskip}
\begin{corollary}\label{dog}
Suppose that $c>c_0$. Then the probability of disclosure\begin{itemize}\vspace{-0.3cm}
\item[(i)] increases in the noise $s$ under $\overline{\tau}(c)$  and, under $\underline{\tau}(c)$, decreases in $s$; \vspace{-0.3cm}
\item[(ii)] increases in the uncertainty $\sigma$ under $\overline{\tau}(c)$ when $\overline{\tau}(c)>0$ or $\sigma<s$, and, under $\underline{\tau}(c)$, decreases whenever $\sigma<s$.\footnote{The probability of disclosure under $\underline{\tau}(c)$ eventually equals one if $c$ is below the limiting unravelling bound, and eventually increases strictly in $\sigma$ if $c$ is above it.}
\end{itemize}\vspace{-0.3cm}
In addition, under $\underline{\tau}(c)$, the equilibrium converges to full disclosure if $s$ or $\sigma$ converge to zero or $\sigma$ becomes large, and the probability of disclosure under any equilibrium with noise is greater than without noise.
\end{corollary}

\begin{figure}[H]
\centering 
\begin{minipage}{0.48\textwidth}
    \centering
    \includegraphics[width=\textwidth]{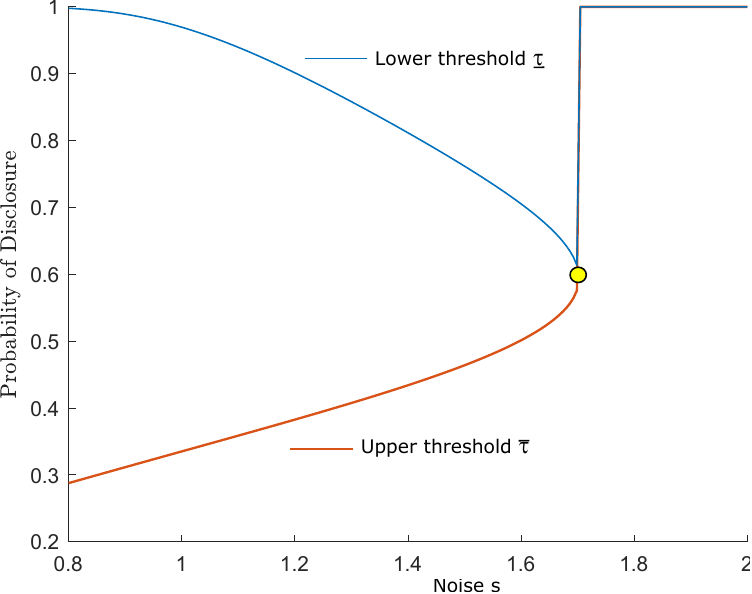}
    \caption*{with $c=1.5$ and $\sigma=1$}
\end{minipage}
\hfill
\begin{minipage}{0.48\textwidth}
    \centering
    \includegraphics[width=\textwidth]{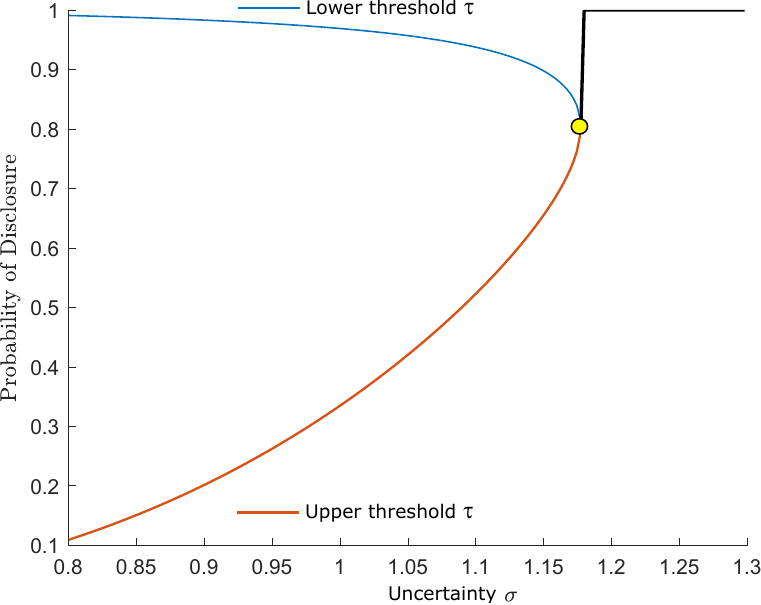}
    \caption*{with $c=1.5$ and $s=1$}
\end{minipage}
\caption{Comparative statics of the disclosure probability $\Pr(\tilde{v} \ge \tau(c))$ under the lower- and upper-threshold equilibria.}\label{LB1}\vspace{-0.3cm}  
\end{figure}

We plot the comparative statics of the probability of disclosure with respect to the signal noise $s$ and the volatility of the fundamental $\sigma$ in Figure \ref{LB1}. More noise in disclosure increases the payoff to the marginal discloser by pooling it with a wider range of values $v \geq \tau(c)$. This increases the net benefit of disclosure which, all else equal, has an effect similar to a reduction in the disclosure cost. To remain consistent with equilibrium, this effect must be offset against a lower informational value of disclosure. Under the upper equilibrium $\overline{\tau}(c)$, this adjustment occurs through less precise disclosures, via a lower threshold and a higher probability of disclosure. Vice-versa, under the lower equilibrium $\underline{\tau}(c)$, the adjustment occurs from a less precise non-disclosure,  which is achieved with a higher probability of non-disclosure. Surprisingly, even a small amount of noise can dramatically alter the predictions of the model: as $s\rightarrow 0$, which means that the noisy model is arbitrarily close to the classic benchmark without noise, the equilibrium under $\underline{\tau}(c)$ unravels to full disclosure.

For similar reasons, the comparative statics with respect to fundamental uncertainty need not match the prediction in \cite{ver90}, who shows (absent noise) that firms disclose more as $\sigma$ increases. This prediction holds in the upper equilibrium $\overline{\tau}(c)$ when $\overline{\tau}(c)>0$ or $\sigma<s$; under the lower equilibrium $\underline{\tau}(c)$, the effect is ambiguous. An increase in uncertainty reduces the relative cost of disclosure but also diminishes the relative noise per unit of uncertainty. These mechanisms work in opposite directions, generating a non-monotonic effect on the probability of disclosure. The model therefore unravels to full disclosure when uncertainty is either very high or very low, while over the region where uncertainty is below the processing noise the disclosure probability decreases as uncertainty rises. 

\vspace{-0.3cm}
\section{Certification} \label{sec:cert}
   \vspace{-0.3cm}
\subsection{The Certifier's Problem}
\vspace{-0.3cm}
\paragraph{}
A richer two-stage game endogenizes the cost in Section~\ref{sec:analysisF} as the fee charged by a certifier, holding fixed the noise $s$ of the disclosed signal. We treat for now $s$ as an exogenous quality parameter of the certification technology. One possible interpretation is an ESG evaluation that commits to a fee per rated firm, taking as given the certification quality, expressed as the variance of the noise added to the firm's information. We continue to use the language of certification, with the understanding that similar economic intuitions could apply to audit, securities research, or other forms of third-party assurance.\footnote{Many of these settings feature richer institutional environments that we do not intend to fully model here; for example, audits are often mandatory but also contain voluntary aspects (or may be fully voluntary for private firms) and may be subject to standards, credit and equity research can be investor-paid or part of a new issuance (e.g., an IPO). We do not intend the stylized model to be a complete or realistic model of these settings, though the conceptual trade-offs are plausibly relevant to them as well.}

The model is identical to Section~\ref{sec:analysisF}, with the following addition in Figure~\ref{tim1}.
 At a prior stage, a single certifier publicly commits to a
fee $c$
to be charged per certified firm. After observing the fee, the firm privately observes $\tilde v$ and decides whether to disclose. If the firm pays the fee $c$, the certifier publicly issues $\tilde x = \tilde v + \tilde\varepsilon$ with $\tilde\varepsilon \sim N(0, s^2)$. The certifier's payoff is $c (1-\Phi(\tau/\sigma))$, where $\tau$ is the equilibrium disclosure threshold induced by $c$. 

Proposition~\ref{mosttwo} continues to characterize the set of equilibria, with a unique full-disclosure equilibrium for $c < c_0$ and generically two interior threshold equilibria $\underline{\tau}(c)<\overline{\tau}(c)$ for $c > c_0$. Consistent with equilibrium, defining $\tau(c)\in \{-\infty,\underline{\tau}(c),\overline{\tau}(c)\}$ as the equilibrium belief, we assume that the certifier maximizes profits $c (1- \Phi(\tau(c)/\sigma))$. Multiple subgame equilibria complicate the certifier's problem, because the certifier's payoff depends on which equilibrium is selected in $\tau(c)$.
\vspace{-0.3cm}
\begin{definition} \label{def:lc}
An equilibrium $(c^*, \tau^*)$ is a least-cost certification equilibrium if there is no other equilibrium with $(c', \tau')$ such that $c^* (1- \Phi(\tau^*/\sigma))>c' (1- \Phi(\tau'/\sigma))$. If several equilibria attain this least cost, the criterion selects the one at which every deviation by the certifier strictly lowers its profit.
\end{definition}
\vspace{-0.2cm}

Because the certification serves as a signal to separate types but does not create value, we assume that the firms do not engage in excess separation or opt for excess fees, unless dictated by equilibrium.\footnote{If firms realized productive benefits from information, the criterion would have to be amended to include the benefits of production, as in \cite{sha94}, but the basic intuitions that rule out excessive fees or undesirable equilibria would plausibly apply as well. We state the model without an ex-ante value of information because, in our setting, the insight about excess production of information will be similar to this earlier literature and would simply burden our main argument.} Of course, just like any refinement, the validity of the criterion is partly an empirical question, and the Online Appendix examines two alternative selection criteria that can yield equilibria more favorable to the certifier.\footnote{Both alternatives require an exogenous cap $\overline{c}$ on admissible fees. The continuity criterion rules out jumps between the two thresholds. The forward-induction criterion additionally requires a profitable deviation to be followed by an equilibrium continuation that makes it profitable, whenever one exists. Under their respective sufficient conditions in the Online Appendix, both select the cap, $c^*=\overline{c}$, followed by the high-disclosure threshold $\underline{\tau}(\overline{c})$, which maximizes certifier revenue among equilibria subject to that cap. The forward-induction criterion is motivated by the best-rationalization principle, under which unexpected moves are interpreted, whenever possible, as choices of rational and strategically sophisticated players (\citealt{battigalli2003}). Strong rationalizability builds on this principle without imposing equilibrium and delivers unraveling in disclosure games with verifiable messages (\citealt{battigalli2006,battigalli2019}). Our criterion adds a coordination restriction on continuation equilibria, which we impose directly and do not derive from their solution concepts.} Nevertheless, the criterion has several attractive theoretical features. Conceptually, it disciplines off-equilibrium beliefs by ruling out a solution where investors (mis)coordinate on excessively skeptical beliefs that force firms to certify too much against their own interest; in fact, the certifier would be able to unboundedly increase the fee under $\underline{\tau}(c)$. In other words, the criterion is conservative in the sense of being unfavorable to the certifier, because by always selecting the equilibrium with the least scope for fee increases it produces lower-bound profits and avoids attributing market power to the certifier that is not supported by the disclosure game itself.

 \vspace{-0.3cm}
\subsection{Demand for Certification and\ Fees}
\vspace{-0.3cm}
\paragraph{}

\begin{figure}[H]
    \centering
    \includegraphics[width=.74\linewidth]{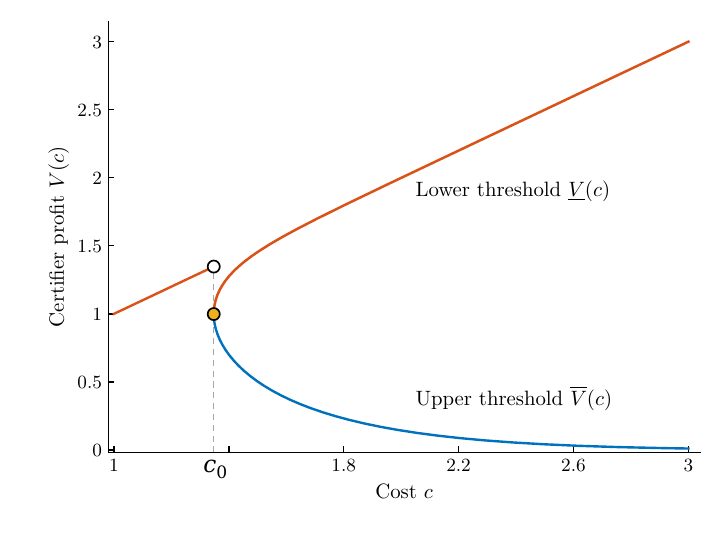}
  \caption{Certifier profit $V(c)$ ($\sigma=s=1$).} \label{plot:profit_1}\vspace{-0.1cm}
\end{figure}

To solve the full game, we proceed by backward induction, moving next to the certifier's optimal choice of fee taking the equilibrium as given and optimizing over the fee $c$ to solve \vspace{-0.4cm}
\begin{equation*}
V(c^*;\tau(c^*))=\max_{c\geq 0}\ c(1-\Phi(\frac{\tau(c)}{\sigma})),
\vspace{-0.3cm}\end{equation*}
where $\tau(c)$ is the belief. The certifier's profit is plotted in Figure \ref{plot:profit_1}, as a function of the chosen fee. For $c<c_0$, the unique equilibrium entails full disclosure, with profit $V(c;-\infty)=c$ increasing in $c$. At $c=c_0$, profit falls discontinuously to $V(c_0;\tau(c_0))=c_0(1-\Phi(\frac{\tau(c_0)}{\sigma}))$, at the lowest fee compatible with an interior solution. Thus, no fee $c\leq c_0$ can solve the certifier's problem, as we explain below. 

Any fee $c<c_0$ within the full-disclosure region is such that $V(c;-\infty)=c$, increasing in the fee because all firms certify at these fee levels. Were such a fee $c$ optimal, the profit could be further increased by increasing $c$ slightly toward $c_0$, e.g., $c'=(c+c_0)/2$, contradicting the optimality of any choice $c<c_0$ and implying that $c_0$ is a lower bound on the equilibrium fee. Yet, $c=c_0$ cannot be an equilibrium either, because  the interior equilibrium implies a discrete loss of revenue as $V(c_0;\tau(c_0))=c_0(1-\Phi(\frac{\tau(c_0)}{\sigma}))<c_0$; as a result, some fee strictly below $c_0$ is always preferred by the certifier to $c_0$, e.g., $c'=\frac{1}{2}(V(c_0;\tau(c_0))+c_0)$, contradicting the optimality of $c_0$.

Thus, a solution to the certifier's problem must be found at $c>c_0$. Consider the two candidate solutions involving one of the two thresholds: \vspace{-0.3cm}
\begin{equation*}
\underline{V}(c)\equiv V(c;\underline{\tau}(c))=c(1-\Phi(\frac{\underline{\tau}(c)}{\sigma})), \hspace{.5cm} \overline{V}(c)\equiv  V(c;\overline{\tau}(c))=c(1-\Phi(\frac{\overline{\tau}(c)}{\sigma})),\vspace{-0.5cm}
\end{equation*}  
with the convention that $\underline{V}(c)=\overline{V}(c)=V(c;\tau(c))$ when $c\leq c_0$.\ 

\begin{lemma}
\label{mouse}
If $c\geq c_0$, $\underline{V}(c)$ is increasing in $c$ with $\lim_{c\rightarrow \infty} \underline{V}(c)=\infty$ and  $\overline{V}(c)$ is decreasing in $c$ when $c$ is sufficiently large,  with $\lim_{c\rightarrow \infty} \overline{V}(c)=0$.
\end{lemma}
The two thresholds have different implications for the certifier's strategy. When anticipating the lower threshold $\underline{\tau}(c)$, the certifier strictly prefers to increase the fee, because this increases the demand for disclosure by shifting the threshold further into the lower tail (by Proposition~\ref{mosttwo}). By increasing the fee, the certifier sustains increasingly skeptical market beliefs, under which choosing not to disclose indicates highly unfavorable lower-tail outcomes.
Under the upper threshold $\overline{\tau}(c)$, a higher fee conveys, conversely, that only firms with highly favorable information are willing to pay for certification. The higher the fee, the more favorable the disclosed information must be, whereas the information conveyed by non-disclosure remains comparably flat, and hence the lower the probability of disclosure. A standard continuity argument implies the existence of a maximum certifier profit $\overline{V}^{**}=\max_{c\geq c_0} \overline{V}(c)$, with a profit-maximizing fee $c^{**}$ and threshold $\overline{\tau}(c^{**})$. For the selection below when $\overline V^{**}\geq c_0$, we assume that this maximizer is unique.

To characterize the least-cost equilibrium, we need to first describe the full set of equilibria. Note that any equilibrium must be such that $V(c^*;\tau(c^*))\ge c_0$ since the certifier can achieve a profit arbitrarily close to $c_0$ with $c<c_0$. This leaves us with two types of equilibria. In the first (i), the certifier chooses $c^{**}$ and the disclosure subgame resolves at the upper-threshold continuation equilibrium $\overline{\tau}(c^{**})$. In the second (ii), there is a continuum of equilibria in which the certifier chooses a fee $c^*$ with $\underline{V}(c^*)\geq \max\{c_0,\overline{V}^{**}\}$, the exact condition for no deviation to be profitable, followed on the equilibrium path by the lower-threshold continuation equilibrium $\underline{\tau}(c^*)$. When $\overline{V}^{**}\leq c_0$, the cheapest such fee is $\hat{c}$, defined by $\underline{V}(\hat{c})=\hat{c}\,(1-\Phi(\underline{\tau}(\hat{c})/\sigma))= c_0$, which earns the certifier exactly $c_0$.\footnote{In (ii), the off-equilibrium belief for the certifier not to increase the fee above $c^*$ is that $\tau(c)=\overline{\tau}(c)$ for any $c>c^*$.} 

\begin{figure}[H]
\centering
\begin{minipage}{0.48\textwidth}
    \centering
    \includegraphics[width=\textwidth]{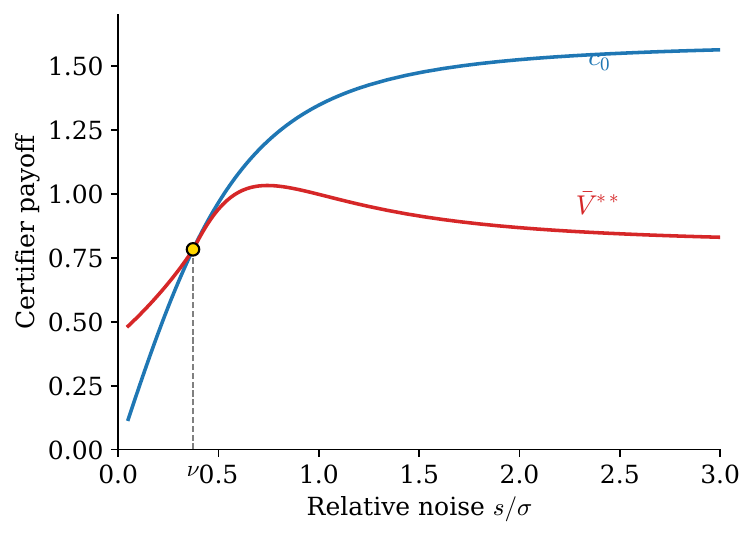}
    \caption*{Certifier payoff curves. }
\end{minipage}
\hfill
\begin{minipage}{0.48\textwidth}
    \centering
    \includegraphics[width=\textwidth]{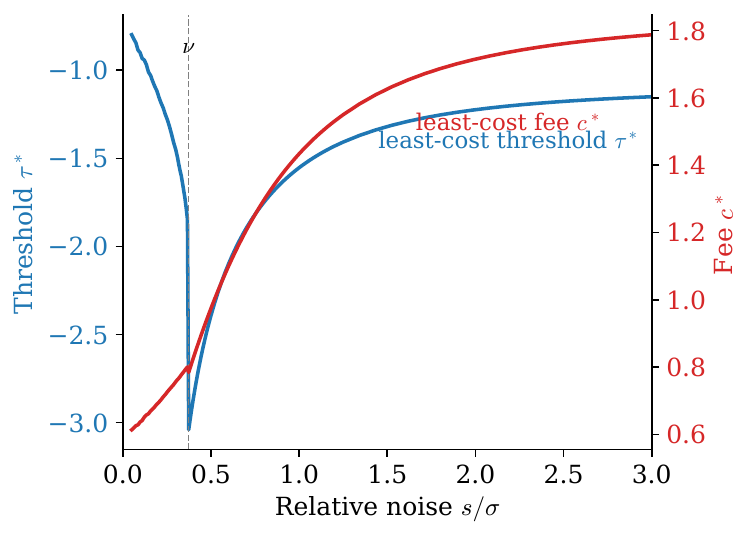}
    \caption*{Least-cost equilibrium $\tau^*$ and $c^*$}
\end{minipage}
\caption{Left: the two payoff curves $c_0$ and $\overline{V}^{**}$ as functions of $s/\sigma$, crossing at the cutoff $\nu\approx0.372846$. Right: the least-cost equilibrium threshold $\tau^*$ (left axis, blue) and fee $c^*$ (right axis, red) as functions of $s/\sigma$, with the discontinuous jump at $\nu$. Both panels set $\sigma=1$.}\label{plot:cert}\vspace{-0.3cm}
\end{figure}

Among these equilibria, (i) can exist only if $\overline{V}^{**}\geq c_0$, which the proof below shows is equivalent to the noise $s$ being sufficiently small. Intuitively, a very noisy certification implies that unravelling can be achieved at a very high fee which always dominates an upper-tail partial certification solution where some firms prefer not to certify.\ Vice-versa, in the continuum (ii), the least-cost equilibrium must choose the lowest fee consistent with equilibrium, since any higher choice of $c$ implies greater profit to the certifier. When $\overline{V}^{**}<c_0$, that fee is $\hat{c}$; when $\overline{V}^{**}\geq c_0$, it earns exactly $\overline{V}^{**}$ and ties with (i), leaving the certifier exactly indifferent to a deviation toward $c^{**}$ and selecting (i). Figure~\ref{plot:cert} (left panel) plots the two payoff curves $\overline{V}^{**}$ and $c_0$ as functions of $s/\sigma$ and shows their unique crossing at~$\nu$, computed numerically from $\overline V^{**}=c_0$.

\begin{proposition}
\label{finalized}
Suppose that $\overline V$ has a unique maximizer when $s/\sigma\leq\nu$. If $s/\sigma>\nu$, where $\nu\simeq 0.37$, the least-cost equilibrium is such that the certifier chooses $c^*=\hat{c}$ and the certification threshold is $\tau^*=\underline{\tau}(\hat{c})$, while, if $s/\sigma\leq\nu$, the certifier chooses $c^*=c^{**}$ and the certification threshold is $\tau^*=\overline{\tau}(c^{**})$. The certifier achieves a profit $\max\{c_0,\overline{V}^{**}\}$.
\end{proposition}

Proposition \ref{finalized} offers a simple characterization of the certification game, with the two payoff curves and the implied equilibrium fee and threshold plotted against $s/\sigma$ in Figure~\ref{plot:cert}. Even when firms coordinate on the least-cost certification equilibrium, the game may generate costly and widely adopted noisy ratings. Certification noise affects not only the informativeness of the rating but also the certifier's strategic pricing problem. When $s/\sigma$ is sufficiently high, $\overline{V}^{**}<c_0$. Any higher-threshold outcome then gives the certifier less than $c_0$, whereas a fee sufficiently close to $c_0$ from below induces full disclosure and earns more. Under the disclosure-equilibrium concept in Definition~\ref{d1}, no equilibrium of the certification game can therefore have a higher-threshold outcome in this region. Noisier certification therefore need not discipline the rating market, and by moving the equilibrium from one threshold to the other it implies wider dissemination of certification.

Note that without the least-cost refinement, the certification game may support still higher fees and wider adoption. In particular, ESG ratings, among other ratings, are often criticized for noise, disagreement, and lack of standardization despite their widespread use. In the model, rating noise affects not only informativeness but also fees and adoption, so noisy ratings can remain widely used even when their information quality is limited.

Interestingly, the result also connects to the main insight of \cite{liz99}, who shows that an unregulated certifier could set a sufficiently noisy certification to increase demand and ultimately extract the full surplus. As Corollary \ref{dog} shows, however, this demand mechanism need not operate under $\underline{\tau}(c)$ because the threshold is increasing in $s$, so greater noise reduces demand for certification. Further, Proposition \ref{finalized} shows that the lower-threshold equilibrium obtains for sufficiently high noise. Nevertheless, a limited form of Lizzeri's surplus-extraction conclusion survives. By Proposition \ref{finalized}, the certifier earns $c_0$, and Corollary \ref{C1} shows that $c_0$ increases with the noise.
The extraction has a ceiling, however. As $s$ grows large, the disclosed signal carries no information beyond the decision to certify, so the marginal discloser's benefit converges to $\mathbb{E}(\tilde{v}|\tilde{v}\geq \tau)-\mathbb{E}(\tilde{v}|\tilde{v}<\tau)$, whose minimum over thresholds equals $4 \sigma/\sqrt{2 \pi }$. Proposition \ref{sima} later shows that $c_0$ converges to this value. Therefore, while the certifier does indeed prefer more noise, full surplus extraction does not obtain and, in fact, the act of certification remains somewhat informative by separating types with a cutoff located around the unconditional mean (the full characterization coincides with the model with noise privately chosen by the firm and is given in Section \ref{sec:noise}).

\section{Discussion and Extensions}
\label{extensions}
\vspace{-0.3cm}
\subsection{Bounded Support and General Distributions} \label{sec:bound}
\vspace{-0.3cm}
\paragraph{}
Until now, we specialized the analysis to normal distributions for tractability, which allows us to use projection theorems and obtain closed-form expressions for posterior beliefs. Conceptually, however, neither normality nor unbounded support is essential. In particular, the key arguments from Section~\ref{sec:analysisF} rely only on the fact that disclosure generates a noisy but informative signal that is more likely under higher firm values, and on the way in which truncation of the prior modifies investors' beliefs after disclosure or non-disclosure.\footnote{A potential concern with the normal benchmark is that unbounded support can generate counterintuitive implications such as unbounded certification fees in the absence of an exogenous upper bound on fees. These should be interpreted as ``very large'' fees in a stylized model rather than literally infinite fees. Still, it is useful to show that our equilibrium characterization does not rely on unbounded support. In particular, if the fundamentals are bounded, the value of separating from the worst type is itself bounded, so that beyond some fee level no firm is willing to purchase certification and the unique equilibrium is no disclosure.}

In this subsection we allow for a general (not necessarily normal) prior and impose milder regularity conditions that guarantee the shape properties used in Section~\ref{sec:analysisF}. Throughout, $\tilde{v}$ denotes the firm's private information, $\tilde{\varepsilon}$ the independent certification noise, and $\tilde{x}=\tilde{v}+\tilde{\varepsilon}$ the publicly observed signal upon disclosure. Given a conjectured threshold strategy $d(v)=\mathbf{1}\{v\ge\tau\}$, prices continue to be determined by Bayes' rule as in~(\ref{Bayes2}): $P(x)=\E(\tilde{v}| \tilde{v}\ge\tau,x)$, and $P(\emptyset)=\E(\tilde{v}| \tilde{v}<\tau)$, and an interior threshold~$\tau$ is a (candidate) equilibrium cutoff if and only if the marginal firm is indifferent, $\E P(\tau+\tilde{\varepsilon}) - c = P(\emptyset)$, where, as before, we suppress the dependence of $P(\cdot)$ on the conjectured threshold whenever this does not create confusion.

Two primitive assumptions deliver an expected payoff of the marginal discloser that is increasing in the threshold. First, $\tilde{v}$ is absolutely integrable with a continuous strictly positive density $f_v(.)$ and support in  $[\underline{v},\overline{v}]$, such that $\underline{v}$ and $\overline{v}$ are finite. Second, the noise $\tilde{\varepsilon}$ is independent of $\tilde{v}$, has a strictly positive, continuous, log-concave density $f_\varepsilon$ on $\mathbb{R}$, and mean zero; therefore, the signal structure $x=\tilde{v}+\tilde{\varepsilon}$ satisfies the monotone likelihood ratio property (MLRP). It follows that the equilibria still feature a threshold disclosure strategy.\footnote{This result, under bounded support and a general distribution, is formally established in Lemma~9 of the Online Appendix.} Letting the marginal discloser's benefit from disclosure be $\psi(\tau) \equiv \E P(\tau+\tilde{\varepsilon}) -P(\emptyset)$, which is strictly positive for every interior $\tau$, an interior equilibrium threshold $\tau$ is characterized by $\psi(\tau)=c$. The price $\psi(\tau)$ is continuous on $[\underline{v},\overline{v}]$ and hence attains a finite minimum and maximum, with \vspace{-.5cm}
$$0<c_0 \;\equiv\; \min_{\tau\in[\underline{v},\overline{v}]} \psi(\tau)
\;\le\;
\max_{\tau\in[\underline{v},\overline{v}]} \psi(\tau)
\;\equiv\; c_1
\;<\;\infty. \vspace{-.5cm}$$ 

\begin{proposition}\label{mosttwo_2}
The disclosure equilibrium for a given fee $c$ has the following properties:
\begin{itemize}\vspace{-.2cm}
\item[(i)] if $c<c_0$, the unique subgame equilibrium is full disclosure, i.e., $\tau(c)=\underline{v}$;\vspace{-.2cm}
\item[(ii)] if $c\in [c_0,c_1]$, there exists at least one subgame equilibrium; any interior equilibrium threshold $\tau^*$ solves $\psi(\tau^*)=c$;\vspace{-.2cm}
\item[(iii)] if $c> c_1$, the unique subgame equilibrium is no disclosure, i.e., $\tau(c)=\overline{v}$; \vspace{-.2cm}
\item[(iv)] if, in addition, $\tau_0 \in (\underline{v},\overline{v})$ is a local minimizer of $\psi$ and $c \in (\psi(\tau_0),\allowbreak \min\{\psi(\overline{v}),\psi(\underline{v})\})$, then there exist at least two distinct equilibrium thresholds $\underline{\tau}(c)<\tau_0< \overline{\tau}(c)$ satisfying $\psi(\overline{\tau}(c)) = \psi(\underline{\tau}(c)) = c$.\footnote{What the proposition, as stated, does \textit{not} say is whether the region with multiple equilibria is non-empty; the stated assumptions alone do not ensure that the minimum is strictly below both endpoint values. However, we show in the Online Appendix that the region is non-empty with unbounded support and regular tail behavior of the prior, where the benefit diverges in both tails and every cost above $c_0$ supports at least two equilibria (Proposition 20), and for a normal prior truncated at a sufficiently low point (Proposition 21).}
\end{itemize}
\end{proposition}
Proposition~\ref{mosttwo_2} shows that the qualitative structure of equilibria in the disclosure subgame is robust to relaxing normality. The only property of the primitives that matters is the range of the disclosure benefit $\psi(\tau)$ to the marginal discloser, which is determined by the distribution of $\tilde{v}$ and the informativeness of the noisy signal $\tilde{x}=\tilde{v}+\tilde{\varepsilon}$. For low fees $c<c_0$, the payoff to disclosing is always strictly greater than non-disclosure, so that the unique equilibrium unravels to full disclosure ($\tau(c)= \underline{v}$). At the other extreme, for potentially high fees $c>c_1$, under bounded support, the cost of disclosure may be higher than any benefit and a non-disclosure equilibrium ensues ($\tau(c)= \overline{v}$). In-between these values, multiple partial-disclosure equilibria may occur. If an interior minimizer $\tau_0$ satisfies $\psi(\tau_0)<\min\{\psi(\underline{v}),\psi(\overline{v})\}$, there is a non-empty range of fees with at least two equilibria.

We illustrate the role of a lower bound by considering as a special case a truncation at $\eta$ of a normal distribution $N(0,\sigma^2)$. The lower bound $\eta$, which may be positive or negative relative to the underlying mean, can represent a limited-liability floor. It is clear that the price in equation (\ref{EQ}) remains unchanged because the lower truncation does not affect the conditional distribution with $v\geq \tau$. What does change is the fact that $\tau\geq \eta$ \textit{and} the non-disclosure price must now be stated in terms of the truncated distribution:
\begin{equation*}
P(\emptyset)=\mathbb{E}(\tilde{v}|\tilde{v}\in [\eta,\tau]) = \sigma \frac{\phi(\frac{\eta}{\sigma}) - \phi(\frac{\tau}{\sigma})}{\Phi(\frac{\tau}{\sigma}) - \Phi(\frac{\eta}{\sigma})}.
\end{equation*}
In Figure \ref{LBT} (left), this yields similar intuitions except that the non-disclosure price is now flatter and, if $\eta$ is large enough, there can only be a unique equilibrium. 

An immediate corollary is that, relative to the classical noiseless benchmark, noisy information processing enlarges the set of disclosure costs for which the game unravels to full disclosure. Under perfect disclosure, the marginal discloser receives exactly $\tau$. As the threshold approaches full disclosure, the net benefit from disclosure goes to $-c$. Then no positive cost can sustain full disclosure. With noisy processing, the marginal discloser is pooled with higher types, raising the expected disclosure payoff above the realized type and creating a region of positive costs under which full disclosure arises. 
\begin{corollary} \label{cor:noise}
Let $c_0^s \equiv \sup\{c\ge 0: \text{ there exists an equilibrium in which }\tau(c) = \underline{v}\}$ be the maximal cost at which full disclosure is sustained under noise $\tilde{\varepsilon} \sim N(0, s^2)$. Then $c_0^0 = 0 < c_0^s$ \ for all $s > 0$.
\end{corollary}

Under bounded support, $c_1<\infty$ and disclosure eventually shuts down once $c > c_1$ by Proposition \ref{mosttwo_2}. In the unbounded normal baseline, Proposition \ref{mosttwo} implies that the lower-threshold equilibrium exists for arbitrarily large fees and satisfies $\underline{\tau}(c) \to -\infty$ as $c \to \infty$. Bounded support therefore affects equilibrium outcomes in the tail at large fees while leaving unchanged the local forces that generate multiplicity and the comparative statics of the thresholds. Boundedness limits only how far the mechanism can operate, by imposing a finite cap $c_1$ on the maximum disclosure benefit.

\begin{proposition} \label{prop:bounded-cs}
Suppose that $\tau_0 \in (\underline{v},\overline{v})$ is a local minimizer of $\psi$. For $c \in (\psi(\tau_0),\allowbreak \min\{\psi(\overline{v}),\psi(\underline{v})\})$, the equation $\psi(\tau) = c$ admits two distinct solutions $\underline{\tau}(c)<\tau_0< \overline{\tau}(c)$, where $\psi(\underline{\tau}(c))=\psi(\overline{\tau}(c))=c$ and there is no other solution to $\psi(\tau)=c$ in $(\underline{\tau}(c), \overline{\tau}(c))$. Then $\underline{\tau}(c)$ decreases with $c$ and $\overline{\tau}(c)$ increases with $c$ for $c \in (\psi(\tau_0),\allowbreak \min\{\psi(\overline{v}),\psi(\underline{v})\})$.
\end{proposition}

By Proposition \ref{mosttwo_2}, there are two equilibria near the interior minimizer $\tau_0$ when the condition in part (iv) holds: $\underline{\tau}(c)$ is the solution to $\psi(\tau)=c$ immediately to the left of $\tau_0$, while $\overline{\tau}(c)$ is the solution immediately to the right. Similar to the baseline, as $c$ increases above $\psi(\tau_0)$, the interval $\{\tau:\psi(\tau)<c\}$ around $\tau_0$ expands, and its left endpoint moves left while its right endpoint moves right. Thus, under the endpoint condition in Proposition~\ref{prop:bounded-cs}, the lower-threshold equilibrium exhibits more disclosure as fees increase over the stated range, while the upper-threshold equilibrium exhibits less.\footnote{The endpoint condition ensures that this region of fees is non-empty. See also Proposition~22 in the Online Appendix for an illustration. For a truncated normal with lower truncation at $\eta$, lowering $\eta$ sufficiently allows the model to approximate the unbounded baseline over any fixed compact range of fees strictly above $c_0$. That is, as $\eta$ decreases, the truncated model converges to the unbounded model. As $\eta$ increases, however, the non-disclosure price flattens and multiplicity may disappear. Once the threshold approaches the lower bound, the market can no longer make non-disclosure arbitrarily punitive. Thus, the lower threshold cannot fall indefinitely and, at sufficiently high fees, the unique equilibrium is the upper threshold (Proposition 21).}

\subsection{Uncertain Information Endowment}

\vspace{-.3cm}
\paragraph{}
In an alternative environment, certification is costless but firms face uncertainty regarding their information endowment \citep{dye85}. The assumptions of the baseline model remain unchanged, except that (i) the cost is set to $c = 0$, and (ii) the firm may be uninformed or unable to disclose with probability $q \in (0,1)$. In that case, if the firm is unable to disclose, we impose $d(v)=0$; otherwise, with complementary probability $1 - q$, it chooses $d(v)\in\{0,1\}$.

Since disclosure is costless, the indifference condition becomes $\mathbb{E}[P(\tau + \tilde{\varepsilon})] = P(\emptyset)$. The expected price conditional on disclosure, $\mathbb{E}[P(\tau + \tilde{\varepsilon})]$, is identical to expression (\ref{EQ}). Holding fixed the conjectured threshold, an observed disclosure indicates that the firm is able to disclose, and thus the Bayesian updating is the same as in the baseline model. This function increases monotonically in $\tau$ from $-\infty$ to $\infty$.\footnote{The uniqueness of the equilibrium requires only that $\mathbb{E}[P(\tau + \tilde{\varepsilon})]$ be monotonic, which holds under any signal structure satisfying the monotone likelihood ratio property.}

\begin{figure}[H]
\centering
\begin{minipage}[t]{0.48\textwidth}
    \centering
    \includegraphics[width=\textwidth]{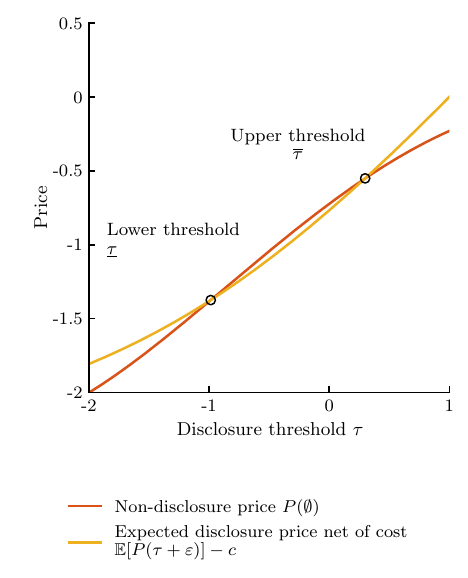}
    \caption*{Truncated normal model ($\eta=-2$, $\sigma=1$, $s=2$, $c=1.5$)}
\end{minipage}
\hfill
\begin{minipage}[t]{0.48\textwidth}
    \centering
    \includegraphics[width=\textwidth]{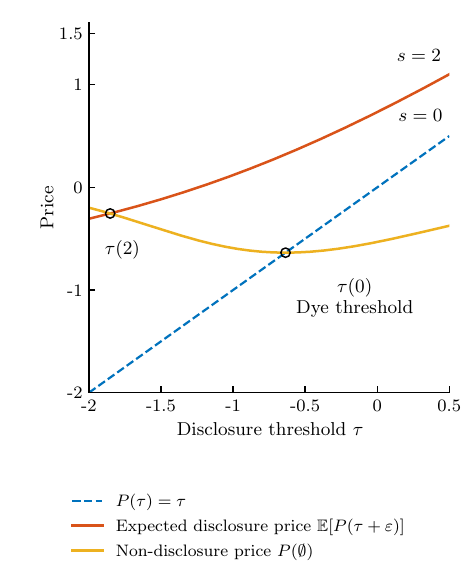}
    \caption*{Dye model ($\sigma=1$, $q=0.2$, $c=0$): noisy disclosure ($s=2$) and the noiseless benchmark ($s=0$)}
\end{minipage}
\caption{Disclosure thresholds for other models for $P(\emptyset)$.}\label{LBT}
\vspace{-.5cm}\end{figure}

The non-disclosure price  $P(\emptyset)$ must incorporate the probability that the firm is unable to disclose. As shown in \citet{junkwo88}, this expectation is\vspace{-0.3cm}
\begin{equation}\label{eq:pempty-dye-noise}
P(\emptyset) = \frac{q \mathbb{E}(\tilde{v}) + (1 - q) \Phi(\tau/\sigma) \mathbb{E}(\tilde{v} | \tilde{v} \leq \tau)}{q + (1 - q) \Phi(\tau/\sigma) }. \vspace{-0.3cm}
\end{equation}
\citet{achdemkre11} show that this non-disclosure price is U-shaped, attaining its minimum at $\tau(0)$, defined as the equilibrium in \citet{junkwo88} when there is no noise ($s = 0$).
\vspace{-0.3cm}
\begin{proposition}\label{dye1}
There is a unique equilibrium given by a threshold $\tau(s)\leq \tau(0)$. The probability of disclosure is increasing in $s$ and decreasing in $\sigma$.
\end{proposition}
In Figure \ref{LBT} (right), the model now features a unique equilibrium with partial disclosure, and the probability of disclosure is monotonically increasing in the level of noise. A new result in this class of models is that, unlike in \citet{junkwo88}, the probability of disclosure depends on fundamental uncertainty $\sigma$ \textit{when noise is present}. That is, in the benchmark model without noise, and under the assumption of a constant $q$, the disclosure probability would not vary with $\sigma$ \citep{per25}. With noisy processing, however, that irrelevance breaks down, and the probability of disclosure decreases in environments with greater fundamental uncertainty. That is, the uncertainty mutes the effect of noise, making it relatively easier to separate out the marginal discloser, thereby weakening incentives to disclose. This comparative static is opposite to that in \citet{ver90}, where higher uncertainty leads to more disclosure in the absence of noise and may help distinguish between types of frictions. \citet{banerjee2024} similarly emphasize that the nature of the friction affects the comparative statics of disclosure probability, although from a different channel.\footnote{The key distinction in \citet{banerjee2024} lies in the shape of beliefs under non-disclosure, whereas our mechanism centers on the interaction between fundamental uncertainty and noise that affects the pooling benefit of disclosure.}

In the Online Appendix, we further examine the joint role of $q$ and a strictly positive disclosure cost~$c$. The marginal-discloser indifference condition then becomes $\mathbb{E}[P(\tau+\tilde\varepsilon)\mid\tau]-c=P(\emptyset)(\tau)$, with $P(\emptyset)$ as in~(\ref{eq:pempty-dye-noise}). Proposition~14 there shows that, over any compact range of intermediate costs, at least three equilibria coexist when exogenous non-disclosure is rare ($q$ small), and that a unique threshold obtains for every $c$ once $q$ is large enough. 
\vspace{-0.3cm}
\subsection{Firm Preferences over Noise}\label{sec:noise}
\vspace{-0.3cm}
\paragraph{}
So far, we have treated the processing noise $s$ as an inherent friction tied to the nature of information communicated by the firm, e.g., complexity, credibility, or processing uncertainty. In the variation considered here, the firm endogenously controls the level of noise. This is, admittedly, an extreme assumption but serves to illustrate, in the most direct manner possible, the firm's incentives to strategically influence how effectively information is processed. There are many possible choices over timing and observability so, to make a baseline choice, we assume that $s$ is privately known and chosen by the firm at the start of the game before it observes its fundamental value $v$. We explain this choice by first discussing alternative information structures.

A different timing would be to choose the noise after fundamentals $v$ are observed. We do not formally solve for this ex-post environment, because it is already discussed in prior literature (\citealt{aghsmi23}) and in our setting, would not yield incremental implications beyond incentives for firms with bad news to strategically obfuscate.\footnote{In practice, it is an empirical question which timing is most descriptive, and the answer to this question may depend on types of obfuscation. The firm may omit strategically important details or bury them in footnotes, making processing more complicated. Alternatively, the firm may spend more time collecting and organizing its information to make it more readable, which would occur early in its reporting period before all uncertainty is resolved.}

Another modeling choice is whether $s$ is observable, which may be plausible in certain settings but implicitly involves an additional source of information. Because there is no practical disclosure of $s$ in the real world, it is an empirical question as to whether investors are fully aware of this uncertainty, or completely unaware, with the reality likely somewhere in-between.\ For presentation purposes, however, observability of $s$ requires yet another secondary modelling choice as to whether $s$ is observed before or after $c$ is observed. So, for expositional purposes, we defer a complete discussion until the intuitions without observability are fully expressed.

In what follows, we thus assume that the firm chooses $s$ ex ante, before the certifier sets its fee, and that the certifier and investors do not observe $s$, so that the certifier and investors act on a common conjecture $\hat{s}$, with choices in the extended set $(0,\infty]$, where $s=\infty$ denotes a report that is uninformative beyond the decision to certify. We need only adjust the prices $P(x)$ in (\ref{eq:price}) to be a function of a market belief $\hat{s}$ about noise, hereafter, denoted  $\hat{P}(x)$. Then, the firm's optimal choice of noise is to maximize \vspace{-0.4cm}
\begin{equation*}
s^*\in \arg\max_{s} \hspace{.2cm}V(s)\equiv \mathbb{E}\left(\max\left(P(\emptyset),\mathbb{E}_\varepsilon \hat{P}(v+\varepsilon)-c\right)\right),
\vspace{-0.4cm}\end{equation*}
where, in the above expression, only $\varepsilon$ depends on the actual $s$ while beliefs depend on the conjectured noise $\hat{s}$. Then, a policy that maximizes $\mathbb{E}_\varepsilon P(v+\varepsilon|\hat{s})$ for any $v$ must necessarily maximize the firm's expected payoff.
\vspace{-0.1cm}
\begin{lemma}\label{leb}
With a finite conjectured threshold and finite conjectured noise $\hat{s}$, $\mathbb{E}_\varepsilon P(v+\varepsilon|\hat{s})$ is strictly increasing in $s$ for every $v$, and $V(s)$ is increasing in $s$.
\end{lemma}
\vspace{-0.3cm}
Because the certifier's equilibrium fee exceeds the unravelling bound and induces a finite threshold, Lemma \ref{leb} implies that the firm always gains from more noise, and we look at the limiting case with $s=\hat{s}\rightarrow \infty$, in which case the certified message no longer contains useful information beyond the fact that $v\geq \tau$, so that the indifference condition reduces to $P(\emptyset)=\mathbb{E}(\tilde{v}|\tilde{v}< \tau)=\mathbb{E}(\tilde{v}|\tilde{v}\geq \tau)-c$,
which can be rewritten as \vspace{-0.3cm}
\begin{equation}\label{cbeau}
c= \mathbb{E}[v | v \geq \tau] - \mathbb{E}[v| v <\tau]
= \frac{\sigma\, \phi\!\left(\tfrac{\tau}{\sigma}\right)}
       {\Phi\!\left(\tfrac{\tau}{\sigma}\right)(1 - \Phi(\tfrac{\tau}{\sigma}))}.\vspace{-0.8cm}
\end{equation}
\begin{lemma}\label{TO}
The function $\mathbb{E}[v | v \geq \tau] - \mathbb{E}[v| v <\tau]$ is U-shaped in $\tau$ with a minimum at $\tau=0$ and unbounded range $[4 \sigma/\sqrt{2 \pi},\infty)$.
\end{lemma}

Using Lemma \ref{TO}, we are now equipped to characterize the solution of the game when the firm privately controls its level of noise, demonstrating, as is intuitive, that this information environment is more costly to the firm as it may incur high certification costs.\footnote{This limit, and the implied equilibrium results, do not reconcile with the results in \cite{marsri15} under their pointwise linear approximation of the expectation, further demonstrating that the approximation is improper even if the noise is large. We instead take the limit of the equilibrium indifference condition as $s\rightarrow \infty$.}
\begin{proposition}\label{sima}
If  $s=\infty$, $c_0=4 \sigma/\sqrt{2 \pi}$ and
\begin{itemize}\vspace{-0.3cm}
\item[(i)] if $c<c_0$, the equilibrium unravels to full disclosure;\vspace{-0.3cm}
\item[(ii)] if $c=c_0$, there is a unique equilibrium with $\tau=\mathbb{E}(v)$ such that half of the firms disclose;\vspace{-0.3cm}
\item[(iii)] if $c>c_0$, there are two distinct equilibria $\underline{\tau}(c)<0<\overline{\tau}(c)$, with $|\underline{\tau}(c)|,\overline{\tau}(c)\rightarrow \infty$ as $c \to \infty$.
\end{itemize}
\end{proposition}
It may seem surprising that the firm pays a cost for an uninformative signal but the intuition is similar to \cite{liz99}; namely, the firm that does not make the disclosure must be perceived as a low type, which penalizes non-purchase and creates an incentive to buy the certification. We can also explicitly characterize the optimal fee under $\overline{\tau}(c)$. Using the value of $c$ from (\ref{cbeau}), the profit to the certifier is \vspace{-0.3cm}
$$\overline{\Pi}(c)=c (1-\Phi(\frac{\overline{\tau} (c)}{\sigma}))=\frac{\sigma\, \phi\!\left(\tfrac{\overline{\tau} (c)}{\sigma}\right)}
       {\Phi\!\left(\tfrac{\overline{\tau} (c)}{\sigma}\right)}.\vspace{-0.3cm}$$
       
       The term on the right-hand side is decreasing in $\overline{\tau}(c)$, so the most the certifier can earn against the upper threshold is $\overline{\Pi}(c_0)=\sigma\phi(0)/\Phi(0)=c_0/2$, which is less than the profit $c_0$ available from unravelling. As in Proposition~\ref{finalized} for $s/\sigma>\nu$, the least-cost certification equilibrium therefore has the lower threshold, at the fee $\hat{c}$ for which $\underline{V}(\hat{c})=c_0$, and (\ref{cbeau}) gives $z^*$ below.\vspace{-0.1cm}
\begin{proposition}
\label{eqb}
With endogenous noise, the least-cost certification equilibrium has $s^*=\infty$, threshold $\underline{\tau}(c^*)=\sigma z^*$ and fee $c^*=c_0/(1-\Phi(z^*))$, where $z^*<0$ is the unique solution of $\phi(z)/\Phi(z)=4/\sqrt{2\pi}$.
\end{proposition} \vspace{-0.1cm}
Numerically, $z^*\simeq-1.09$, so about 86 percent of firms pay a fee of about $1.85\sigma$ for a message that conveys nothing beyond the decision to purchase it; the certifier then earns a large fraction of the value $c_0^*\approx 1.59 \sigma$, equal to the profit it could approach by pricing just below the unravelling bound.\footnote{The assumption that $s$ is unobserved is important because it allows the firm to manipulate market expectations by choosing noisier disclosure. If $s$ is observable, this channel disappears: for any given $c$, the firm's expected market value remains $\mathbb{E}(\tilde v)=0$, so it chooses $s$ only to reduce the expected certification cost. If $s$ is chosen after $c$, the firm's preferred noise then depends on which equilibrium is anticipated: under $\overline{\tau}(c,s)$ it prefers less noise, while under $\underline{\tau}(c,s)$ it prefers more noise. If instead $s$ is observable and chosen before $c$, choosing $s=0$ allows the firm to eliminate the multiplicity and rule out the lower-threshold equilibrium. This last conclusion requires that zero noise be feasible; with a positive lower bound on noise, the same equilibrium-dependent forces reappear.
}

\vspace{-0.2cm}
\subsection{Competition via other information sources}

\vspace{-0.3cm}
\paragraph{}

A natural concern is that the certifier's role may disappear in the presence of competing information providers. To consider an intentionally severe form of competition, suppose that, before the certification game begins, a competitive fringe produces at zero cost a public signal $\tilde{x}_2=\tilde{v}+\tilde{\varepsilon}_2$.
The signal is automatically released and observed by the firm, the certifier, and investors, so that its acquisition or disclosure is not itself a strategic choice. We allow the noise in this signal to be correlated with the noise in the certifier's signal. Specifically, suppose that the two noise terms are jointly normally distributed with mean zero and
\[
\operatorname{Var}(\tilde{\varepsilon}_2)=s_2^2,
\qquad
\operatorname{Corr}(\tilde{\varepsilon},\tilde{\varepsilon}_2)=\rho_{\varepsilon, \varepsilon_2}\in(-1,1),
\]
while \(\tilde{v}\) is independent of
\((\tilde{\varepsilon},\tilde{\varepsilon}_2)\). Thus, the competing signal is informative about the fundamental and, when
\(\rho_{\varepsilon, \varepsilon_2}\neq 0\), also informative about the noise in the certifier's signal.

At first glance, this correlation may appear to change the game. After observing its value \(v\) and the public signal \(x_2\), the firm can infer
\(\varepsilon_2=x_2-v\), which provides information about the certification noise \(\varepsilon\). Nevertheless, this does not give the firm a second independent dimension of private information. Conditional on the public realization \(x_2\), the inferred value \(\varepsilon_2=x_2-v\) is an affine function of \(v\). Hence, conditional on \(x_2\), the firm's private information remains one-dimensional and the certification signal can be orthogonalized to recover the baseline structure.

To see this, one can define the public posterior means
\[
\mu_v(x_2)\equiv
\mathbb{E}(\tilde{v}\mid \tilde{x}_2=x_2)
=
\frac{\sigma^2}{\sigma^2+s_2^2}x_2, \quad
\mu_x(x_2)\equiv
\mathbb{E}(\tilde{x}\mid \tilde{x}_2=x_2)
=
\frac{\sigma^2+\rho_{\varepsilon, \varepsilon_2} s s_2}{\sigma^2+s_2^2}x_2.
\]
For $\kappa\equiv 1-\rho_{\varepsilon, \varepsilon_2}\frac{s}{s_2}\ne 0$, 
transformed fundamental and certification signals can be written as
\begin{equation*}
\tilde{v}'\equiv
\tilde{v}-\mu_v(\tilde{x}_2),
\qquad
\tilde{x}'\equiv
\frac{\tilde{x}-\mu_x(\tilde{x}_2)}{\kappa},
\end{equation*}
implying that $\tilde{x}'=\tilde{v}'+\tilde{\varepsilon}'$, where 
$\tilde{\varepsilon}'
\equiv
\frac{\tilde{\varepsilon}-\rho_{\varepsilon, \varepsilon_2}(s/s_2)\tilde{\varepsilon}_2}{\kappa}$ is the renormalized noise term. The effective variances in the transformed game are
\begin{equation*}
{\sigma'}^2
\equiv
\operatorname{Var}(\tilde{v}'\mid\tilde{x}_2)
=
\frac{\sigma^2s_2^2}{\sigma^2+s_2^2},
\qquad
{s'}^2
\equiv
\operatorname{Var}(\tilde{\varepsilon}')
=
\frac{s^2(1-\rho_{\varepsilon, \varepsilon_2}^2)}
{\left(1-\rho_{\varepsilon, \varepsilon_2} s/s_2\right)^2}.
\end{equation*}Conditioning on
\(\tilde{x}_2\) changes both the residual uncertainty about fundamentals and the relevant noise.

This argument only establishes robustness to potential competition, but not the incremental effect of competition. Although a model of competition may require a full model of strategic interaction, some partial intuition as to why competition may not necessarily reduce price may still be obtained. Suppose that there are $n$ certifiers, indexed by $i$, each charging a fee $c_i$ and issuing a (for simplicity) symmetric noisy signal $x_i=v+\varepsilon_i$, where $\varepsilon_i$ are uncorrelated white noise with variance $s^2$. We consider first the firm's choice and conjecture an equilibrium in which the (symmetric) strategy is to buy all signals when $v\geq \tau(\sum c_i)$ and buy no signal otherwise. These equilibria can always be sustained as long as selectively buying signals would involve unfavorable expectations. Selective buying equilibria are harder to sustain because buying fewer signals reveals weaker fundamentals, offsetting the potential effect of signal noise. Also, on a symmetric equilibrium path, certifiers will choose the same fee. 

Under this conjecture, the firm's problem is identical to the baseline, with two adjustments. First, the residual noise falls to $s'=s/\sqrt{n}$, and second, the relevant cost is $C=\sum c_i$. The main insights from the baseline are unchanged, and can then be obtained from the comparative statics on $s'$ and $C$.\ In particular, holding the total fee fixed, more competition lowers $s'$ and therefore, by Corollary~\ref{dog}, tends to decrease disclosure under $\overline{\tau}$ and increase it under $\underline{\tau}$.

Moving to the certifiers' problem, under $\underline{\tau}$, any increase in the fee will continue to increase the probability of disclosure, so incentives to increase the fee are unchanged in this equilibrium.\ Under $\overline{\tau}$, the objective function is slightly changed, because a competitor receives $c_i (1-\Phi(\overline{\tau}(c_i +\sum_{j\ne i} c_j)/\sigma))$, whose derivative in $c_i$ is \textit{higher} than in the baseline objective $C (1-\Phi(\overline{\tau}(C)/\sigma))$, because each individual certifier receives only $c_i<C$ per certification and internalizes only part of the demand loss caused by raising its fee. Competition therefore need not reduce fees.

\subsection{Risk-Averse Investors } \label{sec:hetinv}
\vspace{-0.3cm}
\paragraph{}

With risk-neutral investors, prices equal posterior means, so investors are indifferent across disclosure equilibria in an asset-demand sense. Mean-variance preferences may break that indifference. Suppose that a continuum of investors $i\in[0,1]$ shares risk aversion $\gamma>0$, the risky asset is in unit net supply, and every investor sees the same public information generated by the firm's disclosure decision. Conditional on public information $I$, investor $i$ chooses a position $q_i$ to maximize
\vspace{-0.2cm}
\begin{equation}\label{eq:mv-demand}
q_i\bigl(\E[\tilde v\mid I]-P(I)\bigr)-\frac{\gamma}{2}q_i^2\operatorname{Var}(\tilde v\mid I).
\vspace{-0.2cm}
\end{equation}
Because investors have identical posterior beliefs, market clearing implies $q_i=1$ and therefore
\vspace{-0.2cm}
\begin{equation*}
P_\gamma(I)=\E[\tilde v\mid I]-\gamma\operatorname{Var}(\tilde v\mid I).
\vspace{-0.2cm}
\end{equation*}
Risk aversion therefore changes prices through the residual-risk discount but does not change investors' information.

For a conjectured threshold $\tau$, we write $B_0(\tau)\equiv\E[P(\tau+\tilde\varepsilon;\tau)\mid\tilde v=\tau]-P(\emptyset;\tau)$ for the marginal firm's gross benefit from disclosure at the risk-neutral prices of Section~\ref{sec:analysisF}, whose minimum is the unravelling bound $c_0$, and $B_\gamma(\tau)$ for the same benefit at the risk-adjusted prices defined above. Differentiating the two risk-neutral prices shows that $\sigma^2B_0'(\tau)$ equals the posterior variance after silence minus the marginal discloser's expected posterior variance after disclosure, and the corresponding risk discounts add $\gamma$ times that difference to the disclosure benefit (the derivation is in the Online Appendix), so that
\vspace{-0.2cm}
\begin{equation}\label{eq:B-gamma}
B_\gamma(\tau)=B_0(\tau)+\gamma\sigma^2B_0'(\tau).
\vspace{-0.2cm}
\end{equation}
By~\eqref{eq:B-gamma}, any finite candidate threshold satisfies $B_\gamma(\tau)=c$.
\begin{proposition}\label{prop:hetinv}
Let $c_0(\gamma)\equiv\min_{\tau}B_\gamma(\tau)$ and let $\tau_0$ be the minimizer of $B_0$.
\begin{itemize}\vspace{-0.25cm}
\item[(i)] $c_0(\gamma)$ is finite and strictly decreasing in $\gamma$, with $c_0(0)=c_0$, and, for every $\gamma>0$, every minimizer of $B_\gamma$ lies below $\tau_0$;\vspace{-0.2cm}
\item[(ii)] if $\gamma<\overline\gamma$ for some $\overline\gamma > 0$, then $B_\gamma(\tau)=c$ has no solution for $c<c_0(\gamma)$ and at least two for $c>c_0(\gamma)$, each a threshold equilibrium;\vspace{-0.2cm}
\item[(iii)] at any fixed cost $c>c_0$, for $\gamma$ small enough there are exactly two equilibria, each equal to its risk-neutral counterpart minus $\gamma\sigma^2+o(\gamma)$.
\end{itemize}\vspace{-0.3cm}
\end{proposition}
The bound $\overline\gamma$ (defined in the Online Appendix) ensures increasing prices. Because $B_0'$ is bounded, $B_\gamma$ keeps the tails of the risk-neutral benefit, so multiplicity survives risk aversion under the bound in part (ii). In terms of first order terms, risk aversion shifts the entire disclosure-benefit curve to the left by $\gamma\sigma^2$, which is why both thresholds fall and disclosure rises in both equilibria while the unravelling bound moves only according to second order terms.

Which equilibrium mean-variance investors prefer turns on how much uncertainty each one leaves in the market. Let $I_\tau$ denote the public information generated by threshold $\tau$, so that $I_\tau=\emptyset$ after non-disclosure and $I_\tau=x$ after disclosure, and define the expected residual posterior variance $\mathcal V(\tau)\equiv\E[\operatorname{Var}(\tilde v\mid I_\tau)]$. We evaluate investor surplus by the ex-ante average of the conditional certainty equivalents in~\eqref{eq:mv-demand}.

\begin{proposition}\label{prop:invpref}
For $\gamma>0$, investors' expected conditional certainty-equivalent trading surplus under threshold $\tau$ is
$S_I(\tau)=\frac{\gamma}{2}\mathcal V(\tau)$. Hence, for any two threshold equilibria at a common cost, investors prefer the one with the larger $\mathcal V$.
\end{proposition}

Conditional on $I_\tau$, the price discounts expected value by $\gamma\operatorname{Var}(\tilde v\mid I_\tau)$, while an investor holding one unit carries a risk cost of only half that amount. Investors therefore prefer the equilibrium with greater residual risk. In our setting, that ranking is not determined by disclosure frequency because disclosure is noisy and silence is informative through truncation.

Total surplus $W(\tau;c)$, the sum of the firm's ex-ante payoff and investors' expected conditional certainty-equivalent surplus, nets out the market price, so that, after the normalization $\E(\tilde v)=0$, it equals $-\tfrac{\gamma}{2}\mathcal V(\tau)-c\,(1-\Phi(\tau/\sigma))$. The derivation, together with the case in which the fee accrues to a certifier inside the model, is in the Online Appendix. Unlike investor surplus, total surplus decreases with residual risk.

\begin{figure}[H]\centering
\includegraphics[width=\linewidth]{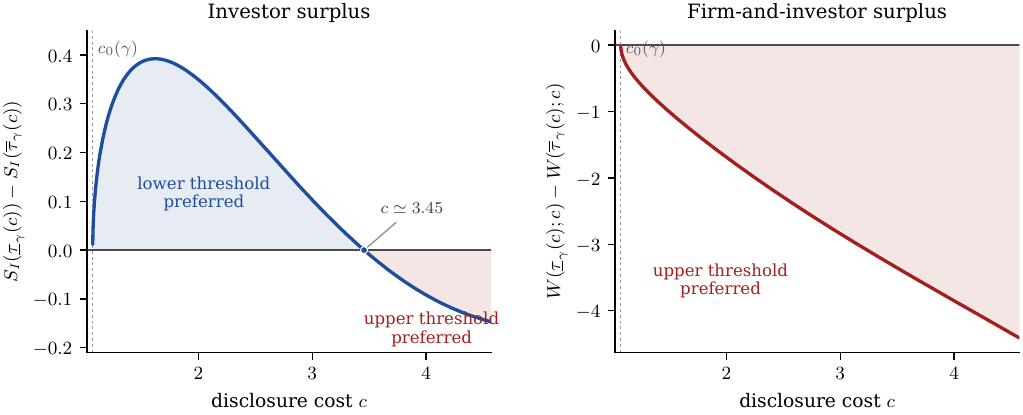}
\caption{\textit{Ranking of the two threshold equilibria for $\sigma=1$, $s=2$, and $\gamma=1.75$.} The left panel plots the investor-surplus difference $S_I(\underline{\tau}_\gamma(c))-S_I(\overline{\tau}_\gamma(c))$, and the right panel the total-surplus difference $W(\underline{\tau}_\gamma(c);c)-W(\overline{\tau}_\gamma(c);c)$, both against the disclosure cost $c\geq c_0(\gamma)\simeq 1.072$. Positive values indicate that the lower-threshold equilibrium is preferred. Investors prefer the lower threshold up to $c\simeq 3.45$, whereas total surplus favors the upper threshold over the plotted range.}
\label{plot:hetinv}\vspace{-.5cm}
\end{figure}

How informative certification is determines which equilibrium leaves more residual variance. In the uninformative case $s=\infty$ of Section~\ref{sec:noise}, the decision to certify is the only information investors receive, so $\mathcal V(\tau)=\sigma^2\,(1-\phi(\tau/\sigma)[\lambda(\tau/\sigma)+\lambda(-\tau/\sigma)])$ is lowest at $\tau=0$, where that decision is most informative, and rises with $|\tau|/\sigma$.
\vspace{-0.2cm}
\begin{proposition}\label{prop:hetinv-local}
Let $s=\infty$ and $\gamma\sigma<4/\sqrt{2\pi}$. Then $0<c_0(\gamma)<c_0$ and, for every $c>c_0(\gamma)$, investors prefer the threshold equilibrium with the most disclosure, while the firm and total surplus rank that equilibrium below every other. If instead $s<\infty$ and $\gamma<\overline\gamma$, investors, the firm, and total surplus all prefer the threshold equilibrium with the least disclosure once $c$ is large enough.
\end{proposition}
\vspace{-0.2cm}
With uninformative certification, the risk adjustment lowers the disclosure benefit by more at low thresholds and thereby places the equilibrium with more disclosure further from the prior mean than the equilibrium with less. The certification decision in the more distant equilibrium reveals less, which investors prefer because the risk discount they collect in the price exceeds the risk they are exposed to. The firm pays that discount together with the certification cost, so the firm and total surplus prefer the equilibrium with less disclosure. Informative certification changes the comparison at high costs, where near-universal disclosure reveals fundamentals up to the processing noise while near-universal silence reveals nothing, and investors then prefer the opaque equilibrium, as in \citet{gutkadkan06}.
At $(\sigma,s,\gamma)=(1,2,1.75)$, $\gamma<\overline\gamma\simeq1.84$ and $c_0(\gamma)\simeq1.072$. Numerical evaluation gives exactly two finite roots of $B_\gamma(\tau)=c$ over the range plotted in Figure~\ref{plot:hetinv}, which traces both payoff differences. Investors prefer the lower threshold up to $c\simeq3.45$ and the upper threshold beyond; further, total surplus favors the upper threshold throughout.

\section{Empirical Implications}\label{sec:empirical}
\vspace{-0.3cm}
\paragraph{}
The two main applications introduced above, the proprietary-cost hypothesis and the ESG-rating industry, together with the comparative statics derived in Sections~\ref{sec:analysisF} and~\ref{sec:cert}, deliver empirical predictions that standard models of voluntary disclosure do not produce. The classic \citet{ver83} prediction is that higher proprietary costs unambiguously reduce voluntary disclosure. Empirical tests, however, deliver mixed results, with the cost-disclosure correlation switching sign across samples and across proxies (see the discussion and references in \citealt{beycohlys10}). Proposition~\ref{mosttwo} and Corollary~\ref{dog} rationalize these patterns. In the low-disclosure regime (upper threshold $\overline\tau(c)$), the sign of the cost-disclosure correlation is the conventional one, so that higher costs reduce disclosure. In the high-disclosure regime (lower threshold $\underline\tau(c)$), the sign is \emph{reversed}, and higher costs are associated with \emph{more} disclosure, because in equilibrium the increased disclosure benefit needed to offset the higher cost must come from a more punitive inference at non-disclosure, which requires lowering $\tau$. When the cost proxies competitive pressure, the same logic implies that competition can either raise or lower voluntary disclosure depending on the prevailing regime, one candidate explanation for the mixed evidence on competition and disclosure reviewed by \citet{lansul14}. 

Empirical implementation therefore requires information about the fundamental and noise distributions, or context identifying whether the information environment features a high or low prevailing disclosure regime. Conditional on this determination, the cost-disclosure correlation should be \emph{positive} on the lower threshold and \emph{negative} on the upper threshold. Yet, because the disclosure regime is typically unobserved at the firm-quarter level, the conditional test may not always be feasible. We therefore provide a joint test that does not require knowing which equilibrium is played. By Corollary~\ref{dog}, in any single regime the comparative statics of the disclosure probability are sign-consistent across the three primitives when volatility is below the processing noise, since in the upper-threshold equilibrium disclosure decreases in $c$ and increases in $s$ and $\sigma$, while in the lower-threshold equilibrium disclosure increases in $c$ and decreases in $s$ and $\sigma$. In any regime, therefore, disclosure covaries with the proprietary cost in one direction and with $s$ in the opposite direction, and with $\sigma$ likewise when $\sigma<s$.

This prediction can be separated from the standard \citet{ver83} model as long as $s$, which is not part of that model, is observed, or when the disclosure probability varies positively in $c$ or negatively in $\sigma$. The test is less stringent than the implication of standard proprietary cost theory but nevertheless places testable restrictions on joint relations between disclosure and primitives.

Taking the joint test to data requires proxies for the three primitives. For the proprietary cost, the natural starting point is industry concentration, with the caveat in \citet{lansul14} that concentration built from public firms alone measures competitive pressure with substantial error, so that R\&D intensity and related competition measures are useful complements. Fundamental uncertainty $\sigma$ is the most direct of the three, proxied by the volatility of returns or of realized earnings. Processing noise $s$ is a property of how a disclosure is absorbed rather than of the fundamentals themselves, so its proxies should capture the difficulty of mapping the disclosed information into value, such as the length and linguistic complexity of the filing, the share of qualitative relative to quantitative content, or the dispersion of analyst forecast revisions following the disclosure. Within a regime, the model then predicts that the estimated sign on the cost proxy is opposite to the sign on the processing-noise proxy and, when volatility is below the noise, on the volatility proxy.

A further question in voluntary disclosure theory is whether non-disclosure is due to cost frictions or to practical restrictions on the verifiability of messages. \cite{dye85}, for example, observes that firms cannot credibly disclose the absence of information, which may lead to non-disclosure equilibria that look a priori similar, that is, threshold-like, to equilibria due to costs. Because we do not observe the friction directly and, theoretically, the information state of the manager is assumed to be unobservable, these theories can be difficult to distinguish.

Proposition~\ref{dye1} delivers a sharp contrast between the two main frictions used in the disclosure literature. In the costly disclosure model (Section~\ref{sec:analysisF}), the probability of disclosure increases in fundamental volatility $\sigma$ on the upper threshold when $\overline{\tau}(c)>0$ or $\sigma<s$, and decreases on the lower threshold when $\sigma<s$. In a pure Dye model, the probability of disclosure is not a function of $\sigma$, since volatility simply scales all variables, whereas with noisy processing the probability of disclosure strictly \emph{decreases} in $\sigma$ for any $s>0$. The two frictions therefore yield opposite empirical predictions on the upper threshold under the stated conditions, with the costly-disclosure model predicting $\partial\Pr(\text{disclose})/\partial\sigma>0$ and the noisy Dye--Jung--Kwon (DJK) model predicting $\partial\Pr(\text{disclose})/\partial\sigma<0$.\footnote{Empirical studies sometimes interpret the probability that information is unavailable to be plausibly related to the uncertainty, which implies a similar comparative static (which holds in the original Dye model) that more uncertainty would decrease disclosure.}

In the ESG-rating application, the selected equilibrium switches from the upper-threshold continuation at fee $c^{**}$ to the lower-threshold continuation at fee $\hat c$ as $s/\sigma$ rises through $\nu$. At this switch, the fee falls and adoption rises: both limiting outcomes earn $c_0$, and the lower-threshold continuation attains that revenue at a lower fee because it has greater demand at a given fee. The high-noise regime can nevertheless sustain substantial fees and broad adoption because silence conveys unfavorable information.

Empirical comparisons across industries or jurisdictions with different rating standards should therefore use the equilibrium fee and adoption schedules in Proposition~\ref{finalized}. They do not imply a universal ordering of fees and adoption by rating quality; greater noise within the high-noise region need not increase adoption. Holding the fee and signal distributions fixed, the lower-threshold continuation has a mean disclosure price closer to the prior mean and a more punitive silence price.

      \vspace{-0.3cm}

\section{Conclusion}\vspace{-0.3cm}
\paragraph{}
This paper examines how noisy information processing interacts with disclosure frictions. A firm privately observes its fundamental information and chooses whether to disclose for a cost to produce a verifiable, but noisily processed, report. Investors price the firm based on their processed report combined with conjectures about the firm's disclosure strategy. Even in this simple environment, adding realistic noise to the report has sharp implications for the existence, structure, and comparative statics of voluntary disclosure equilibria.

Our first main insight is that any amount of processing noise generically overturns the classical result of a unique partial-disclosure equilibrium. When the certifier's fee is low, the unique equilibrium exhibits full disclosure, so even modest noise does not prevent unraveling. As the fee increases beyond a critical cost $c_0$, the model typically admits two distinct threshold equilibria: one with relatively high disclosure and one with relatively low disclosure. This multiplicity arises because noise alters not only how investors interpret disclosures, but also the inference when remaining silent relative to those reports.

Second, the model delivers comparative statics that differ from the benchmark with perfectly precise disclosure. In the upper-threshold equilibrium, disclosure decreases as disclosure becomes more costly or more precise, and increases with fundamental volatility when the threshold is positive or volatility is below the processing noise. In the high-disclosure equilibrium, however, the relationship between disclosure and frictions can be reversed or non-monotonic: disclosure is more frequent when noise is lower or costs are higher, and changes in uncertainty can push the equilibrium toward full disclosure. These results highlight that empirical patterns linking disclosure to frictions may depend critically on which equilibrium the capital market will coordinate on.

Third, by endogenizing the cost as a fee charged by a third party, we link equilibrium selection to the economics of certification. A profit-maximizing certifier chooses fees knowing that the disclosure subgame may admit multiple equilibria. As a result, some parameter configurations lead to high-fee/high-disclosure regimes, whereas others favor low-fee/low-disclosure regimes. This perspective suggests that observed variation in audit or rating fees and in verification intensity across settings need not simply reflect cost heterogeneity; it may also reflect different equilibrium regimes induced by noisy processing.

The analysis also yields testable implications for empirical work in accounting and finance. The model predicts that disclosure behavior can switch regimes as costs, uncertainty, or processing noise change, and provides guidance on how to use the overall level of disclosure to infer which comparative statics should be expected in a given environment. Once the equilibrium is identified, disclosure decreases in costs and increases in noise on the upper threshold, with the opposite signs on the lower threshold. Within either threshold, joint comparative statics can be tested, with the prediction that noise and the variance of fundamentals affect disclosure in the same direction and in the opposite direction to the cost, with the volatility prediction applying when volatility is below noise.

The model is stylized and suggests several avenues for future research. Extending the analysis to dynamic settings, richer heterogeneity in investors' information processing, competition among multiple certifiers, or explicit enforcement and litigation frictions can bring the framework closer to specific institutional questions. Our results indicate that incorporating realistic processing frictions into disclosure models can generate qualitatively new predictions and help reconcile seemingly conflicting empirical findings on the relation between disclosure, verification, and information frictions. We view this as a useful step toward a more complete theory of disclosure in noisy environments.

\vspace{-0.2cm}

\newpage

\begin{figure}[H]
\begin{center}
\psset{linecolor=black,linewidth=1pt}

\psset{linecolor=black,linewidth=1pt}

\begingroup
\setlength{\unitlength}{0.01\linewidth}
\begin{picture}(100,20)(0,-5)
  \put(12,0){\line(1,0){75}}
  \put(12,0){\circle*{0.75}}
  \put(12,-1.2){\makebox(0,0)[t]{\scriptsize $t=1$}}
  \put(12,3){\makebox(0,0)[b]{\scriptsize\parbox{0.23\linewidth}{\centering Certifier publicly commits to fee $c$ (noise $s$ exogenous)}}}
  \put(37,0){\circle*{0.75}}
  \put(37,-1.2){\makebox(0,0)[t]{\scriptsize $t=2$}}
  \put(37,3){\makebox(0,0)[b]{\scriptsize\parbox{0.23\linewidth}{\centering Firm observes $v$ and decides to disclose iff $v \geq \tau$}}}
  \put(62,0){\circle*{0.75}}
  \put(62,-1.2){\makebox(0,0)[t]{\scriptsize $t=3$}}
  \put(62,3){\makebox(0,0)[b]{\scriptsize\parbox{0.23\linewidth}{\centering Certifier issues report $\tilde{x} = \tilde{v} + \tilde{\varepsilon}$, $\tilde\varepsilon \sim N(0,s^2)$}}}
  \put(87,0){\circle*{0.75}}
  \put(87,-1.2){\makebox(0,0)[t]{\scriptsize $t=4$}}
  \put(87,3){\makebox(0,0)[b]{\scriptsize\parbox{0.23\linewidth}{\centering Investors form price $P(x)$ or $P(\emptyset)$ }}}
\end{picture}
\endgroup\end{center}\vspace{-0.5cm}
\caption{Certification game timeline.} \label{tim1}\vspace{-0.5cm}
        \end{figure}

        
\newpage
\section*{Appendix}
\vspace{-0.3cm}
In this appendix, we use $\lambda(z)\equiv \phi(z)/(1-\Phi(z))$ as the inverse Mills ratio, with $\phi(z)$ (resp., $\Phi(z)$) indicating the pdf (resp., cdf) of the standard normal distribution.\ In parts of the proof, we also simplify notations using $\rho\;\equiv\;\frac{s}{\sqrt{\sigma^2+s^2}}\in (0,1)$, $\theta_s(y)\equiv \frac{\tau s/\sigma-\sigma y}{\sqrt{\sigma^{2}+s^{2}}}=\frac{\tau\rho}{\sigma}-y\sqrt{1-\rho^2}$, and $\theta\equiv \theta_s(\tilde Y)$, with $\tilde{Y}$ defined as a standard normal random variable. This appendix contains the proofs of Lemmas \ref{L1} to \ref{hyena} and of Proposition \ref{mosttwo}; all remaining proofs are in the Online Appendix.

\vspace{-0.4cm}

\paragraph{Proof of Lemma \ref{L1}.}
Let $\Delta(\tau;c)\;=\;\E[P(x;c)\mid v=\tau]-c-P(\emptyset;c).$ Since $P(x;c)\geq\tau$ for every $x$, the expected disclosure price diverges as $\tau\to\infty$, while $P(\emptyset;c)\to0$, so $\Delta(\tau;c)\to\infty$. For the other tail, the truncated-normal expression~(\ref{PEX}) gives $\E[P(x;c)\mid v=\tau]=a(s)\tau+b(s)\E[\lambda(\theta_s(\tilde Y))]$. For $\tau\leq0$, the integrand is bounded by the integrable $\lambda(-\sqrt{a(s)}\,\tilde Y)$ and converges pointwise to zero as $\tau\to-\infty$. Dominated convergence therefore gives $\E[P(x;c)\mid v=\tau]=a(s)\tau+o(1)$, whereas \vspace{-0.5cm}

\begin{equation}\label{empty1}
P(\emptyset;c)
=-\sigma\,\lambda\Big(-\frac{\tau}{\sigma}\Big)\sim \tau.
\end{equation} 
Therefore\vspace{-0.3cm}
\begin{equation}\label{eq:Delta-tails}
\lim_{\tau\to-\infty}\Delta(\tau;c)=\lim_{\tau\to+\infty}\Delta(\tau;c)=\infty.\vspace{-0.2cm}
\end{equation}
Either (a) there exists a finite $\tau^\star$ with $\Delta(\tau^\star;c)=0$, which yields an interior threshold equilibrium, or (b) $\Delta(\tau;c)>0$ for all $\tau\in \mathbb{R}$, in which case the equilibrium is unique with $\tau(c)=-\infty$ (full disclosure).
Because it cannot be the case that $\Delta(\tau;c)\leq 0$ for all $\tau$, Definition~\ref{d1}(ii) rules out $\tau(c)=\infty$ (no disclosure).\qed

\vspace{-0.4cm}

\paragraph{Proof of Lemma \ref{zebra}.} 
Under normal updating,\vspace{-0.3cm}
\[
\tilde v\mid x \sim N \Big(a(s)\,x,\,b(s)^2\Big),
\quad \text{where }
a(s)\,=\frac{\sigma^2}{\sigma^2+s^2}\,
\; \text{ and } \;
b(s)=\frac{\sigma s}{\sqrt{\sigma^2+s^2}},
\]\vspace{-0.1cm}
so that $
P(x)\;=\;\E\left[\tilde v \mid \tilde v\ge \tau,\,\tilde x=x\right]
\;=\; a(s)\,x \;+\; b(s)\,\lambda\!\left(\frac{\tau-a(s)\,x}{b(s)}\right).$
For the marginal discloser $\tilde v=\tau$, we have $\tilde x=\tau+\tilde\varepsilon=\tau+s\tilde Y$ with $\tilde Y\sim N(0,1)$. Taking expectations over $\tilde Y$, 
\begin{equation}\label{PEX}\vspace{-0.3cm}
\E\!\left[P(x)\mid \tilde v=\tau\right]
= a(s)\,\tau \;+\; b(s)\,\E\!\left[\lambda\!\left(\theta_s(\tilde{Y})\right)\right],
\end{equation}
where $\theta_s(y)\equiv \frac{\tau s/\sigma-\sigma y}{\sqrt{\sigma^{2}+s^{2}}}$. Differentiating inside the expectation:\vspace{-0.3cm}
\begin{equation}\label{green}\vspace{-0.3cm}
\frac{\partial}{\partial\tau}\,\E\!\left[P(x;c)\mid \tilde v=\tau\right]
= a(s) \;+\; \frac{s^2}{\sigma^2+s^2}\,
\E\!\big[\lambda'(\theta_s(\tilde Y))\big],
\end{equation}
because $b(s)\,\frac{\partial}{\partial \tau} \theta_s(\tilde Y)=\frac{s^2}{\sigma^2+s^2}$. 
Since $\lambda'(z)=\lambda(z)\big(\lambda(z)-z\big)\in(0,1)$ for all $z$ \citep{sampford1953some} and $a(s)=\frac{\sigma^2}{\sigma^2+s^2}\in(0,1)$, it follows that the derivative is in $(0,1)$. \qed

\vspace{-0.4cm}
\paragraph{Proof of Lemma \ref{hyena}.} The statement is (\ref{eq:Delta-tails}), established in the proof of Lemma \ref{L1}. \qed

\vspace{-0.4cm}

\paragraph{Proof of Proposition \ref{mosttwo}.} Differentiating the non-disclosure price (from equation (\ref{empty1})), 
\vspace{-0.2cm}\[
\frac{\partial}{\partial\tau}P(\emptyset)=\lambda'\!\Big(-\frac{\tau}{\sigma}\Big),
\qquad
\frac{\partial^2}{\partial\tau^2}P(\emptyset)=-\frac{1}{\sigma}\,\lambda''\!\Big(-\frac{\tau}{\sigma}\Big)\;<\;0.
\]
From Lemma~\ref{zebra} and equation (\ref{green}), $\frac{\partial}{\partial\tau}\E\!\big[P(\tau+\tilde\varepsilon)\mid \tau\big] \in (0,1)$ and
\vspace{-0.2cm}\begin{eqnarray*}
\frac{\partial^2}{\partial\tau^2}\E\!\big[P(\tau+\tilde\varepsilon)\mid \tau\big]
&=&\frac{s^2}{\sigma^2+s^2}\cdot\frac{s}{\sigma\sqrt{\sigma^2+s^2}}\,
\E\!\big[\lambda''(\theta_s(\tilde Y))\big]\;>\;0,
\end{eqnarray*}
since the inverse Mills ratio is convex. Hence $\E[P(\tau+\tilde\varepsilon)\mid\tau]$ is strictly convex, while $P(\emptyset)$ is strictly concave. Therefore, $\Delta(\cdot;c)$ is strictly convex in $\tau$.
By Lemma~\ref{hyena},  $\Delta(\cdot;c)$ is strictly convex and diverges to $\infty$ at both tails, so it has either no finite root, a unique tangency, that is, a double root, or exactly two roots.

Further, strict convexity yields a unique minimizer $\tau^*$ of $\Delta(\cdot;c)$, characterized by the first-order condition $\frac{\partial}{\partial\tau}\Delta(\tau^*;c)=0$.
Moreover, $\lim_{\tau\to-\infty}\frac{\partial}{\partial\tau}\Delta(\tau;c)=a(s)-1<0$ implies that $\tau^*$ is finite.
We show next that $\tau^*<0$ by evaluating $\frac{\partial}{\partial\tau}\Delta(\tau;c)$ at $\tau=0$. First, for the non-disclosure price, $\frac{\partial}{\partial \tau}P(\emptyset)\Big|_{\tau=0}=\lambda'(0)=\frac{2}{\pi}$. Second, for the disclosure expected payoff, let us define:
\vspace{-0.2cm}\[
H(a) \equiv\frac{\partial}{\partial\tau}\E\!\big[P(\tau+\tilde\varepsilon)\mid\tau\big]\Big|_{\tau=0}
= a + (1-a)\,\E\!\big[\lambda'(-\sqrt{a}\,\tilde Y)\big].
\]
We need to show that \(H'(a)>0\) on \([0,1]\), since it then immediately implies \(H(a)>H(0)=\lambda'(0)=2/\pi\) and so \(\frac{\partial}{\partial\tau}\Delta(0;c)>0\). To show this, we use Stein's Lemma and the identity
\vspace{-0.2cm}\[
\lambda'''(z)=(2\lambda(z)-z)\lambda''(z)+2\bigl(\lambda'(z)-1\bigr)\lambda'(z),
\]
with $\lambda''(.)>0$ and $2\lambda(z)-z>0$ (since $\lambda'(z)=\lambda(z)(\lambda(z)-z)>0$). This implies:\vspace{-0.2cm}
\begin{eqnarray*}
H'(a) & = & 1 - \E\!\big[\lambda'(-\sqrt{a}\,\tilde Y)\big]
  + \frac{1-a}{2}\,\E\!\big[\lambda'''(-\sqrt{a}\,\tilde Y)\big] \\ & \ge & 1 - \E\!\big[\lambda'(-\sqrt{a}\,\tilde Y)\big]
  + \left(1-a\right) \E\!\big[\left(\lambda'(-\sqrt{a}\,\tilde Y) - 1\right) \lambda'(-\sqrt{a}\,\tilde Y)\big] \\ & \ge & 1 - \E\!\big[\lambda'(-\sqrt{a}\,\tilde Y)\big]
  + \left(1-a\right) \left(\left(\E\!\big[\lambda'(-\sqrt{a}\,\tilde Y)\big]\right)^2 - \E\!\big[\lambda'(-\sqrt{a}\,\tilde Y)\big] \right) \\ & = & \left(1 - \E\!\big[\lambda'(-\sqrt{a}\,\tilde Y)\big] \right) \left(1 - \left(1-a\right) \E\!\big[\lambda'(-\sqrt{a}\,\tilde Y)\big] \right)>0, 
\end{eqnarray*}
where the last inequality follows from \(0<\lambda'(z)<1\). 

Because $P(x)\geq \tau$ while $P(\emptyset)<\tau$, the benefit $\Delta(\tau;0)$ is strictly positive at every finite threshold, including the minimizer. Then, $c_0>0$ is uniquely defined as $\Delta(\tau^*;c_0)=0$, and the main claim in the Proposition follows immediately from noting that $\Delta(\tau;c)=0$ has no solution (resp., two solutions)
if $c<c_0$ ($c>c_0$), and a single solution $\tau(c_0)=\tau^*$ if $c=c_0$.
Furthermore, because $\Delta(\cdot;c)$ is strictly convex and minimized at $\tau^*<0$, the two roots at $c>c_0$ must satisfy $\underline{\tau}(c)\;<\;\tau^*\;<\;\overline{\tau}(c)$. At either root $\tau(c)$, the implicit function theorem yields\vspace{-0.2cm}
\begin{equation*}
\tau'(c)
=-\frac{\partial\Delta/\partial c}{\partial\Delta/\partial\tau}
=\frac{1}{\,\frac{\partial}{\partial\tau}\E[P(\tau+\tilde\varepsilon)\mid\tau]-\frac{\partial}{\partial \tau}P(\emptyset)\,}
=\frac{1}{\,\partial\Delta/\partial\tau\,}.\vspace{-0.2cm}
\end{equation*}
Since $\Delta(\cdot;c)$ is strictly convex, at the left (right) root, we have $\partial\Delta/\partial\tau<0$ (resp. $>0$). Hence $\underline{\tau}'(c)<0<\overline{\tau}'(c)$.
The same implicit-function argument yields continuity of $\underline{\tau}(\cdot)$ and $\overline{\tau}(\cdot)$ in $c$ for $c>c_0$ and, since $\lim_{|\tau|\rightarrow \infty}\Delta(\tau;c_0)=\infty$, it must be that 
$\underline{\tau}(c)\downarrow-\infty$ and $\overline{\tau}(c)\uparrow\infty$ as $c\to\infty$. \qed

\clearpage
\setcounter{proposition}{10}
\setcounter{lemma}{6}
\setcounter{corollary}{3}
\setcounter{definition}{2}
\setcounter{equation}{0}
\setcounter{figure}{0}
\renewcommand{\theequation}{OA.\arabic{equation}}
\renewcommand{\thefigure}{OA.\arabic{figure}}
\renewcommand*{\theHequation}{OA.\arabic{equation}}
\renewcommand*{\theHfigure}{OA.\arabic{figure}}
\section*{Online Appendix}

In this appendix, we use $\lambda(z)\equiv \phi(z)/(1-\Phi(z))$ as the inverse Mills ratio, with $\phi(z)$ (resp., $\Phi(z)$) indicating the pdf (resp., cdf) of the standard normal distribution.\ In parts of the proof, we also simplify notations using $\rho\;\equiv\;\frac{s}{\sqrt{\sigma^2+s^2}}\in (0,1)$, $\theta_s(y)\equiv \frac{\tau s/\sigma-\sigma y}{\sqrt{\sigma^{2}+s^{2}}}=\frac{\tau\rho}{\sigma}-y\sqrt{1-\rho^2}$, and $\theta\equiv \theta_s(\tilde Y)$, with $\tilde{Y}$ defined as a standard normal random variable.

\subsection*{Omitted Proofs of Main-Text Results}

\paragraph{Proof of Corollary 1.} Recall from Proposition 1 that $c_0$ is defined by $\Delta(\tau^*;c_0)=0$ where $\tau^*$ satisfies the first-order condition $\Delta_\tau(\tau^*;.)=0$, i.e., 
\[
\frac{\partial}{\partial\tau}\E\!\big[P(\tau+\tilde\varepsilon)\mid\tau\big]\Big|_{\tau=\tau^*}
\;=\;\frac{\partial}{\partial \tau}P(\emptyset)\Big|_{\tau=\tau^*},
\]
which simplifies to \vspace{-0.2cm}
\begin{equation}\label{FIU}
(1-\rho^2)+\rho^2\,\E[\lambda'(\theta)]=\lambda'(-\frac{\tau^*}{\sigma}).\vspace{-0.2cm}
\end{equation}
Differentiating $c_0$ and applying the chain rule and the Envelope theorem yields, for $\zeta$ set to $s$ or $\sigma$,
\[
\frac{d c_0}{d\zeta}
=\frac{\partial}{\partial \zeta}\,\Big\{\E\!\big[P(\tau+\tilde\varepsilon)\mid\tau\big]-P(\emptyset)\Big\}\Big|_{\tau=\tau^*},
\]
where the non-disclosure price does not depend on $s$, so only the first term remains when $\zeta=s$.
In what follows, to simplify notations, let us rewrite main-text equation (11) as \vspace{-0.2cm}
$$\E\!\big[P(\tau+\tilde\varepsilon)\mid\tau\big]=(1-\rho^2)\,\tau+\sigma\rho\,\E[\lambda(\theta)].$$

\textit{Comparative statics in $s$.} Differentiating,
\begin{eqnarray*}
\frac{\partial}{\partial \rho}\,\E[\lambda(\theta)]
&=&\E\!\Big[\lambda'(\theta)\,\frac{\partial \theta}{\partial \rho}\Big]
=\frac{\tau}{\sigma}\,\E[\lambda'(\theta)]
+\frac{\rho}{\sqrt{1-\rho^2}}\,\E\big[\lambda'(\theta)\,\tilde Y\big]\\
&=& \frac{\tau}{\sigma}\,\E[\lambda'(\theta)]-\rho\,\E[\lambda''(\theta)],
\end{eqnarray*}
where the second equality follows,
by Stein's lemma,
from $
\E\big[\lambda'(\theta)\,\tilde Y\big]
=\E\!\left[\frac{d}{d\tilde Y}\lambda'(\theta)\right]
=-\sqrt{1-\rho^2}\,\E[\lambda''(\theta)],$
implying that
\[
\frac{\partial}{\partial \rho}\,\E[\lambda(\theta)]
=\frac{\tau}{\sigma}\,\E[\lambda'(\theta)]-\rho\,\E[\lambda''(\theta)].
\]
Finally, using the above to differentiate main-text equation (11),  
\begin{equation}\label{HY}
\frac{\partial}{\partial \rho}\,\E[P(\tau+\tilde\varepsilon)\mid\tau]
=\sigma\Big(\,-2\,\tfrac{\tau\rho}{\sigma}+\E[\lambda(\theta)]
+\tfrac{\tau\rho}{\sigma}\,\E[\lambda'(\theta)]
-\rho^2\,\E[\lambda''(\theta)]\Big).
\end{equation}
Since $\theta=\tfrac{\tau\rho}{\sigma}-\sqrt{1-\rho^2}\,\tilde Y$ is linear in $\tilde Y$,
Stein's lemma ($\E[\tilde Y f(\tilde Y)]=\E[f'(\tilde Y)]$) applied to
$f(\tilde Y)=\lambda'(\theta(\tilde Y))$ yields
\begin{equation}\label{eq:theta_lambda1_no_m}
\E[\theta\lambda'(\theta)]
=\tfrac{\tau\rho}{\sigma}\E[\lambda'(\theta)]
+(1-\rho^2)\,\E[\lambda''(\theta)].
\end{equation}
The inverse Mills ratio satisfies $\lambda''(z)=(2\lambda(z)-z)\lambda'(z)-\lambda(z)$.
Substituting into \eqref{eq:theta_lambda1_no_m} returns \vspace{-0.2cm}
\begin{equation}\label{eq:lambda2_no_m}
(2-\rho^2)\,\E[\lambda''(\theta)]
=2\E[\lambda(\theta)\lambda'(\theta)]
-\tfrac{\tau\rho}{\sigma}\E[\lambda'(\theta)]-\E[\lambda(\theta)].\vspace{-0.2cm}
\end{equation}
Substituting \eqref{eq:lambda2_no_m} into (\ref{HY})  yields
\vspace{-0.2cm}\begin{align*}
\frac{1}{\sigma}\,\frac{\partial}{\partial\rho}\E[P(\tau+\tilde\varepsilon)\mid\tau]
&=-2\,\tfrac{\tau\rho}{\sigma}+\E[\lambda(\theta)]
+\tfrac{\tau\rho}{\sigma}\E[\lambda'(\theta)]
-\rho^2\,\E[\lambda''(\theta)]\\[3pt]
&=2\Big(\E[\lambda(\theta)]-\tfrac{\tau\rho}{\sigma}
-\E[\lambda(\theta)\lambda'(\theta)]
+\tfrac{\tau\rho}{\sigma}\E[\lambda'(\theta)]
+(1-\rho^2)\E[\lambda''(\theta)]\Big).
\end{align*}
Expanding the product, $(\lambda(\theta)-\theta)(1-\lambda'(\theta))
=\lambda(\theta)-\theta-\lambda(\theta)\lambda'(\theta)+\theta\lambda'(\theta).$
Taking expectations and using $\E[\theta]=\tfrac{\tau\rho}{\sigma}$ yields
\vspace{-0.4cm}\[
\E[(1-\lambda'(\theta))(\lambda(\theta)-\theta)]
=\E[\lambda(\theta)]-\tfrac{\tau\rho}{\sigma}
-\E[\lambda(\theta)\lambda'(\theta)]
+\tfrac{\tau\rho}{\sigma}\E[\lambda'(\theta)]
+(1-\rho^2)\E[\lambda''(\theta)].\vspace{-0.2cm}
\]
\vspace{-0.2cm}Comparing terms, we obtain the simplified form\begin{equation}\label{eq:dEdrho_compact_no_m}
\frac{\partial}{\partial\rho}\E[P(\tau+\tilde\varepsilon)\mid\tau]
=2\sigma\,\E\!\big[(1-\lambda'(\theta))(\lambda(\theta)-\theta)\big].
\end{equation}
Since $0<\lambda'(z)<1$ and $\lambda(z)>z$ for all $z$, the integrand in
\eqref{eq:dEdrho_compact_no_m} is strictly positive pointwise. Hence
$\frac{\partial}{\partial\rho}\E[P(\tau+\tilde\varepsilon)\mid\tau]\;>\;0.$
Given that $\frac{d\rho}{ds}
=\frac{\sigma^2}{(\sigma^2+s^2)^{3/2}}\;>\;0,$
the chain rule implies
\vspace{-0.2cm}\begin{equation}\label{vip}
\frac{d c_0}{ds}
=\frac{\partial}{\partial s}\,\E[P(\tau+\tilde\varepsilon)\mid\tau]
\Big|_{\tau=\tau^*}=\frac{\partial}{\partial \rho}\,\E[P(\tau+\tilde\varepsilon)\mid\tau]\Big|_{\tau=\tau^*}\frac{d\rho}{ds}\;>\;0.
\end{equation}

As $s\downarrow 0$, $\rho\to 0$ and, at every fixed $\tau$, $\E[P(\tau+\tilde\varepsilon)\mid\tau]\to \tau$, so the benefit at $\tau$ converges to $\tau+\sigma\lambda(-\tau/\sigma)$. The limit function is strictly increasing, since $1-\lambda'(-\tau/\sigma)>0$, and converges to zero as $\tau\to-\infty$. Fix $\epsilon>0$ and a threshold $T$ with $T+\sigma\lambda(-T/\sigma)<\epsilon/2$. Then $0<c_0\leq \E[P(T+\tilde\varepsilon)\mid T]-P(\emptyset;T)<\epsilon$ for all $s$ small enough, so $\lim_{s\downarrow 0}c_0=0$.

\textit{Comparative statics in $\sigma$.} From the same Envelope argument,\vspace{-0.2cm} \[\vspace{-0.2cm}
\frac{dc_0}{d\sigma}
=\tau\,\frac{\partial a}{\partial\sigma}
+\frac{\partial b}{\partial\sigma}\,\E[\lambda(\theta)]
+b\,\frac{\partial}{\partial\sigma}\E[\lambda(\theta)]
-\frac{\partial}{\partial\sigma}\big(-\sigma\,\lambda(-\frac{\tau}{\sigma})\big)\Big|_{\tau=\tau^*}.
\]
Direct differentiation further yields:\vspace{-0.2cm}\[\vspace{-0.2cm}
\frac{\partial a}{\partial\sigma}=\frac{2\sigma s^2}{(\sigma^2+s^2)^2},\qquad
\frac{\partial b}{\partial\sigma}=\frac{s^3}{(\sigma^2+s^2)^{3/2}},
\]
\[
\frac{\partial}{\partial\sigma}\E[\lambda(\theta)]
=\E\!\left[\lambda'(\theta)\,\frac{\partial\theta}{\partial\sigma}\right],
\qquad
\frac{\partial}{\partial\sigma}(-\sigma\lambda(-\frac{\tau}{\sigma}))
=-\lambda(-\frac{\tau}{\sigma})-\frac{\tau}{\sigma}\lambda'(-\frac{\tau}{\sigma}).
\]
We can then write\vspace{-0.2cm}
\begin{align*}
\tau^*\,\frac{\partial a}{\partial\sigma}
=2\frac{\tau^*}{\sigma}(1-\rho^2)\rho^2 \qquad \text{and} \qquad
\frac{\partial b}{\partial\sigma}\,\E[\lambda(\theta)]
=\rho^3\,\E[\lambda(\theta)].
\end{align*}
By $\theta=\rho\frac{\tau}{\sigma}-\sqrt{1-\rho^2}\,\tilde{Y}$, holding $\tau$ fixed,\vspace{-0.2cm}
\[
\frac{\partial\theta}{\partial\sigma}
=-\frac{\frac{\tau}{\sigma}\,\rho(2-\rho^2)}{\sigma}-\frac{\rho^2\sqrt{1-\rho^2}}{\sigma}\,\tilde{Y}.
\]
So that
\[
b\,\E\!\Big[\lambda'(\theta)\,\frac{\partial\theta}{\partial\sigma}\Big]
=-\,\frac{\tau^*}{\sigma}\rho^2(2-\rho^2)\,\E[\lambda'(\theta)]
-\rho^3\sqrt{1-\rho^2}\,\E[\lambda'(\theta)\,\tilde Y].
\]
By Stein's lemma,\vspace{-0.2cm}
\[
\E[\lambda'(\theta)\,\tilde Y]
=\E\!\left[\frac{d}{d\tilde Y}\lambda'(\theta(Y))\right]
=-\sqrt{1-\rho^2}\,\E[\lambda''(\theta)].
\]
Then $b\,\E\!\Big[\lambda'(\theta)\,\frac{\partial\theta}{\partial\sigma}\Big]=-\,\frac{\tau^*}{\sigma}\,\rho^2(2-\rho^2)\,\E[\lambda'(\theta)]
+\rho^3(1-\rho^2)\,\E[\lambda''(\theta)].$
\smallskip
Adding the terms together, 
\begin{align*}
\frac{dc_0}{d\sigma}
&= 2\frac{\tau^*}{\sigma}(1-\rho^2)\rho^2
  + \rho^3\,\E[\lambda(\theta)]
  - \frac{\tau^*}{\sigma}\rho^2(2-\rho^2)\,\E[\lambda'(\theta)]
\\
& \hspace{2cm}\quad
  + \rho^3(1-\rho^2)\,\E[\lambda''(\theta)]
  + \lambda\!\left(-\frac{\tau^{*}}{\sigma}\right)
  + \frac{\tau^*}{\sigma}\lambda'\!\left(-\frac{\tau^*}{\sigma}\right).
\end{align*}
Using (\ref{FIU}), this expression
becomes
\begin{align}
\frac{dc_0}{d\sigma}
&=
\underbrace{\Big(
  \lambda\!\left(-\frac{\tau^*}{\sigma}\right)
  + 2\frac{\tau^*}{\sigma}
  - \lambda'\!\left(-\frac{\tau^*}{\sigma}\right)\frac{\tau^*}{\sigma}
\Big)}_{\text{(i)}}
\underbrace{-\,\rho^4\,\frac{\tau^*}{\sigma}
  \Big(2-\E[\lambda'(\theta)]\Big)}_{\text{(ii)}}
\nonumber\\
&\hspace{2cm}\quad
+\,\underbrace{\rho^3\,\E[\lambda(\theta)]}_{\text{(iii)}}
+\,\underbrace{\rho^3(1-\rho^2)\,\E[\lambda''(\theta)]}_{\text{(iv)}}.\label{eq:dc0dsigma}
\end{align}
Each bracketed term is strictly positive:  
(i) with $z=-\frac{\tau^*}{\sigma}>0$, this term is equal to 
$\lambda(z)-2z+z\,\lambda'(z) = \left(\lambda(z)-z \right) - z + z \lambda'(z) > \left(\lambda(z)-z \right) \lambda'(z) - z + z \lambda'(z) = \left(\lambda(z)-z \right)^2 \lambda(z) - z + z \lambda'(z) = \left(\lambda''(z) - \lambda(z) \left(\lambda'(z) - 1 \right)\right) - z + z \lambda'(z) = \lambda''(z) + \left(\lambda(z) - z \right)\left(1 - \lambda'(z) \right) > 0$, where the inequalities follow from $\lambda'(z) < 1$, $\lambda(z) > z$, and strict convexity of $\lambda(z)$; 
(ii) $\frac{\tau^*}{\sigma}<0$, $\rho\in(0,1)$, and $0<\E[\lambda'(\theta)]<1$ imply (ii)$>0$;  
(iii) $\lambda>0$ implies (iii)$>0$;  
(iv) $\lambda''>0$ and $\rho\in(0,1)$ imply (iv)$>0$.  

Finally, if $\sigma\downarrow 0$, evaluate the benefit at the fixed threshold $\tau=0$, where $\theta=-\sqrt{1-\rho^2}\,\tilde Y$, so that
\vspace{-0.2cm}\begin{equation}
0\;<\;c_0(\sigma)\;\leq\;\sigma\rho\,\E\big[\lambda(-\sqrt{1-\rho^2}\,\tilde Y)\big]+\sigma\,\lambda(0)\;\longrightarrow\;0,
\tag*{\qedsymbol}
\end{equation}
since the expectation stays bounded as $\rho\to 1$.

\vspace{-0.4cm} \paragraph{Proof of Corollary 2.} Suppose that $c>c_0$, so that there are two subgame equilibria $\underline{\tau}$ and $\overline{\tau}$.

\textit{Comparative statics in $s$.} In equation (\ref{vip}), we proved that\vspace{-0.2cm}\begin{equation*}
        0 < \frac{\partial}{\partial s}\,\E[P(\tau+\tilde\varepsilon)\mid\tau]  = \frac{\partial \Delta}{\partial s}. \vspace{-0.2cm}
\end{equation*}
        Therefore $\Delta$ is increasing in $s$. Recall from the proof of Proposition 1 that $\frac{\partial \Delta}{\partial \tau}\Big|_{\tau = \underline{\tau}} < 0$ and $\frac{\partial \Delta}{\partial \tau}\Big|_{\tau = \overline{\tau}} > 0$. The implicit function theorem then gives\vspace{-0.4cm}
        \begin{equation*}
                 \qquad \frac{d \tau}{d s}\Big|_{\tau = \overline{\tau}} < 0<\frac{d \tau}{d s}\Big|_{\tau = \underline{\tau}}.
        \end{equation*}

Finally, as $s\downarrow 0$, the left side of (\ref{FIU}) converges to one, so $\lambda'(-\tau^*/\sigma)\to 1$ and the minimizer $\tau^*$ diverges to $-\infty$. Since $\underline{\tau}(c)<\tau^*$, the probability of disclosure under the lower threshold converges to one.

\textit{Comparative statics in $\sigma$.} First, we study the comparative statics for $\underline{\tau}$\ (which is known to be negative from $\underline{\tau}(c)\leq\tau(c_0)<0$ in Corollary 1) when $\sigma\downarrow 0.$ Then, both $a(s)$ and $b(s)$ converge to zero in Lemma 2 and, moreover, $\theta_s(y)\rightarrow -\infty$, which implies that\vspace{-0.2cm}\[\vspace{-0.2cm}
\E\big[P(\tau+\tilde\varepsilon)| \tau\big]
= a(s)\,\tau + b(s)\,\E[\lambda(\theta_s(\tilde Y))]
\rightarrow\;0.
\]
For the non-disclosure price, an expansion of the Mills ratio yields
\vspace{-0.2cm}
\[
P(\emptyset)=-\sigma\,\lambda\!\Big(-\frac{\tau}{\sigma}\Big)
=-\sigma\big(-\frac{\tau}{\sigma}+(-\frac{\tau}{\sigma})^{-1}+o((-\frac{\tau}{\sigma})^{-1})\big)
\rightarrow \tau.
\vspace{-0.2cm}\]
The limit $\Delta(\tau;c,0)\equiv\lim_{\sigma\downarrow0}\Delta(\tau;c,\sigma)=-c-\tau$ is strictly decreasing with unique root $-c$. Because $\Delta(\cdot;c,\sigma)$ converges to this limit uniformly on compact subsets of $(-\infty,0)$, is strictly convex, and has a minimizer $\tau^*\rightarrow 0$ (from the first-order condition (\ref{FIU}), which implies $\tau^*/\sigma\to0$ as $\rho\to1$), its unique root below the minimizer satisfies $\underline{\tau}(c)\rightarrow -c$. The probability of disclosure $1-\Phi\Big(\frac{\underline{\tau}(c)}{\sigma}\Big)
\sim \Phi\!\Big(\frac{c}{\sigma}\Big)
\;\rightarrow\;1$. The decreasing sign for $\sigma$ sufficiently small follows from Proposition~\ref{LM} below.

Second, for $\overline{\tau}$, we differentiate the two prices in $\sigma$ at a fixed $\tau\geq 0$. For every real $w$, since $\lambda'(w)=\lambda(w)(\lambda(w)-w)$,
\vspace{-0.2cm}\[\vspace{-0.2cm}
\lambda(w)-w\,\lambda'(w)=\lambda(w)\big(1-w(\lambda(w)-w)\big)>0,
\]
where the inequality is immediate for $w\leq 0$ and follows for $w>0$ from $w(\lambda(w)-w)<\lambda(w)(\lambda(w)-w)=\lambda'(w)<1$. With $w=-\tau/\sigma$, the non-disclosure price satisfies
\vspace{-0.2cm}\[\vspace{-0.2cm}
\frac{\partial}{\partial \sigma}P(\emptyset)=\frac{\partial}{\partial \sigma}\Big(-\sigma\,\lambda\big(-\frac{\tau}{\sigma}\big)\Big)=-\big(\lambda(w)-w\,\lambda'(w)\big)<0.
\]
For the disclosed price, differentiating main-text equation (11) at fixed $\tau$ and $s$, with $\frac{d\rho}{d\sigma}=-\frac{\rho(1-\rho^2)}{\sigma}$, $\frac{d(\sigma\rho)}{d\sigma}=\rho^3$, and Stein's lemma for $\E[\lambda'(\theta)\tilde Y]$, yields
\begin{align*}
\frac{\partial}{\partial \sigma}\E[P(\tau+\tilde\varepsilon)\mid\tau]
&=2\rho^2(1-\rho^2)\frac{\tau}{\sigma}+\rho^3\,\E[\lambda(\theta)]
\\ &\quad-\rho^2(2-\rho^2)\frac{\tau}{\sigma}\,\E[\lambda'(\theta)]
+\rho^3(1-\rho^2)\,\E[\lambda''(\theta)]\\
&=2\rho^2(1-\rho^2)\frac{\tau}{\sigma}\big(1-\E[\lambda'(\theta)]\big)
+\rho^3\,\E\big[\lambda(\theta)-\theta\lambda'(\theta)\big]
\\ &\quad+2\rho^3(1-\rho^2)\,\E[\lambda''(\theta)],
\end{align*}
where the second equality substitutes (\ref{eq:theta_lambda1_no_m}). When $\tau\geq 0$, the first term is nonnegative because $\lambda'<1$, the second is strictly positive by the pointwise inequality above, and the third is strictly positive by convexity of $\lambda$. Hence $\partial\Delta/\partial\sigma>0$ at every $\tau\geq 0$ and, in particular, at $\overline{\tau}(c)>0$, where $\partial\Delta/\partial\tau>0$, so the implicit function theorem gives $d\overline{\tau}/d\sigma<0$. For $\overline{\tau}(c)>0$, a falling threshold and a rising $\sigma$ both lower the ratio $\overline{\tau}(c)/\sigma$, so the probability of disclosure $1-\Phi(\overline{\tau}(c)/\sigma)$ is increasing. Proposition \ref{LM} below extends both branch signs to every $\sigma\leq s$, without requiring $\overline{\tau}(c)>0$. If $\sigma \rightarrow \infty$, the probability of disclosure converges to one when $s=0$ (\citealt{ver90}) and, by part~(i), is larger for any $s>0$, so it must also converge to one for any $s>0$.\qed


\paragraph{Proof of Lemma 4.}\ Suppose $c>c_0$. The claim for $\underline{V}$ is immediate because $\underline{\tau}(c)$ decreases in $c$, so both factors of $c(1-\Phi(\underline{\tau}(c)/\sigma))$ increase, and the product diverges as the second factor tends to one. We next show that $\overline V$ is eventually decreasing in $c$. By 
the implicit function theorem,  \vspace{-0.3cm}
\[ 
\overline{V}'(c)=1-\Phi(\overline{\tau}/\sigma)-\frac{c}{\sigma}\,\phi(\overline{\tau}/\sigma)\,\overline{\tau}'(c),
\qquad
\overline{\tau}'(c)=\frac{1}{\frac{\partial{\Delta(\tau,c)}}{\partial{\tau}}|_{\tau=\overline{\tau}}}.  \vspace{-0.1cm}
\]
On the upper root, we know that $\overline{\tau}'(c)>0$. Then, from  $1-\Phi(\overline{\tau}/\sigma)=\phi(\overline{\tau}/\sigma)/\lambda(\overline{\tau}/\sigma)$,  
\begin{equation}\label{eq:Vprime}
\overline{V}'(c)\;=\;\frac{\phi(\overline{\tau}/\sigma)}{\lambda(\overline{\tau}/\sigma)}\,
\Bigg[\,1-\frac{c}{\sigma}\,\frac{\lambda(\overline{\tau}/\sigma)}{{\frac{\partial{\Delta(\tau,c)}}{\partial{\tau}}|_{\tau=\overline{\tau}}}}\,\Bigg].
\end{equation}
The sign of $V'(c)$ is the sign of the bracket.
Because $\overline{\tau}(c)/\sigma\ge \overline{\tau}(c_0)/\sigma$ and $\lambda$ is increasing,
\[
\frac{c}{\sigma}\,\frac{\lambda(\overline{\tau}/\sigma)}{\frac{\partial{\Delta(\tau,c)}}{\partial{\tau}}|_{\tau=\overline{\tau}}}\ge\;
\frac{c_0}{\sigma}\,\frac{\lambda(\overline{\tau}(c_0)/\sigma)}{\sup_{c\ge c_0} \frac{\partial{\Delta(\tau,c)}}{\partial{\tau}}|_{\tau=\overline{\tau}}}.
\]
Given that  $0<\lambda'(\cdot)<1$, \[\frac{\partial{\Delta(\tau,c)}}{\partial{\tau}}|_{\tau=\overline{\tau}}=a+(1-a)\E[\lambda'(\cdot)]-\lambda'(-\overline{\tau}(c)/\sigma)
\;\le\; a+(1-a)\cdot 1 \;=\;1,
\]
where $a$ is from lemma 2 (omitting the dependence on $s$ to simplify notations). Since $\overline{\tau}(c)\uparrow\infty$ by Proposition 1 and $\lambda$ is unbounded above, the ratio $\frac{c}{\sigma}\lambda(\overline{\tau}(c)/\sigma)$ diverges, so the bracket in (\ref{eq:Vprime}) is negative for $c$ sufficiently large, establishing the claim. The same bound shows that $\overline{V}$ is decreasing on all of $[c_0,\infty)$ whenever $\frac{c_0}{\sigma}\,\lambda\big(\overline{\tau}(c_0)/\sigma\big)>1$, a condition used only by the alternative selection criteria in the Online Appendix, where it is restated and applied.

Now we prove that the profit of the certifier tends to zero when considering $\overline{\tau}\rightarrow \infty$ for $c$ sufficiently large. Recall the indifference condition
\begin{equation}\label{eq:indiff}
\underbrace{\mathbb{E}\bigl[P(\tau+\tilde{\varepsilon})\mid \tau\bigr]}_{=a(s)\,\tau + b(s)\,\mathbb{E}\bigl[\lambda(\theta_s(\tilde Y))\bigr]}-c
\;=\;P(\emptyset;\tau)
\;=\;\;-\sigma\,\lambda\!\Bigl(-\tfrac{\tau}{\sigma}\Bigr).
\end{equation}
As $\tau\to\infty$, 
$\theta_s(\tilde Y)=\frac{s}{\sigma\sqrt{\sigma^2+s^2}}\tau + O(1)$,
and since $\lambda(x)/x\to 1$ as $x\to\infty$,
\[
\mathbb{E}\bigl[\lambda(\theta_s(\tilde Y))\bigr]
=\frac{s}{\sigma\sqrt{\sigma^2+s^2}}\,\tau + o(\tau),
\]
implying that
\begin{equation}\label{eq:EP-asym}
\mathbb{E}\bigl[P(\tau+\tilde{\varepsilon})\mid \tau\bigr]
= \Bigl(\frac{\sigma^2}{\sigma^2+s^2}+\frac{s^2}{\sigma^2+s^2}\Bigr)\tau + o(\tau)
= \tau + o(\tau).\qquad 
\end{equation}
Moreover, $\lambda(-x)\to 0$ as $x\to\infty$, so
\begin{equation}\label{eq:P0-asym}
P(\emptyset;\tau)=-\sigma\,\lambda\!\Bigl(-\tfrac{\tau}{\sigma}\Bigr)=o(\tau).
\end{equation}

It then follows from the indifference condition jointly with \eqref{eq:EP-asym} and \eqref{eq:P0-asym} that $\overline{\tau}(c)/c\to 1$. Hence $\overline{\tau}(c)\geq c/2$ for all $c$ large, and
\[
0\leq \overline{V}(c)=c\Big(1-\Phi\Big(\frac{\overline{\tau}(c)}{\sigma}\Big)\Big)\leq c\Big(1-\Phi\Big(\frac{c}{2\sigma}\Big)\Big)\longrightarrow 0.
\]\qed\\
\textbf{Proof of Proposition 2:} It remains to compare the best upper-threshold profit $\overline{V}^{**}$ with the certifier's outside payoff $c_0$.  Let
$
H(\tau,\rho)
\equiv
(1-\rho^2)\frac{\tau}{\sigma}
+\rho\E\left[\lambda(\theta)\right]
+\lambda\left(-\frac{\tau}{\sigma}\right).$
By Lemma~2, the marginal-discloser indifference condition can be written as $c/\sigma=H(\tau,\rho)$. \\Let $\tau_0(\rho)$ be the unique minimizer of $H(\cdot,\rho)$. Then $c_0=\sigma H(\tau_0(\rho),\rho)$ and
\[
\overline{V}^{**}
=
\sigma\sup_{\tau>\tau_0(\rho)}
H(\tau,\rho)
\left(1-\Phi\left(\frac{\tau}{\sigma}\right)\right).
\]
Every $\tau>\tau_0(\rho)$ corresponds to the upper root at fee $c=\sigma H(\tau,\rho)$, and every upper-root fee is obtained in this way.

We first record the primitive normal facts used below. The inverse Mills ratio satisfies $\lambda(u)>u$, $0<\lambda'(u)<1$, and $\lambda''(u)>0$. Moreover,
\begin{align*}
H_{\tau\tau}(\tau,\rho)
&=
\frac{1}{\sigma^2}
\left(
\rho^3\E\left[\lambda''(\theta)\right]
+\lambda''\left(-\frac{\tau}{\sigma}\right)
\right)>0,\\
H_\rho(\tau,\rho)
&=
2\E\left[
(1-\lambda'(\theta))(\lambda(\theta)-\theta)
\right]>0,
\end{align*}
where the second equality follows by differentiating with respect to $\rho$ and applying Stein's lemma. Moreover,
$H_{\tau\rho}(\tau,\rho)
=
-\frac{2\rho}{\sigma}
\E\left(
\lambda''(\theta)(\lambda(\theta)-\theta)
+(1-\lambda'(\theta))^2
\right)<0.$
Thus $H(\cdot,\rho)$ is strictly convex, $\tau_0(\rho)$ is unique, and $H_\rho(\tau,\rho)$ is strictly decreasing in $\tau$. By the envelope and implicit-function theorems,
\[
\frac{d}{d\rho}H(\tau_0(\rho),\rho)
=
H_\rho(\tau_0(\rho),\rho)>0,
\qquad
\tau_0'(\rho)
=
-\frac{H_{\tau\rho}(\tau_0(\rho),\rho)}
{H_{\tau\tau}(\tau_0(\rho),\rho)}>0.
\]

We next establish single crossing. For every $\tau>\tau_0(\rho)$,
\begin{align*}
\frac{\partial}{\partial\rho}
\bigg\{
&H(\tau,\rho)
\left(1-\Phi\left(\frac{\tau}{\sigma}\right)\right)
-H(\tau_0(\rho),\rho)
\bigg\}\\
&=
H_\rho(\tau,\rho)
\left(1-\Phi\left(\frac{\tau}{\sigma}\right)\right)
-H_\rho(\tau_0(\rho),\rho)<0,
\end{align*}
because $H_\rho(\tau,\rho)<H_\rho(\tau_0(\rho),\rho)$ and $0<1-\Phi(\tau/\sigma)<1$.

Suppose that $\overline{V}^{**}\leq c_0$ at $\rho_1$, and let $\rho_2>\rho_1$. Since $\tau_0(\rho)$ is increasing, every $\tau>\tau_0(\rho_2)$ satisfies $\tau>\tau_0(\rho)$ for all $\rho\in[\rho_1,\rho_2]$. Integrating the preceding inequality yields
\begin{align*}
&H(\tau,\rho_2)
\left(1-\Phi\left(\frac{\tau}{\sigma}\right)\right)
-H(\tau_0(\rho_2),\rho_2)\\
&\quad<
H(\tau,\rho_1)
\left(1-\Phi\left(\frac{\tau}{\sigma}\right)\right)
-H(\tau_0(\rho_1),\rho_1)
\leq 0.
\end{align*}
As $\tau\downarrow\tau_0(\rho_2)$, the left-hand side converges to
\[
-H(\tau_0(\rho_2),\rho_2)
\Phi\left(\frac{\tau_0(\rho_2)}{\sigma}\right)<0,
\]
while as $\tau\to\infty$ it converges to $-H(\tau_0(\rho_2),\rho_2)<0$, since
\[
H(\tau,\rho_2)
\left(1-\Phi\left(\frac{\tau}{\sigma}\right)\right)
\longrightarrow0.
\]
Hence $\overline{V}^{**}<c_0$ at $\rho_2$. Thus, once the best upper-threshold profit is weakly below $c_0$, it remains strictly below $c_0$ at every higher noise level.

It remains to check the endpoint signs. As $\rho\downarrow0$,
\[
H(\tau,\rho)
\longrightarrow
\frac{\tau}{\sigma}
+\lambda\left(-\frac{\tau}{\sigma}\right)
\]
locally uniformly. The limiting function has infimum zero, approached as $\tau\to-\infty$. Hence $H(\tau_0(\rho),\rho)\to0$ and $\tau_0(\rho)\to-\infty$. For small $\rho$, $\tau=0$ is therefore an upper root, so
\[
\liminf_{\rho\downarrow0}
\frac{\overline{V}^{**}}{\sigma}
\geq
\lim_{\rho\downarrow0}
\frac{H(0,\rho)}{2}
=
\frac{\lambda(0)}{2}>0.
\]
Therefore $\overline{V}^{**}>c_0$ for sufficiently small $\rho$.

Conversely, as $\rho\uparrow1$,
\[
H(\tau,\rho)
\longrightarrow
\lambda\left(\frac{\tau}{\sigma}\right)
+\lambda\left(-\frac{\tau}{\sigma}\right),
\]
whose unique minimizer is $\tau=0$. Thus $c_0/\sigma\to2\lambda(0)$. At an upper root, $\tau\geq0$ in the limit, and
\begin{align*}
&
\left(
\lambda\left(\frac{\tau}{\sigma}\right)
+\lambda\left(-\frac{\tau}{\sigma}\right)
\right)
\left(1-\Phi\left(\frac{\tau}{\sigma}\right)\right)\\
&\qquad=
\frac{\phi(\tau/\sigma)}{\Phi(\tau/\sigma)}.
\end{align*}
The right-hand side is maximized at $\tau=0$, with value $\lambda(0)$. A standard maximum-theorem argument, using the uniform tail bound
\[
H(\tau,\rho)
\left(1-\Phi\left(\frac{\tau}{\sigma}\right)\right)
\longrightarrow0,
\]
gives $\overline{V}^{**}/\sigma\to\lambda(0)$ and therefore $\overline{V}^{**}-c_0\to-\sigma\lambda(0)<0$.

The difference $\overline{V}^{**}-c_0$ is continuous in $\rho$ by the same maximum-theorem argument on the moving domain $\tau>\tau_0(\rho)$. The endpoint signs and the absorbing property above imply a unique zero $\rho^*\in(0,1)$. Since
\[
\frac{s}{\sigma}
=
\frac{\rho}{\sqrt{1-\rho^2}}
\]
is strictly increasing in $\rho$, this zero is equivalently a unique cutoff in relative noise,
\[
\nu
\equiv
\frac{\rho^*}{\sqrt{1-(\rho^*)^2}}
\simeq0.372846.
\]
Thus $\overline{V}^{**}>c_0$ if $s/\sigma<\nu$, $\overline{V}^{**}=c_0$ if $s/\sigma=\nu$, and $\overline{V}^{**}<c_0$ if $s/\sigma>\nu$. The equilibrium characterization follows. A lower-threshold equilibrium at fee $c^*$ exists exactly when no deviation is profitable, which requires $\underline{V}(c^*)\geq\max\{c_0,\overline{V}^{**}\}$, since a deviation to a fee below $c_0$ earns that fee and a deviation to any other fee is met by the upper-threshold continuation and earns at most $\overline{V}^{**}$. When $\overline{V}^{**}<c_0$, the upper-root candidate $(c^{**},\overline{\tau}(c^{**}))$ is not an equilibrium, because a fee just below $c_0$ yields more than $\overline{V}^{**}$, and the least-cost equilibrium is the lowest lower-root fee $\hat c$, with $\underline{V}(\hat c)=c_0$. Every deviation from $\hat c$ then returns strictly less than $c_0$, so the selection in Definition 2 is satisfied. When $\overline{V}^{**}\geq c_0$, the candidate $(c^{**},\overline{\tau}(c^{**}))$ is an equilibrium with profit $\overline{V}^{**}$ and, because $\underline{V}$ is continuous and increasing to infinity by Lemma 4, exactly one fee $\tilde c$ satisfies $\underline{V}(\tilde c)=\overline{V}^{**}$, so one lower-threshold equilibrium attains the same least cost. Since $c^{**}>c_0$ and $\underline V(c^{**})>\overline V(c^{**})$, we have $\tilde c<c^{**}$. At $\tilde c$ the certifier is exactly indifferent to deviating to $c^{**}$, whereas at $c^{**}$ every deviation earns strictly less, provided the maximizer of $\overline{V}$ is unique. This remains true when $\overline V^{**}=c_0$, because each fee below $c_0$ earns strictly less than $c_0$. Under this uniqueness condition, the strictness requirement in Definition 2 selects $(c^{**},\overline{\tau}(c^{**}))$. The conclusion for $s/\sigma\leq\nu$ is stated under this condition; the strictness criterion does not select among tied upper-branch maximizers. Hence the least-cost equilibrium is $(\hat c,\underline{\tau}(\hat c))$ if $s/\sigma>\nu$ and $(c^{**},\overline{\tau}(c^{**}))$ if $s/\sigma\leq\nu$. \hfill$\Box$

The strictness tie-break avoids relying on the certifier's exact indifference between distinct fees. At the tying lower-threshold outcome, a deviation to $c^{**}$ earns the same profit under the upper continuation and more under the lower one. Assigning any positive probability to the latter makes this deviation profitable. This motivates the tie-break; Definition~2 imposes the selection directly.


\paragraph{Proof of Proposition 3.} 
If $c<c_0=\min \psi$, then $\psi(\tau)>c$ for every interior $\tau$. Hence the marginal discloser strictly prefers disclosure at every interior threshold and, because $\psi(\overline{v})\geq c_0>c$, at the no-disclosure corner $\tau=\overline{v}$, so the unique subgame equilibrium is full disclosure. For $c\in(c_0,c_1)$, the intermediate value theorem gives an interior root of $\psi(\tau)=c$, hence an interior equilibrium. At $c=c_0$ or $c=c_1$, any interior root likewise gives an equilibrium. If $c=c_0$ and there is no interior root, then $\psi(\tau)>c_0$ at every interior threshold and $\psi(\underline v)\geq c_0$, so full disclosure satisfies the corner condition. If $c=c_1$ and there is no interior root, then $\psi(\tau)<c_1$ at every interior threshold and $\psi(\overline v)\leq c_1$, so no disclosure satisfies the corner condition. If $c_0=c_1$, the benefit is constant and every interior threshold solves the indifference condition. If $c>c_1=\max \psi$, then $\psi(\tau)<c$ for every interior $\tau$. Thus the marginal discloser strictly prefers non-disclosure at every interior threshold and, because $\psi(\underline{v})\leq c_1<c$, at the full-disclosure corner $\tau=\underline{v}$, so the unique subgame equilibrium is no disclosure.

For part (iv), suppose $\tau_0$ is a local minimizer of $\psi$ and $c\in(\psi(\tau_0),\min\{\psi(\underline{v}),\psi(\overline{v})\})$. Because $c>\psi(\tau_0)$ and $c<\psi(\underline{v})$, continuity of $\psi$ yields at least one solution on the interval $(\underline{v},\tau_0)$. Likewise, because $c<\psi(\overline{v})$, continuity yields at least one solution on $(\tau_0,\overline{v})$. Denote any such solutions by $\underline{\tau}(c)$ and $\overline{\tau}(c)$, which satisfy $\underline{\tau}(c)<\tau_0<\overline{\tau}(c)$ and $\psi(\underline{\tau}(c))=\psi(\overline{\tau}(c))=c$. Hence there are at least two distinct equilibrium thresholds. \qed

\paragraph{Proof of Corollary 3.}
In the noiseless benchmark, the marginal type $\tau$ receives price $\tau$ upon disclosure, so the net disclosure benefit is $\psi^{0}(\tau) \equiv \tau-\mathbb{E}(\tilde{v}| \tilde{v}<\tau)$. As $\tau\downarrow \underline{v}$, $\mathbb{E}(\tilde{v}| \tilde{v}<\tau)$ converges to $\underline{v}$. It follows that $\lim_{\tau\downarrow \underline{v}}\psi^{0}(\tau) = \lim_{\tau\downarrow \underline{v}}\left(\tau - \mathbb{E}(\tilde{v}| \tilde{v} < \tau)\right) = \underline{v}-\underline{v}=
0$. So the threshold type's disclosure benefit is arbitrarily small near full disclosure. Hence, no strictly positive cost can sustain full disclosure, i.e., $c_0^0=0$. In the noisy model, Proposition 3 instead implies that the unraveling cutoff satisfies $c_0^s>0$. Therefore, the maximal cost sustaining full disclosure is strictly higher in the noisy disclosure model than in the noiseless benchmark. \qed

\paragraph{Proof of Proposition 4.}
Let $\underline{\tau}(c) \equiv \sup\{\tau \in [\underline{v}, \tau_0): \psi(\tau) = c \}$ and $\overline{\tau}(c) \equiv \inf\{\tau \in (\tau_0, \overline{v}]: \psi(\tau) = c \}$. The existence of $\underline{\tau}(c)$ and $\overline{\tau}(c)$ follows directly from Proposition 3. Take $\psi(\tau_0) < c < c' < \min \{\psi(\overline{v}), \psi(\underline{v})\}$. Because $\psi(\underline{\tau}(c')) = c' > c > \psi(\tau_0)$, there is at least one solution to $\psi(\tau) = c$ in $(\underline{\tau}(c'), \tau_0)$ by the intermediate value theorem. Then the closest such solution to the left of $\tau_0$ must satisfy $\underline{\tau}(c') < \underline{\tau}(c)$. Similarly, because $ \psi(\tau_0) < c < c' = \psi(\overline{\tau}(c'))$, there is at least one solution to $\psi(\tau) = c$ in $(\tau_0, \overline{\tau}(c'))$. Then the closest solution to the right of $\tau_0$ satisfies $\overline{\tau}(c') > \overline{\tau}(c)$. It follows that $\underline{\tau}(c') < \underline{\tau}(c) < \tau_0 < \overline{\tau}(c) < \overline{\tau}(c')$, which proves the result. \qed

\paragraph{Proof of Proposition 5.} We know from the proof of Corollary 1 that $\mathbb{E}  P(\tau+\tilde{\varepsilon})$ is (strictly) increasing in $s$. This implies the following:

(i) From the fact that $P(\emptyset)<\tau$ for any $\tau>\tau(0)$ (\citealt{achdemkre11}), it must also be true that $P(\emptyset)<\tau<\mathbb{E}  P(\tau+\tilde{\varepsilon})$ for any $s>0$. Therefore, there is no equilibrium with $\tau\geq \tau(0)$.

(ii) We also similarly know that $P(\emptyset)$ is decreasing in $\tau$ for $\tau<\tau(0)$, so that it can intersect at most once with the (increasing) $\mathbb{E}  P(\tau+\tilde{\varepsilon})$. The existence of this unique solution then follows from the intermediate value theorem because (ii.a) $\mathbb{E}  P(\tau+\tilde{\varepsilon})\rightarrow -\infty$ as $\tau\rightarrow -\infty$ and
(ii.b) at 
 $\tau=\tau(0)$, $\mathbb{E}  P(\tau(0)+\tilde{\varepsilon})>\tau(0)=P(\emptyset)$.
 
 The comparative statics in $s$ follows immediately from the monotonicity of $\mathbb{E}  P(\tau+\tilde{\varepsilon})$, while the comparative statics in $\sigma$ can be verified by factoring all expectations in the indifference condition by $\sigma$, so that the standardized threshold $\tau/\sigma$ is solely a function of $s/\sigma$, i.e.,

\begin{equation}
\frac{1}{\sigma}\,[a(s)\tau+b(s)\mathbb{E}(\lambda(\theta_s(Y)))]+\;\frac{(1-q)\phi(\tau/\sigma)}{q+(1-q)\Phi(\tau/\sigma)}=0,
\end{equation}
implying, since $\tau/\sigma$ decreases in $s/\sigma$ by the comparative static in $s$, that an increase in $\sigma$ at fixed $s$ raises $\tau/\sigma$ and lowers the probability of disclosure $(1-q)(1-\Phi(\tau/\sigma))$.\qed

\paragraph{Proof of Lemma 5.} Suppose first that the conjectured threshold $\tau$ and noise $\hat s>0$ are finite. We show that $P(.)$ is then strictly convex.
By the chain rule,
\[
P'(x)
= a(s)
+ b(s)\,\lambda'\!\left(
\frac{\tau - a(s)\,x}{b(s)}
\right)\cdot\frac{-a(s)}{b(s)}
= a(s)\Bigg(
1-\lambda'\!\left(
\frac{\tau - a(s)\,x}{b(s)}
\right)
\Bigg).
\]
Differentiating once more,
\[
P''(x)
= a(s)\left(
-\lambda''\!\left(
\frac{\tau - a(s)\,x}{b(s)}
\right)\cdot\frac{-a(s)}{b(s)}
\right)
= \frac{a(s)^2}{b(s)}\,
\lambda''\!\left(
\frac{\tau - a(s)\,x}{b(s)}
\right).
\]
Given that $
\lambda''(.) > 0$,  $P(x)$ is strictly convex in $x$.
If $\epsilon \sim N(0,s^2)$ and $\epsilon' \sim N(0,(s')^2)$, $s>s'$ implies that $\epsilon$ is a mean-preserving spread of $\epsilon'$.\ Hence, from\ Jensen's inequality, $\mathbb{E}_{\epsilon'} P(v+\epsilon'|\hat{s})<\mathbb{E}_\epsilon P(v+\epsilon|\hat{s})$, the strict increase claimed for a finite conjectured threshold and noise. If the conjectured continuation is instead full disclosure, the truncation term vanishes, the pricing rule $P(x)=a(\hat{s})\,x$ is affine, and $\mathbb{E}_\epsilon P(v+\epsilon)=a(\hat{s})\,v$ for every actual $s$, so the expected disclosed price does not depend on the actual noise. Under $\hat{s}=\infty$, the report is ignored and the disclosed price is likewise independent of actual noise.\qed

\paragraph{Proof of Lemma 6.}
Using properties of truncated normal distributions,\begin{eqnarray*}
D(\tau)
&=&\sigma\left[\lambda\!\left(\frac{\tau}{\sigma}\right)
+\lambda\!\left(-\frac{\tau}{\sigma}\right)\right]\\
D'(\tau)&=& \lambda'\!\left(\frac{\tau}{\sigma}\right)
- \lambda'\!\left(-\frac{\tau}{\sigma}\right),
\end{eqnarray*}
so that $D'(.)$ has the sign of $\tau$. Thus $D$ is strictly decreasing on $(-\infty,0)$ and strictly
increasing on $(0,\infty)$, with a unique minimum at $\tau=0$ achieving $D(0)= \frac{4\sigma}{\sqrt{2\pi}}$. Finally, for $\tau>0$, $\mathbb{E}[v\,|\,v\ge\tau]\ge\tau$ while $\mathbb{E}[v\,|\,v<\tau]<\mathbb{E}[v]=0$, hence $D(\tau)>\tau\;\rightarrow\;\infty$ as $\tau\to\infty$, and the left tail follows from $D(-\tau)=D(\tau)$ by symmetry.\qed

\paragraph{Proof of Proposition 7.} In the certification game, no equilibrium fee lies at or below the unravelling bound of the conjectured subgame, because any lower fee can be raised profitably and the bound itself entails a discrete loss of demand, as shown in Section IV. Every candidate equilibrium at a finite conjecture $\hat{s}$ therefore has a finite on-path threshold, and Lemma 5 gives the firm a strictly profitable increase in its actual noise, so no finite conjecture is a rational-expectations fixed point. Under the limiting conjecture $\hat{s}=\infty$ the report is ignored and the expected price no longer depends on the actual noise, so $\hat{s}=s^*=\infty$ is the unique fixed point in the compactified choice set, and Proposition 6 describes the subgame equilibria. Against the upper threshold, main-text equation (6) gives the certifier's profit $\overline{\Pi}(c)=\sigma\phi(\overline{\tau}(c)/\sigma)/\Phi(\overline{\tau}(c)/\sigma)$, which decreases in $c$ because $\overline{\tau}(c)$ increases in $c$ and $\phi(z)/\Phi(z)$ decreases in $z$, so $\overline{\Pi}(c)\leq\overline{\Pi}(c_0)=\sigma\phi(0)/\Phi(0)=c_0/2$. A fee just below $c_0$ earns almost $c_0$, so no equilibrium has the upper threshold on its path, and in any equilibrium with the lower threshold the profit $\underline{V}(c)=c(1-\Phi(\underline{\tau}(c)/\sigma))$ is at least $c_0$, while a deviation to a higher fee is met by the upper threshold and earns at most $c_0/2$. Both factors of $\underline{V}(c)$ increase in $c$, so the least-cost equilibrium is the lowest such fee, with $\underline{V}(c^*)=c_0$. By main-text equation (6), $\underline{V}(c)=\sigma\phi(z)/\Phi(z)$ with $z=\underline{\tau}(c)/\sigma$, so $\underline{V}(c^*)=c_0$ reads $\phi(z^*)/\Phi(z^*)=4/\sqrt{2\pi}$. The left side decreases from $\infty$ to $0$ in $z$ and equals $2/\sqrt{2\pi}$ at $z=0$, so $z^*$ is unique and negative, and $c^*=c_0/(1-\Phi(z^*))$ follows from $c^*(1-\Phi(z^*))=c_0$.\qed

\subsection*{Additional Proofs from Section III}

\begin{lemma}
\label{san}
Let two subgame equilibria be equivalent if they prescribe the same disclosure choices almost surely. Then, any subgame equilibrium is equivalent to a threshold equilibrium, i.e., such that the firm discloses if and only if $v\geq \tau$.
\end{lemma}

\paragraph{Proof of Lemma \ref{san}.} The statement is obvious for full-disclosure or non-disclosure (setting $|\tau|=\infty$). For any disclosure set $A$ with non-zero probability, define 
\[
P(x;c)\;=\;\E(\tilde v |\tilde v \in A,\,\tilde x=x), 
\qquad
P(\emptyset;c)\;=\;\E(\tilde v \mid \tilde v \notin A).
\]

Since $\tilde x |\tilde v \sim N(\tilde v,s^2)$, 
the conditional densities $\{f(x| v)\}_v$ form a location family with the 
monotone likelihood ratio property (MLRP) in $x$: if $v_1>v_2$, then 
$f(x| v_1)/f(x|v_2)$ increases in $x$. By Bayes' rule, this property carries over to the posterior. Hence, under the $A$, the pricing function $P(x;c)$ is strictly increasing in $x$. This shows that if $v \in A$, i.e., 
\[\Delta(v)\;\equiv\;\E_x\left[P(x;c)\mid v\right]\;-\;c\;-\;P(\emptyset;c)\geq 0,\]
then, for any $v'>v$, $\Delta(v')>\Delta(v)$ so that the $v'\in A$.\qed

\begin{proposition}\label{no1}
As $\sigma$ or $s$ become large, $c_0$ converges to a finite limit; in particular, $\lim_{s\uparrow\infty}c_0=2\sqrt{2/\pi}\,\sigma$.
\end{proposition}

\paragraph{Proof of Proposition \ref{no1}.} As $s\uparrow\infty$, $a(s)\to 0$, $b(s)\to\sigma$, and $\theta_s(\tilde Y)\to \tau/\sigma$ almost surely, so that, by dominated convergence in main-text equation (4), the marginal discloser's benefit converges pointwise to $\psi_\infty(\tau)\equiv \sigma\big(\lambda(\tfrac{\tau}{\sigma})+\lambda(-\tfrac{\tau}{\sigma})\big)$. The benefit is increasing in $s$ for every $\tau$, as shown in the proof of Corollary 2, so $c_0$ is increasing in $s$ and any minimizer at noise $s$ lies in the compact set on which the benefit at a fixed smaller noise is at most $\psi_\infty(0)$. On that set, monotone convergence of continuous functions to a continuous limit is uniform by Dini's theorem, so the minimum converges to the minimum of the limit. Because $\lambda'$ is increasing, $\psi_\infty'(\tau)=\lambda'(\tfrac{\tau}{\sigma})-\lambda'(-\tfrac{\tau}{\sigma})$ has the sign of $\tau$, so $\psi_\infty$ is uniquely minimized at $\tau=0$. Hence
\[
c_\infty^{s}(\sigma)\equiv \lim_{s\uparrow\infty}c_0=\psi_\infty(0)=2\sigma\,\lambda(0)=2\sqrt{\tfrac{2}{\pi}}\,\sigma,
\]
with the minimizing threshold converging to zero.

As $\sigma\uparrow\infty$, $\rho\to 0$, so that (\ref{FIU}) simplifies to
\begin{equation}\label{eq:FOC-limit}
1-\lambda'(-\frac{\tau^*}{\sigma})
\;=\;\rho^2\Big(1-\E\!\big[\lambda'(\rho\frac{\tau^*}{\sigma}-\sqrt{1-\rho^2}\,\tilde Y)\big]\Big).
\end{equation}
Asymptotics for the inverse Mills ratio imply
\[
1-\lambda'(z)=Var(X\mid X>z)\sim z^{-2}\qquad(z\to\infty),
\]
and the right side of \eqref{eq:FOC-limit} is at most $\rho^2\to 0$, so $\lambda'(-\tau^*/\sigma)\to 1$ and $-\tau^*/\sigma\to\infty$, whence
\[
1-\lambda'(-\frac{\tau^*}{\sigma})\sim (\frac{\sigma}{\tau^*})^2,
\]

Multiplying both sides of~\eqref{eq:FOC-limit} by $(\frac{\tau^*}{\sigma})^2$ yields
\[
\;(\rho \frac{\tau^*}{\sigma})^{2}\Big(1-\E[\lambda'(\rho\frac{\tau^*}{\sigma}-\sqrt{1-\rho^2}\,\tilde Y)]\Big)\rightarrow 1.
\]
Because the bracket is at most one, $\liminf(-\rho\tau^*/\sigma)\geq 1$, and no subsequence diverges, since the bracket would then converge to one and the product would diverge. Every limit point $k$ of $-\rho\tau^*/\sigma$ therefore solves the fixed point
\[
k^{2}\Big(1-\E\big[\lambda'(-\tilde{Y}-k)\big]\Big)=1,
\]
whose left side rises strictly from zero to infinity, with derivative $2k(1-\E[\lambda'(-\tilde Y-k)])+k^{2}\E[\lambda''(-\tilde Y-k)]>0$, so the solution is unique and $-\rho\tau^*/\sigma\to k$.
Next, to compute the limit of $c_0$, we decompose
\[
c_0=\left(\sigma \frac{\tau^*}{\sigma}+\sigma\,\lambda(-\frac{\tau^*}{\sigma})\right)
\;-\;\sigma\rho^2\frac{\tau^*}{\sigma}
\;+\;\sigma\rho\,\E[\lambda(\theta)].
\]
Using the inverse Mills expansion $\lambda(z)=z+z^{-1}+o(z^{-1})$ as $z\to\infty$, we obtain
\[
\sigma\Big(\frac{\tau^*}{\sigma}+\lambda(-\frac{\tau^*}{\sigma})\Big)
=\sigma\Big(-\frac{\sigma}{\tau^*}+o\big(\frac{\sigma}{|\tau^*|}\big)\Big)
\ \rightarrow\ \frac{s}{k},
\]
Moreover,
\begin{eqnarray*}-\sigma\rho^2 \frac{\tau^*}{\sigma}=-(\sigma\rho)\,(\rho\frac{\tau^*}{\sigma}) &\rightarrow& s k\\
\sigma\rho\,\E\big[\lambda(\rho\frac{\tau^*}{\sigma}-\sqrt{1-\rho^2}\,\tilde Y)\big]\ &\rightarrow& s\,\E[\lambda(-\tilde{Y}-k)].
\end{eqnarray*}
And it follows that 
\begin{equation}\label{csib}
c_\infty^\sigma(s)\equiv \lim_{\sigma\uparrow\infty} c_0
=s\!\left(k+\frac{1}{k}+\E_{}[\lambda(-\tilde{Y}-k)]\right).
\end{equation} \qed

\begin{proposition}
\label{largs}
Let $c_\infty^\sigma(s)$ be the limit in (\ref{csib}). If $c<c_\infty^\sigma(s)$, the probability of disclosure equals one for all $\sigma$ sufficiently large. If $c>c_\infty^\sigma(s)$, the probability of disclosure under $\underline{\tau}(c,\sigma)$ converges to one and is strictly increasing in $\sigma$ for $\sigma$ sufficiently large.
\end{proposition}

\paragraph{Proof of Proposition \ref{largs}.} If $c<c_\infty^\sigma(s)$, then $c<c_0$ for all $\sigma$ large enough, because $c_0$ increases in $\sigma$ to $c_\infty^\sigma(s)$ (Corollary 1 and Proposition \ref{no1}), and the unique subgame equilibrium is full disclosure. Suppose instead that $c>c_\infty^\sigma(s)$, so that two subgame equilibria exist at every $\sigma$. Along the lower equilibrium, $\underline{\tau}(c,\sigma)<\tau^*$ and, from the proof of Proposition \ref{no1}, $-\rho\tau^*/\sigma\to k$ with $\rho\to 0$, so $\underline{\tau}(c,\sigma)/\sigma\to-\infty$ as $\sigma\uparrow\infty$. Thus, with $\tau=\underline{\tau}(c,\sigma)$,
\[
-\frac{\tau}{\sigma}\;\rightarrow\;+\infty,
\qquad
\sigma\,\lambda\,\Big(-\frac{\tau}{\sigma}\Big)
=-\tau-\frac{\sigma^{2}}{\tau}+o\!\Big(\frac{\sigma^{2}}{\tau}\Big)
\quad(\sigma\uparrow\infty),
\]
by the inverse Mills expansion. With $L\equiv-\tau/\sigma^{2}>0$, substituting the expansion together with $\rho\to0$, $\sigma\rho\to s$, and $\theta\to-sL-\tilde Y$ into the indifference condition $\Delta(\tau;c)=0$ and passing to the limit shows that every limit point of $-\underline{\tau}(c,\sigma)/\sigma^{2}$ solves
\[
c=F(L)\equiv s^{2}L+\frac{1}{L}+s\,\E\big[\lambda(-sL-\tilde Y)\big].
\]
These limit points exist and stay positive: $-\tau/\sigma^2> -\tau^*/\sigma^2\to k/s>0$, while the exact benefit satisfies $c> -\rho^2\tau$, so $L<c/(\sigma^2\rho^2)\to c/s^2$. The function $F$ is strictly convex, since $F''(L)=2/L^{3}+s^{3}\,\E[\lambda''(-sL-\tilde Y)]>0$, and diverges at $0$ and at $\infty$. Its first-order condition $(sL)^{2}\big(1-\E[\lambda'(-sL-\tilde Y)]\big)=1$ is the fixed point that defines $k$ in the proof of Proposition \ref{no1}, so the minimum of $F$ is $F(k/s)=c_\infty^\sigma(s)$. For $c>c_\infty^\sigma(s)$, the equation $F(L)=c$ therefore has exactly two roots. Because $\underline{\tau}(c,\sigma)<\tau^*$ and $-\tau^*/\sigma^{2}\to k/s$, the limit is the larger root, denoted $\underline{L}(c)>k/s$, and
\[
\frac{\underline{\tau}(c,\sigma)}{\sigma^{2}}
\;\rightarrow\;
-\underline{L}(c)<0
\qquad(\sigma\uparrow\infty).
\]
Consequently,
\[
\frac{\underline{\tau}(c,\sigma)}{\sigma}
\;\sim\;
-\underline{L}(c)\,\sigma
\;\rightarrow\;-\infty,
\qquad
1-\Phi\!\Big(\frac{\underline{\tau}(c,\sigma)}{\sigma}\Big)
\;\rightarrow\;1
\qquad(\sigma\uparrow\infty).
\]
Moreover, the implicit function theorem gives $\partial\underline{\tau}/\partial\sigma=-\Delta_\sigma/\Delta_\tau$ at the lower root. Evaluating both partial derivatives along $\tau=-\underline{L}(c)\,\sigma^{2}(1+o(1))$, using the $\sigma$-derivative decomposition displayed in the proof of Corollary 2, the tail expansions $\lambda(w)-w\lambda'(w)=2/w+o(1/w)$ and $1-\lambda'(w)=w^{-2}+o(w^{-2})$ at $w=-\tau/\sigma$, and $\theta\to-s\underline{L}(c)-\tilde Y$, yields
\[
\Delta_\sigma=-\frac{2\underline{L}(c)}{\sigma}\,F'(\underline{L}(c))\,(1+o(1)),
\qquad
\Delta_\tau=-\frac{F'(\underline{L}(c))}{\sigma^{2}}\,(1+o(1)),
\]
with $F'(\underline{L}(c))>0$ at the larger root, so that
\[
\partial\underline{\tau}(c,\sigma)/\partial{\sigma}
=-2\,\underline{L}(c)\,\sigma\,(1+o(1)),
\]
and hence
\[
\frac{d}{d\sigma}\Big(\frac{\underline{\tau}(c,\sigma)}{\sigma}\Big)
=\frac{\sigma\,\partial\underline{\tau}(c,\sigma)/\partial{\sigma}
-\underline{\tau}(c,\sigma)}{\sigma^{2}}
\;\rightarrow\;
-\underline{L}(c)
\;<\;0
\qquad(\sigma\uparrow\infty).
\]
Therefore, for all $\sigma$ large enough,
\[
\frac{d}{d\sigma}\Big(\frac{\underline{\tau}(c,\sigma)}{\sigma}\Big)<0,
\]
and the derivative of the probability of disclosure is
\[
-\phi\!\Big(\frac{\underline{\tau}(c,\sigma)}{\sigma}\Big)\,
\frac{d}{d\sigma}\Big(\frac{\underline{\tau}(c,\sigma)}{\sigma}\Big)
>0
\quad\text{for $\sigma$ large}.
\]\qed

\begin{proposition}\label{LM}
Suppose that $c>c_0$ and $\rho\geq 1/\sqrt{2}$ (i.e., $\sigma\leq s$). Holding $s$ and $c$ fixed, the probability of disclosure is decreasing in $\sigma$ under $\underline{\tau}(c)$ and increasing in $\sigma$ under $\overline{\tau}(c)$.
\end{proposition}

\paragraph{Proof of Proposition \ref{LM}.} In what follows, we write $z(\sigma)\equiv \tau/\sigma$ at either equilibrium root and make the dependence of the variables on $\sigma$ explicit when needed for clarity.
By lemma~2, the indifference condition can be written as
\begin{equation}\label{eq:sH=c}
\sigma\,H\big(z(\sigma),\rho(\sigma)\big)
= c,
\end{equation}
where $H(z,\rho)
\;\equiv\; (1-\rho^2)\,z + \rho\,\E\big[\lambda(\theta)\big] + \lambda(-z).$ 
Differentiating the above expression,
\[\vspace{-0.2cm}
0
= \frac{d}{d\sigma}\big[\sigma H(z(\sigma),\rho(\sigma))\big]
= H + \sigma\big(H_z\,z'(\sigma)+H_\rho\,\rho'(\sigma)\big).
\]
Substituting $\rho'(\sigma)
= -\frac{\rho(\sigma)\big(1-\rho(\sigma)^2\big)}{\sigma}$ into the above expression and reorganizing terms:\vspace{-0.2cm}
\begin{equation}\vspace{-0.2cm}\label{eq:zprime}
z'(\sigma)
=-\,\frac{H(z,\rho)-\rho(1-\rho^2)H_\rho(z,\rho)}{\sigma H_z(z,\rho)}.
\end{equation}

We use the following auxiliary lemma.

\begin{lemma}\label{prooflem}
$0<H_\rho(z,\rho)
<2\,\E\!\big[\lambda(\theta)-\theta\big].$
\end{lemma}
\paragraph{Proof of Lemma \ref{prooflem}.} 
Because $H(z,\rho)
= \frac{1}{\sigma}\,\E\big[P(\tau+\tilde\varepsilon)\mid\tau\big] + \lambda(-z)$,

\begin{equation}\label{eq:Hrho}
H_\rho(z,\rho)
= 2\,\E\Big[(1-\lambda'(\theta))\big(\lambda(\theta)-\theta\big)\Big].
\end{equation}
Using the fact that $0<\lambda'(u)<1$ and 
$\lambda(u) > u$, each factor in the expectation is strictly positive and therefore $H_\rho(z,\rho)>0.$ Moreover, since $0<1-\lambda'(\theta)<1$, $H_\rho(z,\rho)<2\,\E\big[\lambda(\theta)-\theta\big]$, completing the proof.\qed 
\vspace{-0.2cm}
\paragraph{}
Recall from Proposition 1 that $H_z(z(\sigma),\rho(\sigma))=\frac{\partial\Delta}{\partial\tau}\big|_{\tau=z\sigma}$ is negative at the lower root and positive at the upper root, so the sign of $z'(\sigma)$ in (\ref{eq:zprime}) is that of $N(z,\rho)\equiv H(z,\rho)-\rho(1-\rho^2)H_\rho(z,\rho)$ at the lower root and its opposite at the upper root. By lemma \ref{prooflem} and $2\rho^2\geq 1$,\vspace{-0.2cm}
\[\vspace{-0.2cm}
N(z,\rho)>H(z,\rho)-2\rho(1-\rho^2)\,\E[\lambda(\theta)-\theta]
=z+\lambda(-z)+\rho(2\rho^2-1)\big(\E[\lambda(\theta)]-\E[\theta]\big)>0,
\]
where the equality collects terms using $\E[\theta]=\rho z$, and the final inequality uses $\lambda(u)>u$ together with $z+\lambda(-z)>0$, which follows from $\E[\tilde Y\mid \tilde Y<z]=-\lambda(-z)<z$. Hence $z'(\sigma)>0$ at the lower root, where the probability of disclosure $1-\Phi(z)$ falls, and $z'(\sigma)<0$ at the upper root, where it rises.\qed

\subsection*{Uncertain Information Endowment with Disclosure Costs}\label{sec:dye-cost-supp}

\paragraph{}
The Uncertain Information Endowment subsection of Section~V.2 sets $c=0$. We now allow $c>0$ alongside the Dye-type uncertainty $q\in(0,1)$ and characterize the interaction of the two frictions. The setup is unchanged except that the firm pays a cost $c$ upon disclosing, so that the marginal-discloser indifference condition becomes
\vspace{-0.2cm}\begin{equation}\label{eq:dye-cost-indiff}
\E[P(\tau+\tilde\varepsilon)\mid \tau]-c=P(\emptyset;\tau),
\vspace{-0.2cm}\end{equation}
with the expected disclosure price as in main-text equation~(4) and $P(\emptyset;\tau)$ as in main-text equation~(5). In standardized form $z\equiv\tau/\sigma$, $r\equiv s/\sigma$, $\kappa\equiv c/\sigma$, and writing
\vspace{-0.2cm}\begin{equation}\label{eq:dye-Psi}
\Psi_{r,q}(z)\equiv h_r(z)+\frac{(1-q)\phi(z)}{q+(1-q)\Phi(z)},
\qquad
h_r(z)\equiv\tfrac{1}{\sigma}\E[P(\tau+\tilde\varepsilon)\mid \tau],
\vspace{-0.2cm}\end{equation}
the finite-threshold equilibria are exactly the roots of $\Psi_{r,q}(z)=\kappa$. The tail behavior of $\Psi_{r,q}$ differs from the pure costly-disclosure model. For any fixed $q>0$, $\Psi_{r,q}(z)\to-\infty$ as $z\to-\infty$ because a very low non-disclosure threshold is mostly attributed to the exogenous inability to disclose, and $\Psi_{r,q}(z)\to+\infty$ as $z\to+\infty$. Hence a finite-threshold root exists for every cost $\kappa\geq 0$.

\begin{proposition}\label{prop:dye-costs}
Fix $r>0$ and let $\delta(q)\equiv q/(1-q)$. For every $q\in(0,1)$ and every $\kappa\geq 0$, $\Psi_{r,q}(z)=\kappa$ admits at least one finite solution. Moreover:
\begin{itemize}\vspace{-0.2cm}
\item[(i)] let $\kappa_0(r)\equiv\min_z\{h_r(z)+\lambda(-z)\}$ and let $K$ be any compact subset of $(\kappa_0(r),\infty)$. There exists $q_K(r)>0$ such that, for every $q\in(0,q_K(r))$ and every $\kappa\in K$, the equation has at least three distinct finite solutions, the two roots inherited from the baseline and at least one additional root in the far left tail;\vspace{-0.2cm}
\item[(ii)] if
\vspace{-0.2cm}\begin{equation}\label{eq:dye-highq}
\frac{1}{\delta(q)\sqrt{2\pi e}}+\frac{1}{2\pi\delta(q)^2}<\frac{1}{1+r^2},\vspace{-0.2cm}
\end{equation}
then $\Psi_{r,q}$ is strictly increasing and the equation has a unique finite solution for every $\kappa\geq 0$. In particular, for each $r>0$ there exists $q_H(r)<1$ such that uniqueness holds for $q\geq q_H(r)$.
\end{itemize}
\end{proposition}

Part~(i) shows that the two baseline thresholds survive and that the Dye tail adds at least one further low threshold when exogenous non-disclosure is rare, so the baseline multiplicity is inherited near $q=0$. Part~(ii) is a primitive sufficient condition under which the Dye term is too small to overturn the monotonicity inherited from $h_r$, and the equilibrium is then unique for every cost. The proposition does not claim an exact global root count at intermediate $q$.

\begin{figure}[H]
\centering
\includegraphics[width=0.62\linewidth]{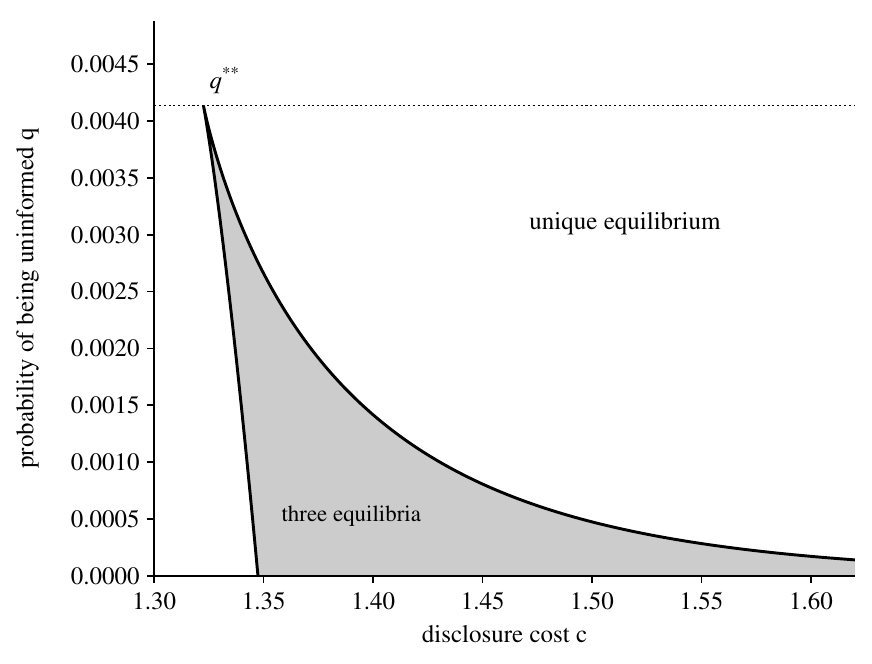}
\caption{Multiplicity wedge of equation (\ref{eq:dye-cost-indiff}) for $\sigma=s=1$. For $q>0$, the interior of the shaded region has three numerically computed equilibria and the unshaded region has one; fold boundaries have two distinct roots. The boundaries meet at $q^{**}\approx 0.004136$ and $c\approx1.322495$. The plot shows a limited range of costs; as $q\downarrow0$, the limiting multiplicity interval is $(c_0,\infty)$.}\label{plot:dye-cost}\vspace{-.3cm}
\end{figure}

Figure~\ref{plot:dye-cost} plots the numerically computed boundary of the multiplicity region in the $(c,q)$ plane for $\sigma=s=1$; within the shaded wedge the computed number of equilibria is exactly three. As $q\downarrow0$, its lower boundary converges to $c_0$ and its upper boundary diverges, consistent with the baseline multiplicity interval $(c_0,\infty)$ of Proposition~1. The numerically computed boundaries meet at $q^{**}\approx 0.004136$ and $c\approx1.322495$; above this value of $q$, the numerical calculation finds a unique equilibrium. The sufficient condition in part~(ii) guarantees uniqueness for $q>0.4612$ when $\sigma=s=1$ and does not characterize the numerical cusp. The picture confirms that the multiplicity which drives the comparative statics of Section~III survives in the Dye+cost environment only for a very thin sliver near $q=0$.

\paragraph{Proof of Proposition \ref{prop:dye-costs}.} Let $D_\delta(z)\equiv\phi(z)/(\delta+\Phi(z))$, with $\delta=q/(1-q)$; then $\Psi_{r,q}(z)=h_r(z)+D_\delta(z)$. Tail behavior: $h_r(z)\sim z/(1+r^2)$ and $D_\delta(z)\to 0$ as $z\to-\infty$, so $\Psi_{r,q}(z)\to-\infty$; at the other tail $h_r(z)\to\infty$ and $D_\delta(z)\to 0$, so $\Psi_{r,q}(z)\to\infty$. Continuity gives at least one root for every $\kappa\geq 0$.

\emph{Part~(ii).} Differentiate: $D_\delta'(z)=-zD_\delta(z)-D_\delta(z)^2$. Since $\Phi(z)+\delta\geq\delta$,
\[
|D_\delta'(z)|\leq |z|\phi(z)/\delta+\phi(z)^2/\delta^2\leq 1/(\delta\sqrt{2\pi e})+1/(2\pi\delta^2),
\]
and $h_r'(z)\geq 1/(1+r^2)$ pointwise. Under~(\ref{eq:dye-highq}), $\Psi_{r,q}'>0$ everywhere, hence strictly monotone and the root is unique. The left-hand side of~(\ref{eq:dye-highq}) vanishes as $q\uparrow 1$, so $q_H(r)<1$ exists.

\emph{Part~(i).} For $q=0$, $D_0(z)=\lambda(-z)$ and $\Psi_{r,0}(z)=h_r(z)+\lambda(-z)$ is the standardized marginal-discloser benefit of the baseline costly-disclosure model, which is strictly convex, divergent in both tails, and has minimum value $\kappa_0(r)$. For every $\kappa\in K$ it therefore has two simple roots, which move continuously with $\kappa$. By compactness of $K$, choose two disjoint bounded neighborhoods containing the two root branches over $K$, on whose boundaries $\Psi_{r,0}(z)-\kappa$ keeps a uniform nonzero sign. On bounded sets, $D_\delta\to D_0$ uniformly as $q\downarrow 0$, so the boundary signs persist and the intermediate value theorem places one root in each neighborhood for every $\kappa\in K$ and every $q$ small enough.

For the third root, choose $L$ to the left of both neighborhoods with $\Psi_{r,0}(L)>\max K+1$, which is possible because $\Psi_{r,0}(z)\to\infty$ as $z\to-\infty$. Pointwise convergence at $L$ gives $\Psi_{r,q}(L)>\max K$ for all $q$ small enough, while, for each fixed $q>0$, $\Psi_{r,q}(z)\to-\infty$ as $z\to-\infty$. The intermediate value theorem then yields a root in $(-\infty,L)$ for every $\kappa\in K$. \qed

\subsection*{Alternative Equilibrium-Selection Criteria for the Certification Game}\label{sec:eqsel-supp}

\paragraph{}
In the main text we resolve the multiplicity of certification equilibria using the least-cost criterion of Definition~2. Proposition~2 then characterizes the unique equilibrium as a function of $s/\sigma$. Other selection criteria have been used elsewhere in the literature and produce predictions that differ from the least-cost one in instructive ways. We analyze two such alternatives, a \emph{continuity refinement} on continuation beliefs and a criterion for selecting continuation equilibria motivated by \emph{forward-induction} reasoning.\footnote{The main text adopts the least-cost criterion directly, and we record the analysis under alternative refinements here.} Both alternatives require an exogenous upper bound on admissible fees, denoted $\overline{c}\in(c_0,\infty)$, which we interpret as a participation or outside-option cap (e.g., above $\overline c$ firms would prefer to remain private).

Throughout this subsection, $\underline{\tau}(c),\overline{\tau}(c)$ denote the two threshold equilibria from Proposition~1 (defined on $c\geq c_0$), and we reuse
\[
\underline{V}(c) \equiv c\,(1-\Phi(\underline{\tau}(c)/\sigma)),\qquad
\overline{V}(c) \equiv c\,(1-\Phi(\overline{\tau}(c)/\sigma)),\qquad
\overline V^{**}=\max_{c\geq c_0}\overline{V}(c),
\]
from Lemma~4 of the main text. The condition
\begin{equation}\label{eq:Vbar-dec}
\frac{c_0}{\sigma}\,
\lambda\!\left(\frac{\tau(c_0)}{\sigma}\right) > 1
\end{equation}
plays a recurring role: when~(\ref{eq:Vbar-dec}) holds, $\overline{V}(c)$ is strictly decreasing on $[c_0,\infty)$ and hence $\overline V^{**}=\overline V(c_0)$. We do not restate Lemma~4 here; its proof appears earlier in this appendix. The existence of the maximizer $\overline V^{**}$ follows from continuity of $\overline V$ on $[c_0,\infty)$ together with $\lim_{c\to\infty}\overline V(c)=0$.

No refinement is imposed at this stage, so the fee-choice game on $c\in[0,\overline c]$ retains its full equilibrium set.

\begin{proposition}\label{opt1}
Suppose $\overline{c}\geq \underline{V}^{-1}(c_0)$ and let $\overline{V}^{**}_{\overline{c}}\equiv\max_{c\in[c_0,\overline{c}]}\overline{V}(c)$. A fee $c^*$ with continuation $\tau(c^*)$ is on the path of an equilibrium of the fee-choice subgame if and only if either (i) $\tau(c^*)=\underline{\tau}(c^*)$ and $\underline{V}(c^*)\geq\max\{c_0,\overline{V}^{**}_{\overline{c}}\}$, or (ii) $\tau(c^*)=\overline{\tau}(c^*)$ and $\overline{V}(c^*)=\overline{V}^{**}_{\overline{c}}\geq c_0$. In either case, assigning the upper-threshold continuation at every other interior fee supports the outcome and, in any equilibrium of type (i), the continuation at fees above $c^*$ is the upper threshold.
\end{proposition}

\begin{corollary}\label{HI}
Under the conditions of Proposition~\ref{opt1}, if (\ref{eq:Vbar-dec}) holds, any equilibrium must satisfy $\tau(c^*)=\underline{\tau}(c^*)$. Otherwise, a fee at the upper threshold is an equilibrium if and only if $\overline{V}(c^*)=\overline{V}^{**}_{\overline{c}}\geq c_0$.
\end{corollary}

For completeness, we record what happens when the cap $\overline c$ is small.

\begin{proposition}\label{opt12}
Suppose that $\overline{c}<\underline{V}^{-1}(c_0)$ and denote $c^*$ the optimal fee chosen by the certifier:
\begin{itemize}\vspace{-0.3cm}
\item[(i)] If $\overline{c}<c_0$, the certifier chooses $c^{*}=\overline{c}$ and achieves full disclosure with profit $\overline{c}$;\vspace{-0.3cm}
\item[(ii)] If $\overline{c}\in [c_0,\underline{V}^{-1}(c_0))$, there exists no equilibrium.\vspace{-0.3cm}
\end{itemize}
\end{proposition}

\paragraph{Proof of Propositions \ref{opt1} and \ref{opt12}, and Corollary \ref{HI}.} Suppose $c\in [0,\overline{c}]$. If $\overline{c}<c_0$, $\tau(c)=-\infty$ features unravelling for any $c$, so the certifier's profit $V(c)=c$ is increasing in $c$ and maximized at $\overline{c}$. If $\overline{c}\in [c_0,\underline{V}^{-1}(c_0))$, $\sup_{c< c_0} V(c)=c_0$, while any $c\geq c_0$ yields a profit $V(c)\leq \underline{V}(c)\leq \underline{V}(\overline{c})<c_0$. Since no fee achieves $V(c)=c_0$, no equilibrium exists.

Suppose next that $\overline{c}\geq \underline{V}^{-1}(c_0)$. Deviations to fees below $c_0$ earn profits arbitrarily close to $c_0$. At a maximizer of $\overline{V}$ on $[c_0,\overline{c}]$, either continuation pays at least $\overline{V}^{**}_{\overline{c}}$, so a deviation there guarantees that amount. Any equilibrium profit is therefore at least $\max\{c_0,\overline{V}^{**}_{\overline{c}}\}$, which gives the necessity of the payoff conditions in (i) and (ii). Conversely, when the on-path profit reaches this bound, assigning the upper-threshold continuation at every other interior fee caps every interior deviation at $\overline{V}^{**}_{\overline{c}}$ and every deviation below $c_0$ strictly below $c_0$, so no deviation is profitable. For the last claim, in an equilibrium of type (i) a lower-threshold continuation at any $c\in(c^*,\overline{c}]$ would earn $\underline{V}(c)>\underline{V}(c^*)$, because $\underline{V}$ is increasing, so the continuation above $c^*$ must be the upper threshold. Finally, if (\ref{eq:Vbar-dec}) holds, then $\overline{V}(c)$ is decreasing on $[c_0,\infty)$, so $\overline{V}^{**}_{\overline{c}}=\overline{V}(c_0)<c_0$ and no type-(ii) equilibrium exists; otherwise a type-(ii) equilibrium requires $\overline{V}(c^*)=\overline{V}^{**}_{\overline{c}}\geq c_0$.\qed

Certain equilibria in Proposition~\ref{opt1} have the feature that, for $c < \overline{c}$, the equilibrium path $\tau(c^*) = \underline{\tau}(c^*)$ must be disciplined by a switch to $\tau(c) = \overline{\tau}(c)$ for $c > c^*$. Our first refinement rules out such ad-hoc discontinuities. Let $\mathcal{C}$, given by $(c_0,\infty)$, denote the set of costs at which the subgame equilibrium is not unique. Assumption~(C) requires that $\tau(c)$ be continuous for any $c$ in the interior of $\mathcal{C}$.

\begin{proposition} \label{prop:unique}
Suppose $\overline{c}\geq\underline V^{-1}(c_0)$ and Assumption~(C) is imposed on off-equilibrium beliefs. If (\ref{eq:Vbar-dec}) holds, the equilibrium is unique with $c^*=\overline{c}$ and $\tau(c^*)=\underline{\tau}(\overline{c})$.
\end{proposition}

\paragraph{Proof of Proposition \ref{prop:unique}.} From Corollary~\ref{HI}, (\ref{eq:Vbar-dec}) implies that $\tau(c^*)=\underline{\tau}(c^*)$ and therefore, from Proposition~\ref{opt1}, for any $c\in (c^*,\overline{c}]$, $\tau(c)=\overline{\tau}(c)$. If $c^*<\overline{c}$, this implies that $\tau(c)$ is discontinuous at $c^*$ even though the subgame equilibrium is not unique at $c^*$, a contradiction to~(C). Hence $c^*=\overline c$.\qed

A limit argument clarifies the regularity built into Assumption~(C). Fix a compact interval $I\subset(c_0,\overline c)$ and a sequence of continuation-threshold conjectures $\mu_n:I\to\mathbb R$. Assume that $\{\mu_n\}$ is equicontinuous: for every $\epsilon>0$ there exists $\delta>0$ such that $|\mu_n(c)-\mu_n(c')|<\epsilon$ for every $n$ whenever $c,c'\in I$ and $|c-c'|<\delta$. This is a uniform restriction on how quickly continuation beliefs can change with the fee; it is not implied by vanishing observation noise alone.

\begin{proposition} \label{prop:foundation}
Suppose that $\mu_n(c)\to\tau(c)$ for every $c\in I$, where $\tau(c)\in\{\underline\tau(c),\overline\tau(c)\}$. Then $\tau$ is continuous on $I$. Moreover, either $\tau(c)=\underline\tau(c)$ for every $c\in I$, or $\tau(c)=\overline\tau(c)$ for every $c\in I$.
\end{proposition}

\paragraph{Proof of Proposition \ref{prop:foundation}.}
Fix $\epsilon>0$ and choose $\delta>0$ from equicontinuity with tolerance $\epsilon/2$. Taking pointwise limits gives $|\tau(c)-\tau(c')|\leq\epsilon/2<\epsilon$ whenever $|c-c'|<\delta$, proving continuity. The function
\[
r(c)=\frac{\tau(c)-\underline\tau(c)}{\overline\tau(c)-\underline\tau(c)}
\]
is therefore continuous and takes values in $\{0,1\}$. Since $I$ is connected, $r$ is constant, proving the branch claim.\qed

The preceding result under Assumption~(C) does not cover the case in which (\ref{eq:Vbar-dec}) fails. As a separate alternative to (C), we consider a forward-induction-motivated criterion that asks whether an equilibrium continuation can rationalize a deviation by the certifier. We first state a restriction requiring the selection of a continuation that makes the deviation profitable when such a continuation exists.\footnote{This condition is motivated by the forward-induction idea of interpreting unexpected choices as rational whenever possible, which underlies the best-rationalization principle in \citet{battigalli2003}. The analysis in \citet{battigalli2003} distinguishes restrictions on first-order beliefs from the implications of rationality and strong belief in rationality, and shows how these restrictions shape behavior. Condition (F), however, is imposed directly on the selection of continuation equilibria, which is not derived from their solution concept and is not identified with $\Delta$-rationalizability of the full certification game.}

For a candidate equilibrium with fee $c^*$ and profit $V^*$, let $E(c)$ be the set of equilibrium thresholds of the disclosure subgame: $E(c)=\{-\infty\}$ for $c<c_0$ and $E(c)=\{\underline\tau(c),\overline\tau(c)\}$ for $c\geq c_0$. Letting a threshold represent its associated continuation equilibrium, we impose the following condition:

\emph{(F) At an off-path fee $c$, if some $t\in E(c)$ gives $V(c;t)>V^*$, the selected continuation $\tau(c)\in E(c)$ must also give $V(c;\tau(c))>V^*$. If no such $t$ exists, there is no additional restriction at $c$.}

In addition to requiring that an unexpected fee be interpreted in light of the certifier's incentive to choose it, condition~(F) imposes a coordination restriction. When an off-path fee admits an equilibrium continuation that makes the deviation strictly profitable for the certifier, the firm and investors are assumed to coordinate on such a continuation. This restriction concerns continuation play, which determines the certifier's ability to rationalize a deviation.

\begin{proposition} \label{prop:forward1}
Suppose $\overline{c}\geq\underline V^{-1}(c_0)$, that $\overline{c}$ does not maximize $\overline{V}$ on $[c_0,\overline{c}]$, and that continuation selection satisfies~(F). Then the unique equilibrium fee and threshold are $c^*=\overline{c}$ and $\tau(c^*)=\underline\tau(\overline c)$.
\end{proposition}

\paragraph{Proof of Proposition \ref{prop:forward1}.}
Let $M=\max_{c\in[c_0,\overline c]}\overline V(c)$. By Proposition~\ref{opt1}, any equilibrium profit satisfies $V^*\geq\max\{c_0,M\}$, and its fee exceeds $c_0$. If $c^*<\overline c$, strict monotonicity of $\underline V$ gives
\[
\overline V(\overline c)<M\leq V^*\leq\underline V(c^*)<\underline V(\overline c).
\]
Thus only the lower continuation at $\overline c$ makes the deviation profitable. Condition~(F) selects it, contradicting optimality of $c^*$. The upper continuation at the cap is also impossible, since its profit is below $M$.

For existence, select the lower continuation at the cap and the upper continuation at every other fee $c\in[c_0,\overline c)$. Any fee below $c_0$ earns less than $c_0\leq\underline V(\overline c)$, and any $c\in[c_0,\overline c)$ earns at most $\underline V(c)<\underline V(\overline c)$. Hence no deviation is profitable under any equilibrium continuation, so~(F) is satisfied and the cap is optimal. Off-path continuations need not be unique.\qed

\subsection*{Additional Proofs from Section V}

The next lemma uses the bounded-support assumptions in Proposition 3; the subsequent propositions explicitly vary the support of the prior.

\begin{lemma} \label{lem:mlrp}
Suppose that $\tilde{v}$ has continuous strictly positive density on $[\underline{v},\overline{v}]$, and $\tilde{\varepsilon}$ is independent of $\tilde{v}$ with a continuous strictly positive log-concave density on $\mathbb{R}$. For any $\tau \in [\underline{v},\overline{v}]$, 
(i) $P(x)$ is increasing in $x$; (ii) $\mathbb{E} \left(P(\tilde{x})|v \right)$ is increasing in $v$; (iii) the firm's disclosure strategy features a threshold in equilibrium; (iv) $\mathbb{E}\left(P\left(\tau + \tilde{\varepsilon} \right)|\tau \right)$ is increasing and continuous in $\tau$. 
\end{lemma}

\paragraph{Proof of Lemma \ref{lem:mlrp}.} 
Because $f_\varepsilon$ is log-concave, $f(x| v)=f_\varepsilon (x-v)$ satisfies MLRP. Equivalently, for any $x_2 > x_1$, $f_\varepsilon(x_2 - v)/f_\varepsilon(x_1 - v)$ is increasing in $v$. Hence, the posterior distribution of $v$ conditional on disclosure and $x$ is ordered by monotone likelihood ratio as $x$ rises. So the posterior mean $P(x) = \mathbb{E} \bigl[\tilde{v}| \tilde{v} \ge \tau, x\bigr]$ is increasing in $x$.

Now fix $\tau$. For any $v_2 > v_1$, $\mathbb{E} \left(P(\tilde{x})|v_2 \right) \ge \mathbb{E} \left(P(\tilde{x})|v_1 \right)$, because $P(x)$ is increasing in $x$ and $v_2 + \tilde{\varepsilon} \ge v_1 + \tilde{\varepsilon}$ pointwise. For $\tau<\overline v$, $P$ is nonconstant: otherwise $\operatorname{Cov}(\tilde v,\tilde x\mid\tilde v\geq\tau)=\operatorname{Var}(\tilde v\mid\tilde v\geq\tau)>0$ would contradict a constant conditional mean. Since $P$ is continuous and nondecreasing and the noise density is positive everywhere, every positive shift of $v$ strictly raises the expectation. Thus $\mathbb{E} \left(P(\tilde{x})|v \right)$ is strictly increasing in $v$ whenever the disclosure pool is nondegenerate. 

Further because the non-disclosure price $P(\emptyset)$ does not depend on the firm's realized type $v$, the net gain from disclosing, $\mathbb{E} \left(P(\tilde{x})|v \right) - c - P(\emptyset)$, is increasing in $v$. The same argument holds for any disclosure set with positive probability, since conditioning the prior on that set preserves the likelihood-ratio ordering and a positive conditional variance. Thus the net gain is strictly increasing whenever the disclosure pool is nondegenerate, and the disclosure set is an upper interval up to an indifferent type of probability zero. The full-disclosure and no-disclosure cases are already threshold strategies.

For the last part, if $\tau_2 > \tau_1$, then $\mathbb{E} \bigl[\tilde{v}| \tilde{v} \ge \tau_2, x\bigr] \ge \mathbb{E} \bigl[\tilde{v}| \tilde{v} \ge \tau_1, x\bigr]$ for all $x$. Because $P(x)$ is increasing in $x$ and $\mathbb{E} \left(P(\tilde{x})|v \right)$ is increasing in $v$, we have \vspace{-.5cm}
\begin{equation*}
    \mathbb{E}\left(P\left(\tau_2 + \tilde{\varepsilon} \right)|\tau_2 \right) \ge \mathbb{E}\left(P\left(\tau_1 + \tilde{\varepsilon} \right)|\tau_2 \right) \ge \mathbb{E}\left(P\left(\tau_1 + \tilde{\varepsilon} \right)|\tau_1 \right).
\end{equation*} 
So $\mathbb{E}\left(P\left(\tau + \tilde{\varepsilon} \right)|\tau \right)$ is increasing in $\tau$. Continuity follows from dominated convergence and continuity of the posterior mean on compact support.
\qed

\begin{proposition}
\label{NN3}
Suppose that the support of $\tilde v$ is unbounded on both sides and that the prior is regular in its lower tail, in the sense that, as $\tau\rightarrow-\infty$, $P(\emptyset)=\tau+o(\tau)$ and $\E[P(\tau+\tilde\varepsilon)\mid\tau]/\tau$ converges to a limit strictly below one. Then $\psi$ diverges in both tails, its minimizer is interior, and, for every $c>c_0$, the equation $\psi(\tau)=c$ has a root on each side of the minimizer.
\end{proposition}

\paragraph{Proof of Proposition \ref{NN3}.} Let $\kappa<1$ denote the limit in the statement. As $\tau\rightarrow-\infty$, $\psi(\tau)=\E[P(\tau+\tilde\varepsilon)\mid\tau]-P(\emptyset)=(\kappa-1)\tau+o(\tau)\rightarrow\infty$. As $\tau\rightarrow\infty$, $P(x)\geq\tau$ for every $x$ while $P(\emptyset)\leq\E(\tilde v)$, so $\psi(\tau)\geq\tau-\E(\tilde v)\rightarrow\infty$. A continuous function that diverges in both tails attains its minimum at an interior point and, for any $c>c_0$, the intermediate value theorem yields a root on each side of the minimizer.\qed

The next proposition specializes the analysis to truncated normals.

\begin{proposition}
\label{NN2}
Suppose that $\tilde{v}$ is a normally distributed random variable $N(0,\sigma^2)$ truncated from below at $\eta$. There  exist three cost levels $c_0^l \leq c_0^m \le c_0^h$, satisfying: 
\begin{itemize}
\item[(i)] if $c < c_0^l$, there is a unique subgame equilibrium and it must be full disclosure, i.e., $\tau(c)=\eta$;
\item[(ii)] if $c_0^l \le c \le c_0^m$, there can be one or more subgame equilibria and at least one features partial disclosure for $c \in (c_0^l, c_0^m]$; 
\item[(iii)] if $c\in (c_0^m, c_0^h)$, there exist at least two subgame equilibria with interior thresholds such that the highest equilibrium threshold $\overline{\tau}(c)$ is increasing in $c$; at $c=c_0^h>c_0^m$, there is at least one partial-disclosure equilibrium;
\item[(iv)] if $c> c_0^h$, there is a unique subgame equilibrium $\overline{\tau}(c).$  
\end{itemize}
Further, if $\eta$ is sufficiently small (large), $c_0^l \le c_0^m < c_0^h$ (resp., $c_0^l = c_0^h$).
\end{proposition}

        \begin{lemma} \label{lem:bounded}
                Let $\phi$ and $\Phi$ denote the standard normal pdf and cdf. For
                \[            P(\varnothing;\tau)\;=\;\sigma\,\frac{\phi\!\left(\frac{\eta}{\sigma}\right)-\phi\!\left(\frac{\tau}{\sigma}\right)}
                {\Phi\!\left(\frac{\tau}{\sigma}\right)-\Phi\!\left(\frac{\eta}{\sigma}\right)}\!,
                \qquad \text{where } \tau\ge \eta,
                \]
                if $\eta\ge 0$, then
                \[
                0<\frac{d}{d\tau}P(\varnothing;\tau)\le \frac12\qquad\text{for all }\ \tau\ge \eta,
                \]
                with equality $\frac{1}{2}$ attained only in the right limit $\tau\downarrow\eta$.
        \end{lemma}
        
        \paragraph{Proof of Lemma \ref{lem:bounded}.}
Write $a=\eta/\sigma$, $t=\tau/\sigma$, and $A=\int_a^t\phi(w)\,dw$. For $t>a$, direct differentiation gives
\[
\frac{dP(\emptyset;\tau)}{d\tau}
=\frac{\phi(t)\int_a^t(t-w)\phi(w)\,dw}{A^2}>0.
\]
If $0\leq a<t$, the standard normal density is strictly decreasing on $(a,t)$, so
\[
A^2
=2\int_a^t\phi(w)\left(\int_w^t\phi(u)\,du\right)dw
>2\phi(t)\int_a^t(t-w)\phi(w)\,dw.
\]
Thus $dP(\emptyset;\tau)/d\tau<1/2$ for $\tau>\eta\geq0$. Expanding the numerator and denominator at $t=a$ gives the right limit $1/2$. \qed

\paragraph{Proof of Proposition \ref{NN2}.}
Write $\psi_\eta(\tau)=\E[P(\tau+\varepsilon)\mid\tau]-P_\eta(\emptyset;\tau)$, extended continuously to $\tau=\eta$. Since $\E[P(\tau+\varepsilon)\mid\tau]>\tau\geq P_\eta(\emptyset;\tau)$, the benefit is strictly positive. Moreover, $\E[P(\tau+\varepsilon)\mid\tau]\sim\tau$ while $P_\eta(\emptyset;\tau)\to\sigma\lambda(\eta/\sigma)$, so $\psi_\eta(\tau)\to\infty$. Hence $c_0^l=\min_{\tau\geq\eta}\psi_\eta(\tau)>0$. If $c<c_0^l$, the unique equilibrium is full disclosure. For $c>c_0^l$, continuity and the upper tail give an interior root, hence a partial-disclosure equilibrium. At $c=c_0^l$, an interior minimizer gives a partial-disclosure equilibrium; if the minimum is attained only at $\eta$, full disclosure satisfies the corner condition.

The normal formulas imply that $\psi_\eta$ is real analytic, including the removable endpoint at $\eta$, and $\psi_\eta'(\tau)\to1$. Thus its stationary points are isolated and lie in a compact interval, so there are finitely many. Let $\mathcal N=\{\tau\geq\eta:\psi_\eta'(\tau)<0\}$. If $\mathcal N$ is empty, $\psi_\eta$ is strictly increasing, because its derivative is nonnegative and has only isolated zeros. Set $c_0^l=c_0^m=c_0^h=\psi_\eta(\eta)$. The equilibrium is full disclosure for $c\leq c_0^l$ and has a unique interior threshold for $c>c_0^l$, proving the claims in this case.

If $\mathcal N$ is nonempty, the derivative is negative somewhere and eventually positive, so there is a finite, nonempty set $\mathcal M$ of interior local minimizers. Analyticity makes each a strict local minimum, although its second derivative may vanish. Define
\[
T=\max\mathcal M,\qquad
c_0^m=\max_{\tau\in\mathcal M}\psi_\eta(\tau),\qquad
c_0^h=\max_{\eta\leq\tau\leq T}\psi_\eta(\tau).
\]
Then $c_0^l\leq c_0^m<c_0^h$: a point immediately to the left of any maximizer over $\mathcal M$ has a strictly larger benefit and lies in $[\eta,T]$. Beyond $T$, the derivative cannot become negative again, since its eventual positivity would produce a later local minimum. Consequently, $\psi_\eta$ is strictly increasing on $[T,\infty)$. For every $c>c_0^m$, there is exactly one root above $T$, because $\psi_\eta(T)\leq c_0^m<c$, and this upper root increases with $c$. If also $c<c_0^h$, continuity gives another root strictly between $\eta$ and $T$: the maximum on $[\eta,T]$ exceeds $c$ and the value at $T$ is below $c$. This proves (iii). At $c=c_0^h$, the upper root remains, and any interior maximizer on $[\eta,T]$ gives an additional equilibrium; a root only at $\eta$ does not, because the benefit is below $c$ at some interior thresholds. For $c>c_0^h$, there are no roots on $[\eta,T]$, so the upper root is the unique equilibrium, proving (iv). The existence argument in the first paragraph proves (ii).

Finally, the derivative of the expected disclosure price increases to one. Choose $\overline\eta\geq0$ large enough that this derivative is at least $3/4$ for $\tau\geq\overline\eta$. If $\eta\geq\overline\eta$, Lemma~\ref{lem:bounded} gives $P_\eta'(\emptyset;\tau)\leq1/2$, so $\psi_\eta'>0$ and $c_0^l=c_0^h$. Conversely, as $\eta\to-\infty$, the explicit non-disclosure price and its derivative converge uniformly on every fixed compact interval to their untruncated counterparts. The derivative of $\psi_\eta$ is therefore negative at any fixed threshold below the baseline minimizer for $\eta$ sufficiently negative. Then $\mathcal M$ is nonempty and $c_0^m<c_0^h$, proving the remaining claim.\qed

The proposition below illustrates the role of boundedness. The lower-tail truncation changes the model only through the non-disclosure price, and this change becomes
arbitrarily small on any fixed range of thresholds as the lower bound moves far enough into the left tail. Hence the main force behind Proposition 1 survives: over an intermediate range of fees, the lower threshold decreases with $c$. What bounded support changes is the behavior at the tail. Once the threshold approaches the lower bound, non-disclosure cannot be made arbitrarily punitive. So the lower-threshold branch eventually disappears and, for sufficiently high fees, the unique equilibrium has a finite upper threshold; disclosure remains positive at every finite fee because the support is unbounded above.

\begin{proposition} \label{prop:approx}
Fix a compact interval of fees $I\subset(c_0,\infty)$ on which the unbounded model has equilibria $\underline{\tau}(c)$ and $\overline{\tau}(c)$. For $\eta$ sufficiently negative, the truncated normal model has equilibria $\underline{\tau}_\eta(c)$ and $\overline{\tau}_\eta(c)$ on $I$, and
\begin{equation*}
\sup_{c \in I} |\underline{\tau}_\eta(c)-\underline{\tau}(c)| \to 0,
\qquad
\sup_{c \in I} |\overline{\tau}_\eta(c)-\overline{\tau}(c)| \to 0.
\end{equation*}
\end{proposition}

\begin{lemma} \label{lem:approx}
Let $\psi_\eta(\tau)$ be the marginal disclosure benefit at the threshold under a truncated normal distribution with lower truncation at $\eta$, and $\psi(\tau)$ be the marginal disclosure benefit in the unbounded normal model. For every compact interval $K \subset \mathbb{R}$,
\begin{equation*}
\sup_{\tau \in K} |\psi_\eta(\tau)-\psi(\tau)| \to 0
\qquad \text{as } \eta \to -\infty.
\end{equation*}
\end{lemma}

\paragraph{Proof of Lemma \ref{lem:approx}.}
Under the truncated normal distribution, the expected price conditional on disclosure is unchanged,
because lower truncation does not affect the conditional distribution given $v \ge \tau$. So the only term that changes is the non-disclosure price $P(\emptyset)$. Recall that $P_\eta(\emptyset) = \sigma
\frac{\phi(\eta/\sigma)-\phi(\tau/\sigma)}
{\Phi(\tau/\sigma)-\Phi(\eta/\sigma)}$. As \(\eta \to -\infty\), both $\phi(\eta/\sigma)$ and $\Phi(\eta/\sigma)$ converge to zero.
Hence, for each fixed $\tau$,
\begin{equation*}
P_\eta(\emptyset) \to -\sigma \frac{\phi(\tau/\sigma)}{\Phi(\tau/\sigma)} = P(\emptyset),
\end{equation*}
where $P(\emptyset)$ is the non-disclosure price in the unbounded model. Furthermore, because $K$ is compact, the denominator $\Phi(\tau/\sigma)$ is bounded away from zero on $K$. Therefore, the convergence is uniform on $K$ and $\sup_{\tau \in K} |\psi_\eta(\tau)-\psi(\tau)| \to 0$. \qed

\paragraph{Proof of Proposition \ref{prop:approx}.}
Because $I$ is compact and lies strictly above $c_0$, the two baseline branches are continuous, stay a positive distance apart, and are simple roots of $\psi(\tau)=c$. Choose disjoint compact neighborhoods of the two branch graphs on whose boundaries $\psi(\tau)-c$ keeps a uniform nonzero sign over $c\in I$. By Lemma \ref{lem:approx}, $\psi_\eta$ preserves these boundary signs for all $\eta$ sufficiently negative, so the intermediate value theorem places a root of $\psi_\eta(\tau)=c$ in each neighborhood for every $c\in I$. Shrinking the neighborhoods yields selections $\underline{\tau}_\eta(c)$ and $\overline{\tau}_\eta(c)$ with the stated uniform convergence. \qed

So a sufficiently low lower bound reproduces the unbounded behavior arbitrarily well.

\subsection*{Additional Material and Proofs for Section V.5}

For a conjectured threshold $\tau$, the posterior variances after a disclosed report $x$ and after non-disclosure follow from the same truncated-normal calculation as the posterior mean in main-text equation~(3):
\begin{equation}\label{eq:ra-variance}
\operatorname{Var}(\tilde v\mid x,\tilde v\geq\tau)
=b(s)^2\left[1-\lambda'\!\left(\frac{\tau-a(s)x}{b(s)}\right)\right],
\qquad
\operatorname{Var}(\tilde v\mid \tilde v<\tau)
=\sigma^2\left[1-\lambda'\!\left(-\frac{\tau}{\sigma}\right)\right].
\vspace{-0.2cm}
\end{equation}

Write $\overline V_D(\tau)\equiv\E[\operatorname{Var}(\tilde v\mid\tilde x,\tilde v\geq\tau)\mid\tilde v=\tau]$ for the marginal discloser's expected posterior variance after disclosure. By Lemma~2, $\E[P(\tau+\tilde\varepsilon;\tau)\mid\tilde v=\tau]=a(s)\tau+b(s)\E[\lambda(\theta_s(\tilde Y))]$, where $\theta_s(\tilde Y)=(\tau-a(s)\tilde x)/b(s)$ for $\tilde x=\tau+\tilde\varepsilon$, $\partial\theta_s(\tilde Y)/\partial\tau=\rho/\sigma$, and $b(s)\rho/\sigma=\rho^2=b(s)^2/\sigma^2$; its derivative is therefore $a(s)+\rho^2\E[\lambda'(\theta_s(\tilde Y))]=1-\overline V_D(\tau)/\sigma^2$ by (\ref{eq:ra-variance}). Likewise $P(\emptyset;\tau)=-\sigma\lambda(-\tau/\sigma)$ has derivative $\lambda'(-\tau/\sigma)=1-\operatorname{Var}(\tilde v\mid\tilde v<\tau)/\sigma^2$. Subtracting,
\begin{equation}\label{eq:variance-slope}
\sigma^2B_0'(\tau)=\operatorname{Var}(\tilde v\mid\tilde v<\tau)-\overline V_D(\tau),
\end{equation}
and since the risk-adjusted prices in Section V.5 subtract $\gamma$ times the respective posterior variances, $B_\gamma(\tau)=B_0(\tau)-\gamma[\overline V_D(\tau)-\operatorname{Var}(\tilde v\mid\tilde v<\tau)]=B_0(\tau)+\gamma\sigma^2B_0'(\tau)$, which is main-text equation (8).

Two further pieces of Section V.5 are collected here. First, as $\gamma\rightarrow\infty$, $c_0(\gamma)\rightarrow-\infty$; the proof is given below with the proof of Proposition~8. The limit should be read with the distinction between roots and equilibria in mind. It says that the indifference equation has solutions at every positive cost, not that those solutions are equilibria, because the bound $\gamma<\overline\gamma$ eventually fails. By part (i) of Proposition~8, $c_0(\gamma)$ falls strictly in $\gamma$ throughout. The infimum defining $\overline\gamma$ is close to $1.647$, so the admissible range of risk aversion scales with $\sqrt{\sigma^2+s^2}/(\sigma s)$ and widens as either the fundamental or the processing noise becomes small.

Second, the residual variance $\mathcal V(\tau)$ decomposes as
\[
\mathcal V(\tau)=
\Phi\!\left(\frac{\tau}{\sigma}\right)
\operatorname{Var}(\tilde v\mid \tilde v<\tau)
+
\left(1-\Phi\!\left(\frac{\tau}{\sigma}\right)\right)
\E\!\left[\operatorname{Var}(\tilde v\mid \tilde x,\tilde v\geq\tau)\mid \tilde v\geq\tau\right].
\]

The firm's side follows from the same decomposition. Averaging the risk-adjusted pricing rule in Section V.5 over public information gives $\E[P_\gamma(I_\tau)]=\E(\tilde v)-\gamma\mathcal V(\tau)$, so the firm's ex-ante payoff is
\[
\E(\tilde v)-\gamma\mathcal V(\tau)
-c\left(1-\Phi\!\left(\frac{\tau}{\sigma}\right)\right).
\]
At $\gamma=0$ this is the baseline ranking, where the firm prefers the upper threshold simply because it discloses less often and so pays less often. Once investors are risk-averse the firm also absorbs the price discount that residual uncertainty creates, and disclosure frequency no longer settles the comparison.

Investor surplus should not, however, be interpreted as aggregate welfare. If $c$ is paid by the firm and is not a transfer to investors, adding the firm's payoff to investors' conditional certainty-equivalent payoff cancels the market price and gives ex-ante total surplus
\vspace{-0.2cm}
\begin{equation}\label{eq:aggregate-welfare}
W(\tau;c)
=\E(\tilde v)-\frac{\gamma}{2}\mathcal V(\tau)
-c\left(1-\Phi\!\left(\frac{\tau}{\sigma}\right)\right).
\vspace{-0.2cm}
\end{equation}
The normalization $\E(\tilde v)=0$ lets us drop the first term when comparing equilibria. We call this total surplus rather than economy-wide welfare, because a proprietary cost may affect parties outside the model. Unlike investor surplus, total surplus decreases with residual risk. Formally, conditional on $I_\tau$ the firm receives $P_\gamma(I_\tau)-cd$ and investors' conditional certainty-equivalent payoff is $\E[\tilde v\mid I_\tau]-P_\gamma(I_\tau)-\frac{\gamma}{2}\operatorname{Var}(\tilde v\mid I_\tau)$; summing cancels the price, and taking expectations with $\Pr(d=1)=1-\Phi(\tau/\sigma)$ gives~\eqref{eq:aggregate-welfare}. Whether the disclosure payment is also a cost depends on who receives it. If $c$ is a fee collected by a certifier included in welfare, the payment is a transfer, the last term of~\eqref{eq:aggregate-welfare} disappears, and total surplus reduces to $-\gamma\mathcal V(\tau)/2$, which ranks the equilibria exactly against investors whenever residual variances differ.

The bound used in Proposition~8 is $\overline\gamma\equiv\frac{\sqrt{\sigma^2+s^2}}{\sigma s}\inf_{z\in\mathbb R}\frac{1-\lambda'(z)}{\lambda''(z)}$. The second-order change in the unravelling bound is $c_0(\gamma)=c_0-\tfrac12\gamma^2\sigma^4B_0''(\tau_0)+o(\gamma^2)$. Both expressions are derived below.

\paragraph{Proof of Proposition 8 and of the large-$\gamma$ limit.}
By main-text equation (8), $B_\gamma=B_0+\gamma\sigma^2B_0'$. The proof of Proposition~1 shows that $B_0$ is smooth and strictly convex with a unique minimizer $\tau_0$ and diverges in both tails, and (\ref{eq:variance-slope}) shows that $B_0'$ is bounded, with $B_0'(\tau)\to-\rho^2$ as $\tau\to-\infty$ and $B_0'(\tau)\to1$ as $\tau\to\infty$. Hence $B_\gamma$ diverges in both tails and attains a finite minimum $c_0(\gamma)$, with $c_0(0)=c_0$.

For part (i), $B_\gamma(\tau_0)=c_0$ and $B_\gamma'(\tau_0)=\gamma\sigma^2B_0''(\tau_0)>0$ for $\gamma>0$, so $c_0(\gamma)<c_0$. At any minimizer $\tau_\gamma$ of $B_\gamma$, $B_0'(\tau_\gamma)=-\gamma\sigma^2B_0''(\tau_\gamma)<0$, so $\tau_\gamma<\tau_0$ by strict convexity of $B_0$. For $0<\gamma_1<\gamma_2$, $c_0(\gamma_2)\leq B_{\gamma_2}(\tau_{\gamma_1})=c_0(\gamma_1)+(\gamma_2-\gamma_1)\sigma^2B_0'(\tau_{\gamma_1})<c_0(\gamma_1)$, which together with $c_0(\gamma)<c_0(0)$ proves strict monotonicity.

For part (ii), if $c<c_0(\gamma)$ then $B_\gamma>c$ everywhere, and if $c>c_0(\gamma)$ then continuity and the tails give a root on each side of a minimizer. For a conjectured threshold $\tau$, set $z=(\tau-a(s)x)/b(s)$. Combining main-text equation~(3), the risk-adjusted pricing rule in Section V.5, and~\eqref{eq:ra-variance} gives
\[
P_\gamma(x;\tau)
=a(s)x+b(s)\lambda(z)-\gamma b(s)^2\bigl[1-\lambda'(z)\bigr],
\]
whose derivative in $x$ is $a(s)[1-\lambda'(z)-\gamma b(s)\lambda''(z)]$. For the normal inverse Mills ratio, $0<\lambda'(z)<1$ and $\lambda''(z)>0$, so the derivative is strictly positive for every $x$ and $\tau$ whenever $\gamma b(s)<\inf_{z}(1-\lambda'(z))/\lambda''(z)$, which is $\gamma<\overline\gamma$. Under this condition $\E[P_\gamma(v+\tilde\varepsilon;\tau)\mid v]$ is strictly increasing in $v$, so if the type $v=\tau$ is indifferent, every $v>\tau$ strictly prefers disclosure and every $v<\tau$ strictly prefers non-disclosure, and every root is a threshold equilibrium. The infimum, about $1.647$, is obtained by minimizing $(1-\lambda'(z))/\lambda''(z)$ over $z$.

For part (iii), fix $c>c_0$. The risk-neutral roots satisfy $B_0'(\underline\tau(c))<0<B_0'(\overline\tau(c))$, so the implicit-function theorem applied to $B_0(\tau)+\gamma\sigma^2B_0'(\tau)=c$ gives a root near each with $\partial\tau/\partial\gamma|_{\gamma=0}=-\sigma^2B_0'(\tau)/B_0'(\tau)=-\sigma^2$. There are no other roots for $\gamma$ small, because outside a compact interval containing both risk-neutral roots in its interior $B_0\geq c+\delta$ for some $\delta>0$ while $|\gamma\sigma^2B_0'|\leq\gamma\sigma^2$, and inside it $B_\gamma\to B_0$ and $B_\gamma'\to B_0'$ uniformly, so $B_\gamma-c$ has exactly one zero near each risk-neutral root and none elsewhere. The same argument applied to $B_\gamma'=B_0'+\gamma\sigma^2B_0''$ at $\tau_0$, where $B_0''(\tau_0)>0$, gives a unique minimizer $\tau_0(\gamma)$ near $\tau_0$ for $\gamma$ small, with $\tau_0'(0)=-\sigma^2$; by the envelope theorem $c_0'(\gamma)=\sigma^2B_0'(\tau_0(\gamma))$, so $c_0'(0)=0$ and $c_0''(0)=\sigma^2B_0''(\tau_0)\tau_0'(0)=-\sigma^4B_0''(\tau_0)$, which is the expansion of $c_0(\gamma)$.

For the large-$\gamma$ limit, $B_0'(\tau)\to-\rho^2<0$ as $\tau\to-\infty$, so at a fixed sufficiently negative $\tau'$, $B_\gamma(\tau')=B_0(\tau')+\gamma\sigma^2B_0'(\tau')\to-\infty$ as $\gamma\to\infty$, and $c_0(\gamma)\leq B_\gamma(\tau')\to-\infty$. \qed

\paragraph{Proof of Proposition 9.}
Fix any realization of the public information $I_\tau$ and write its posterior mean and variance as $m$ and $V$. An investor solves
\[
\max_q\left\{q(m-P)-\frac{\gamma}{2}q^2V\right\}.
\]
Because all investors have the same posterior and the asset is in unit net supply, market clearing gives $q=1$. The risk-adjusted pricing rule in Section V.5 then implies $m-P=\gamma V$, so the investor's conditional certainty-equivalent trading surplus is
\[
\gamma V-\frac{\gamma}{2}V=\frac{\gamma}{2}V.
\]
Taking expectations over $I_\tau$ gives the expected conditional certainty-equivalent trading surplus in main-text Proposition~9. Hence the ranking of any two threshold equilibria is exactly the ranking of their expected residual posterior variances. \qed

\paragraph{Proof of Proposition 10.}
Write $z\equiv\tau/\sigma$, $\kappa\equiv\gamma\sigma$ and $g(z)\equiv\lambda(z)+\lambda(-z)$. With $s=\infty$ the report carries no information, so $\E[P_\gamma(\tilde x;\tau)\mid\tilde v=\tau]=\E[\tilde v\mid\tilde v\geq\tau]-\gamma\operatorname{Var}(\tilde v\mid\tilde v\geq\tau)$. By main-text equation (6), $\E[\tilde v\mid\tilde v\geq\tau]-\E[\tilde v\mid\tilde v<\tau]=\sigma g(z)$, and $\operatorname{Var}(\tilde v\mid\tilde v\geq\tau)-\operatorname{Var}(\tilde v\mid\tilde v<\tau)=\sigma^2[\lambda'(-z)-\lambda'(z)]=-\sigma^2g'(z)$, so
\[
B_\gamma(\tau)=\sigma\,[\,g(z)+\kappa\,g'(z)\,].
\]
Because $\lambda$ is strictly convex with $\lambda'\in(0,1)$, $g$ is symmetric and strictly convex, $g'$ is odd with $g'(z)>0$ for $z>0$, and $|g'|<1$. Hence $c_0(\gamma)\geq\sigma[g(0)-\kappa]=4\sigma/\sqrt{2\pi}-\gamma\sigma^2>0$, and $c_0(\gamma)<B_\gamma(0)=c_0$ because $B_\gamma'(0)=\kappa g''(0)>0$. Next, $\E[\tilde v\mid d=1]=\sigma\lambda(z)$ with probability $1-\Phi(z)$ and $\E[\tilde v\mid d=0]=-\sigma\lambda(-z)$ with probability $\Phi(z)$, so
\[
\mathcal V(\tau)=\sigma^2-\operatorname{Var}\bigl(\E[\tilde v\mid d(\tilde v)]\bigr)=\sigma^2\,[\,1-\phi(z)g(z)\,],
\]
where $\phi(z)g(z)=\phi(z)^2/[\Phi(z)(1-\Phi(z))]$ is symmetric with logarithmic derivative $-2z+\lambda(z)-\lambda(-z)$, which is negative for $z>0$ because $\lambda(z)-z$ decreases and $\lambda(-z)+z$ increases in $z$, both equal to $\lambda(0)$ at zero. Hence $\mathcal V$ is strictly increasing in $|z|$.

Now fix $c>c_0(\gamma)$. Since $B_\gamma\to\infty$ as $|z|\to\infty$, the roots of $B_\gamma(\tau)=c$ form a nonempty compact set; let $\underline\tau$ be the smallest, so that $B_\gamma>c$ on $(-\infty,\underline\tau)$, and let $\tau_j$ be any other root. If $\tau_j>0$, then $B_\gamma(-\tau_j)-B_\gamma(\tau_j)=-2\kappa\sigma g'(\tau_j/\sigma)<0$ gives $B_\gamma(-\tau_j)<c$ and therefore $-\tau_j>\underline\tau$; if $\tau_j\leq0$, then $\underline\tau<\tau_j\leq0$. In both cases $|\underline\tau|>|\tau_j|$, so $\mathcal V(\underline\tau)>\mathcal V(\tau_j)$ and investors prefer $\underline\tau$ by Proposition~9. Averaging the risk-adjusted pricing rule in Section V.5 over public information, the firm's ex-ante payoff is $-\gamma\mathcal V(\tau)-c(1-\Phi(\tau/\sigma))$, and total surplus is (\ref{eq:aggregate-welfare}); both are lower at $\underline\tau$ than at $\tau_j$ because $\mathcal V(\underline\tau)>\mathcal V(\tau_j)$ and $\Phi(\underline\tau/\sigma)<\Phi(\tau_j/\sigma)$.

Finally let $s<\infty$ and $\gamma<\overline\gamma$. From the price derivatives in the proof of main-text Proposition 1, $B_0'(\tau)\to-\rho^2<0$ and $B_0''(\tau)\to 0$ as $\tau\to-\infty$, while $B_0'(\tau)\to 1$ and $B_0''(\tau)\to 0$ as $\tau\to\infty$, so $B_\gamma'=B_0'+\gamma\sigma^2 B_0''$ is negative on $(-\infty,-R]$ and positive on $[R,\infty)$ for some $R>0$. Since $B_\gamma$ is bounded on $[-R,R]$, for every sufficiently large $c$ the equation $B_\gamma(\tau)=c$ has exactly one root in each tail and none in between, so there are exactly two threshold equilibria, with $\underline\tau(c)\to-\infty$ and $\overline\tau(c)\to\infty$. As $\tau\to-\infty$, $\operatorname{Var}(\tilde v\mid\tilde v<\tau)\to0$ and, by (\ref{eq:ra-variance}) and dominated convergence, $\E[\operatorname{Var}(\tilde v\mid\tilde x,\tilde v\geq\tau)\mid\tilde v\geq\tau]\to b(s)^2$; as $\tau\to\infty$, $\operatorname{Var}(\tilde v\mid\tilde v<\tau)\to\sigma^2$ and $\E[\operatorname{Var}(\tilde v\mid\tilde x,\tilde v\geq\tau)\mid\tilde v\geq\tau]\leq\operatorname{Var}(\tilde v\mid\tilde v\geq\tau)\to0$. Hence $\mathcal V(\underline\tau(c))\to b(s)^2$ and $\mathcal V(\overline\tau(c))\to\sigma^2$, with $b(s)^2<\sigma^2$, so investors prefer $\overline\tau(c)$ for $c$ large, and since $\Phi(\overline\tau(c)/\sigma)-\Phi(\underline\tau(c)/\sigma)\to1$, the firm's payoff and total surplus are also lower at $\underline\tau(c)$ than at $\overline\tau(c)$ for $c$ large. \qed

\begingroup\singlespacing

\par\endgroup

\clearpage
\begingroup\singlespacing
\bibliographystyle{cje}
\bibliography{general5}
\par\endgroup
\end{document}